%% file: ms.tex
\documentclass[modern]{aastex62}

\input{cortex}

\input{style}

\begin{document}

\title{%
    \textbf{
        Analytic Light Curves in Reflected Light:%
        \\%
        Phase Curves, Occultations, and Non-Lambertian Scattering for
        Spherical Planets and Moons
    }
}

\author[0000-0002-0296-3826]{Rodrigo Luger}\altaffiliation{Flatiron Fellow}
\email{rluger@flatironinstitute.org}
\affil{Center~for~Computational~Astrophysics, Flatiron~Institute, New~York, NY}
\affil{Virtual~Planetary~Laboratory, University~of~Washington, Seattle, WA}
\author[0000-0002-0802-9145]{Eric Agol}
\affil{Department~of~Astronomy, University~of~Washington, Seattle, WA}
\affil{Virtual~Planetary~Laboratory, University~of~Washington, Seattle, WA}
\author[0000-0001-8630-9794]{Fran Bartoli\'c}
\affil{Centre for Exoplanet Science, University of St. Andrews, St. Andrews, UK}
\affil{Center~for~Computational~Astrophysics, Flatiron~Institute, New~York, NY}
\author[0000-0002-9328-5652]{Daniel Foreman-Mackey}
\affil{Center~for~Computational~Astrophysics, Flatiron~Institute, New~York, NY}


\begin{abstract}
    We derive efficient, closed form, differentiable, and numerically stable
    solutions for the flux measured from a spherical planet or moon seen in reflected
    light, either in or out of occultation.
    Our expressions apply to the computation of scattered
    light phase curves of exoplanets, secondary eclipse light curves in
    the optical, or future measurements of planet-moon and planet-planet
    occultations, as well as to photometry of solar system bodies.
    We derive our solutions for Lambertian bodies illuminated by a point
    source, but extend them to model illumination sources of finite
    angular size and rough surfaces with phase-dependent scattering.
    Our algorithm is implemented in \Python within the open-source
    \starry mapping framework and is designed with efficient gradient-based
    inference in mind.
    The algorithm is ${\sim}4-5$ orders of magnitude faster than direct
    numerical evaluation methods and ${\sim}10$ orders of magnitude more
    precise.
    We show how the techniques developed here may one day lead to the
    construction of two-dimensional maps of terrestrial planet surfaces,
    potentially enabling the detection of continents and oceans on
    exoplanets in the habitable zone.
    \href{https://github.com/rodluger/starrynight}{\color{linkcolor}\faGithub}
\end{abstract}

\section{Introduction}
\label{sec:intro}

Despite recent advances in instrumentation and the dawn of
thirty meter-class telescopes and kilometer-wide interferomer arrays,
extrasolar planets will remain unresolved point sources for decades
to come. Nevertheless, modulations in the light received from these distant
bodies due to their rotation, changing illumination, and eclipses
by their host stars or other bodies in the system can be harnessed to
reconstruct two-dimensional views of their surfaces.
In particular, next-generation space-based telescopes such as the
Large UV/Optical/IR Surveyor (LUVOIR) may enable us to measure
variations in the reflected light signature of terrestrial planets in
the habitable zone, which can be used to map their surfaces and
indirectly infer the presence of clouds, continents, oceans, and perhaps
even life.

There is an extensive literature on techniques for mapping exoplanet surfaces
based on their phase curves
\citep[e.g.,][]{Russell1906,Lacis1972,Knutson2007,Cowan2008,Oakley2009,Berdyugina2017,PaperI,PaperII,Heng2021}
and occultation light curves
\citep[e.g.,][]{Williams2006,Rauscher2007,Majeau2012,deWit2012,Rauscher2018},
both in thermal and reflected (scattered) light.
In particular, much attention has been given to techniques for mapping
Earth-like planets from visible-light reflected phase curves
\citep[e.g.,][]{Ford2001,Kawahara2010,Kawahara2011,Fujii2012,Kawahara2020,Aizawa2020}.
Unlike thermal phase curves, which primarily encode (often degenerate)
information about longitudinal surface brightness variations
\citep{Russell1906}, reflected light curves often contain information
about the full two-dimensional surface albedo distribution
\citep[e.g.,][]{Kawahara2010}.

Occultation light curves in reflected light can encode even more information
about the surface. Thus far, these have been studied primarily within our
solar system. Mutual occultations among the Galilean moons of Jupiter
have been extensively studied to infer surface properties of the moons and to refine
their ephemerides
\citep[e.g.,][]{Arlot1974,Aksnes1984,Arlot2014,deKleer2017,Saquet2018,Morgado2019,Bartolic2021}.
Farther out in the solar system,
mutual occultations of Pluto and Charon in the late 1980s
were used to confirm Charon's existence
\citep{Stern1992},
establish the sizes and orbital parameters of the two bodies
\citep{Tholen1990},
and infer their surface properties
\citep{Marcialis1990}.
In particular, \citet{Dunbar1986} developed an efficient analytic
algorithm to model Pluto-Charon occultation light curves in reflected light
assuming uniform surfaces and used it to infer the two body's
average geometrical albedos. Later,
\citet{Buie1992} used a maximum entropy approach to reconstruct two-dimensional
maps of the two bodies and
\citet{Reinsch1994} analyzed
the complete mutual occultation dataset to infer longitudinal maps of Pluto's
albedo.

Many studies have analyzed real Earth reflected light curves to
infer surface properties of our planet as an exercise in
prepartion for the mapping of exoplanets.
\citet{Cowan2009} and \citet{Cowan2011b} analyzed visible-light disk-integrated light curves
of the Earth taken by the Deep Impact spacecraft to produce
longitudinal maps of the surface, harnessing multi-band observations
to disentangle static surface brightness features from temporally
variable clouds.  A transit of Earth by the Moon was observed
as well \citep{Livengood2011}, although this data has yet to be
exploited for mapping purposes.
More recently, \citet{Jiang2018} and \citet{Fan2019}
used data from the L1-stationed DSCOVR satellite to infer surface
and cloud properties of the Earth, and \citet{Luger2019b} analyzed
background scattered light in TESS photometry to reconstruct a cloud
map of the Earth.

On the open-source software front, \citet{Louden2018} developed
\textsf{spiderman}, an efficient discretization scheme
on the sphere that enables fast computation of exoplanet
phase curves and occultation light curves in \Python.
\citet{Haggard2018} presented
\textsf{EARL} (Exoplanet Analytic Reflected Lightcurves),
a \textsf{Mathematica} code to compute analytic, closed form
solutions for the phase curve of a Lambert sphere, i.e., one that
scatters light isotropically, in the case that the surface albedo
distribution is characterized by either a sum of delta functions
or a sum of spherical harmonics.
\citet{Farr2018} released \textsf{exocartographer}, a Bayesian
framework for doing inference on exoplanet phase curves based on
a HEALPix \citep{healpix} discretization scheme.
Finally, \citet{Luger2019} introduced \starry, a light curve
modeling package that computes thermal phase curves and occultation
light curves, as well as their derivatives, analytically from a spherical
harmonic expansion of the surface brightness.

The present paper is an extension to the \starry algorithm,
adapting it to model phase curves and occultation light curves
in reflected light. The expressions
we derive are analytic: they may all be expressed in closed form
in terms of algebraic operations involving trigonometric functions
and (at times) elliptic integrals. We derive numerically
stable recursion relations for the efficient evaluation of all
expressions and code them within an autodifferentiation framework
to enable the computation of accurate derivatives for use in
gradient-based inference and optimization schemes.
Our code is fully
\href{https://github.com/rodluger/starry}{open-source},
comprehensively
\href{https://dev.azure.com/rodluger/starry/_test/analytics?definitionId=4}{unit-tested},
and supplemented with an extensive
\href{https://starry.readthedocs.io}{API documentation}
and suite of
\href{https://starry.readthedocs.io/en/latest/tutorials/}{tutorials}.
As in all papers in the \starry series, in the caption of each of the figures
we provide links \codeicon\ to the exact \Python scripts that
generated them. Next to many of the equations we also provide
links \prooficon\ to \textsf{Jupyter} notebooks containing
detailed derivations and/or validations.

\section{Overview}
\label{sec:overview}

Our goal in this paper is to derive analytic expressions for the
flux received by a distant observer from a sphere of non-uniform albedo
illuminated by a monochromatic source that may or may not be
occulted by a (possibly different) spherical body. This applies, for example,
to the case of planetary phase curves, secondary
eclipse (occultation) light curves, and moon-moon, planet-moon, and
planet-planet occultations (in the Solar System or not),
all seen in reflected light. We derive all expressions in
the limit that the reflecting
body is Lambertian, i.e., it scatters light isotropically, but we relax
this assumption in later sections.
We model the general case of an
intensity that varies across the surface of the body according to a
spatially-dependent
albedo $A$. Throughout this paper, we will take $A$ to mean the \emph{spherical}
albedo, the fraction of power incident on a body at a given wavelength
that is scattered back out to space (in all directions).
Note that the spherical albedo
is closely related to the Bond albedo: the Bond albedo is the stellar
flux-weighted integral of $A(\lambda)$ over all wavelengths $\lambda$
\citep[see, e.g.,][]{Seager2010}.

As in \citet{Luger2019}, we compute fluxes by first expanding the surface
in terms of spherical harmonics.
While in \citet{Luger2019} we expanded the emissivity of the surface, here
we instead expand the spherical albedo $A$.
Specifically, if $\bvec{y}$ is the vector
of spherical harmonic coefficients describing the albedo anywhere on
the surface and
$\by$ is the spherical harmonics basis (Equation~\ref{eq:by}),
the albedo $A$ at a point $(x, y)$ on the sky-projected disk of the body
is given by the dot product
\begin{align}
    \label{eq:albedo}
    A(x, y) = \by^\top (x, y) \, \bvec{y}
    \quad.
\end{align}
The flux measured from this body is proportional to the surface integral
over the projected disk
of the albedo $A$ times the illumination profile $\mathcal{I}$ of the surface,
given by Lambert's law as
\begin{align}
    \label{eq:LambertsLaw}
    \mathcal{I}(\vartheta_\mathrm{i}) = \mathcal{I}_0 \, \text{max}\big( 0, \cos\vartheta_\mathrm{i} \big)
    \quad,
\end{align}
where $\vartheta_\mathrm{i}$ is the angle between the incident radiation and the surface
normal and $\mathcal{I}_0$ is the peak illumination. We show in
Appendix~\ref{sec:adapting-starry} that in the case of, say, a planet
illuminated by its host star,
\begin{align}
    \mathcal{I}_0 = \frac{f_s}{\pi r_\mathrm{s}^2}
\end{align}
where $r_\mathrm{s}$ is the distance between the planet and the star (in units
of the planet's radius) and $f_s$ is the stellar flux measured at the observer
(in arbitrary units). Following the convention in \citet{Luger2019}, we assume
throughout this paper that $f_s = 1$, so all fluxes are defined as a fraction of
the flux of the illumination source at the observer.

The piecewise nature of the illumination function at the day/night terminator
makes the problem of computing the visible flux particularly difficult; most
studies to date have tackled the problem numerically, either
via Monte Carlo integration \citep[e.g.,][]{Ford2001} or by discretizing the surface
and computing the relevant integrals by summing over the visible
pixels \citep[e.g.,][]{Kawahara2010,Fujii2012}.
Recently, \citet{Haggard2018} developed an analytic
framework for computing light curves of unocculted bodies illuminated
by a point source. In this paper, we re-derive their solution under the
\starry framework and extend it for the first time to the case where
the body is occulted by another spherical body, which may or may not be the
illumination source. We also extend the solution to the case of an extended
illumination source and to non-Lambertian scattering.

In \citet{Luger2019}, we reduced the problem of computing the flux from
an occulted body in thermal (emitted) light to a series of efficient, analytical
operations involving trigonometric functions of the position and size of the
occultor and certain complete elliptic integrals. In the case of reflected
light, however, the change in the limits of integration due to the
unilluminated night side breaks many of the symmetries that simplified
the flux calculation. In particular, the limits of integration now depend
on the solution to a quartic equation specifying the points of intersection
between the occultor and the day/night terminator, and the solution to
those integrals is now a function of \emph{incomplete} elliptic integrals.
The procedure for computing the flux is therefore significantly more complex.
We therefore defer all calculations to the Appendix, and devote the body of
the paper to validating and demonstrating applications of our approach.

This paper is organized as follows.
In \S\ref{sec:validation} we present sample light curves computed
using our algorithm, validate it against numerical integration, and
discuss its performance in terms of computational speed and precision.
In \S\ref{sec:extensions} we extend the model to apply to illumination
sources of finite size and surfaces that scatter light anisotropically.
We discuss implications, applications, and limitations of our model in
\S\ref{sec:discussion} and summarize our findings in \S\ref{sec:conclusions}.
For convenience, Tables~\ref{tab:symbols}--\ref{tab:matrices} at the end
list all symbols and variables used in the text, with descriptions and
links to the equations in which they are defined.

\begin{figure}[t!]
    \begin{centering}
        \includegraphics[width=\linewidth]{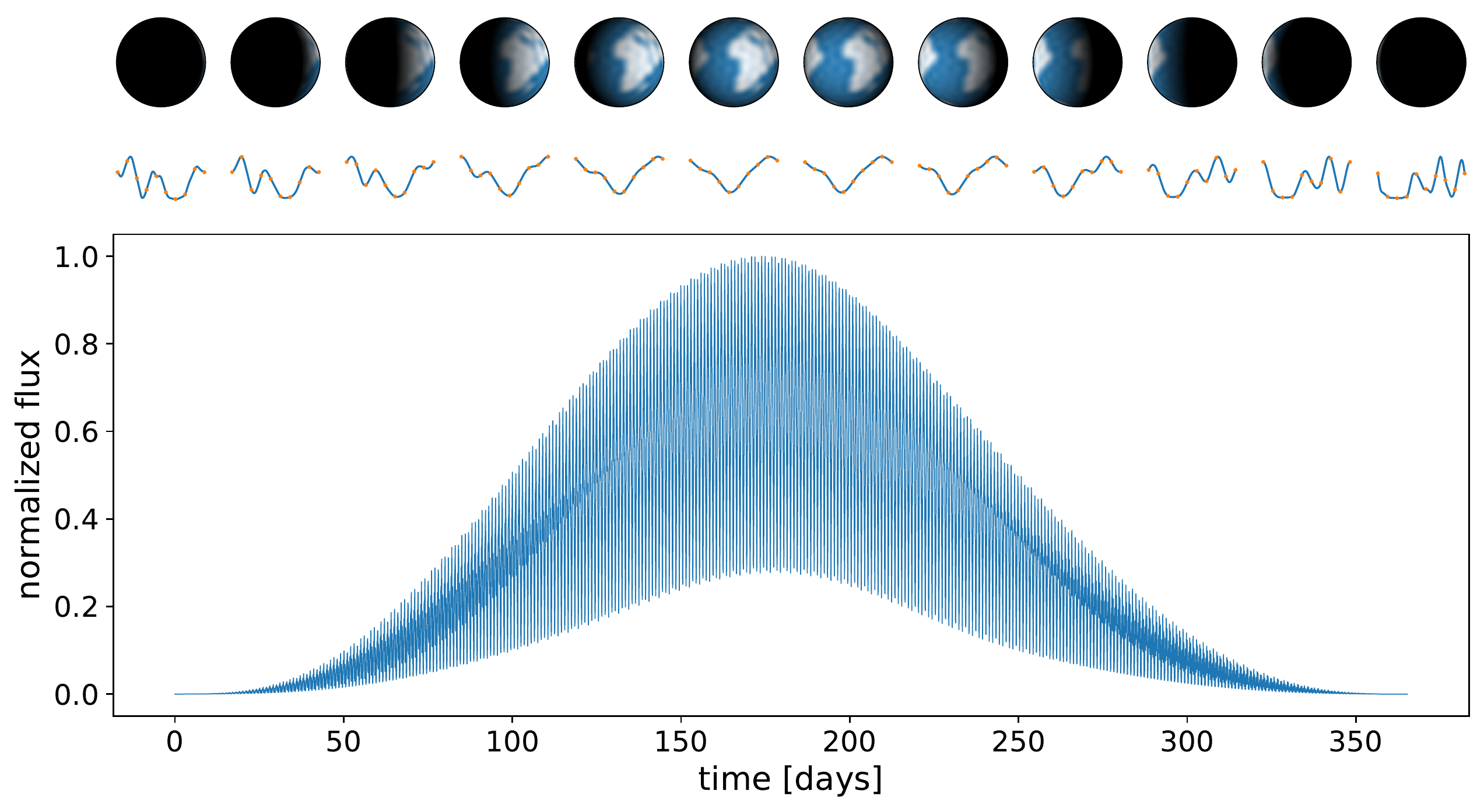}
        \oscaption{earthphase}{%
            Mock reflected light phase curve of the cloudless Earth expanded
            to spherical harmonic degree $l = 25$, viewed along the
            ecliptic. The main plot shows the phase curve over the course of
            one year.
            The images at the top show the corresponding progression of
            the phases of the Earth, from new phase to full phase and back
            to new phase. Below each image we show a normalized 24-hour
            segment of the light curve at that phase (blue).
            Orange dots correspond
            to the flux computed from brute force numerical integration on
            a grid of
            ${\sim}10^5$ points.
            \label{fig:earthphase}
        }
    \end{centering}
\end{figure}

\pagebreak

\section{Reflected light curves in starry}
\label{sec:validation}

\subsection{Sample light curves}
\label{sec:sample}

Figure~\ref{fig:earthphase} shows a sample application of the algorithm
developed in this paper: a reflected light phase curve of the
Earth over the course of one year. The model is computed using the
methodology in Appendix~\ref{sec:solution-no-occ} from an $l=25$
spherical harmonic expansion of the cloudless Earth, where the oceans
are given an albedo of zero and the continents an albedo of unity
(note, however, that since the light curve is normalized, the model does
not depend on the value of the latter).
The Earth is assumed to be a perfect Lambertian scatterer,
so effects like the phase dependence of Rayleigh scattering and
specular reflection (glint) from the oceans are neglected
(but see \S\ref{sec:extensions} for an extension of the model to
non-Lambertian scatterers).
The observer is assumed to be along
the ecliptic, so the illumination source is along the $x-z$ plane of a
right-handed Cartesian coordinate system, with $\hat{z}$ pointing toward the
observer and $\hat{x}$ pointing to the right on the sky. The axis of rotation
of the Earth is therefore tilted clockwise away from
$\hat{y}$ by $23.5^\circ$.
The images at the top show snapshots of the disk of the Earth throughout
the observation; below each one, we plot in blue the normalized phase curve
at that phase over a single rotation. The orange dots correspond to a
brute force numerical solution, obtained by discretizing the disk
on a grid of ${\sim}10^5$ points and summing over the dayside.
The models agree to within the
numerical precision of the brute force solution (about 100 ppm of the
planetary flux in this case).

While the dominant signal in the phase curve is the sine-like envelope
due to the changing phases of the Earth, the local behavior of the light
curve at each phase is complex and varies significantly over the course
of the year. Unlike phase curves in thermal light, which primarily encode
low-order spatial information (since the region of integration is always
the full disk), phase curves in reflected light encode information at
different scales depending on the phase. At crescent phase, the region of the
disk contributing to the total flux is a narrow lune; these measurements
therefore encode information primarily about high-$l$ modes. At full
phase, the region of integration is the full disk, so these measurements
encode information about low-$l$ modes. Furthermore, because of the obliquity
of the Earth,
the orientation of the crescent lune changes relative to features on the
surface over the course of one orbit, changing the relative contribution of
different portions of the surface to the flux and increasing the
overall information content of the observation.
As we will show in \S\ref{sec:information}, the information content of
reflected light phase curves is overwhelmingly higher than that of
phase curves in thermal light, particularly for planets with significant
obliquity.

\begin{figure}[t!]
    \begin{centering}
        \includegraphics[width=\linewidth]{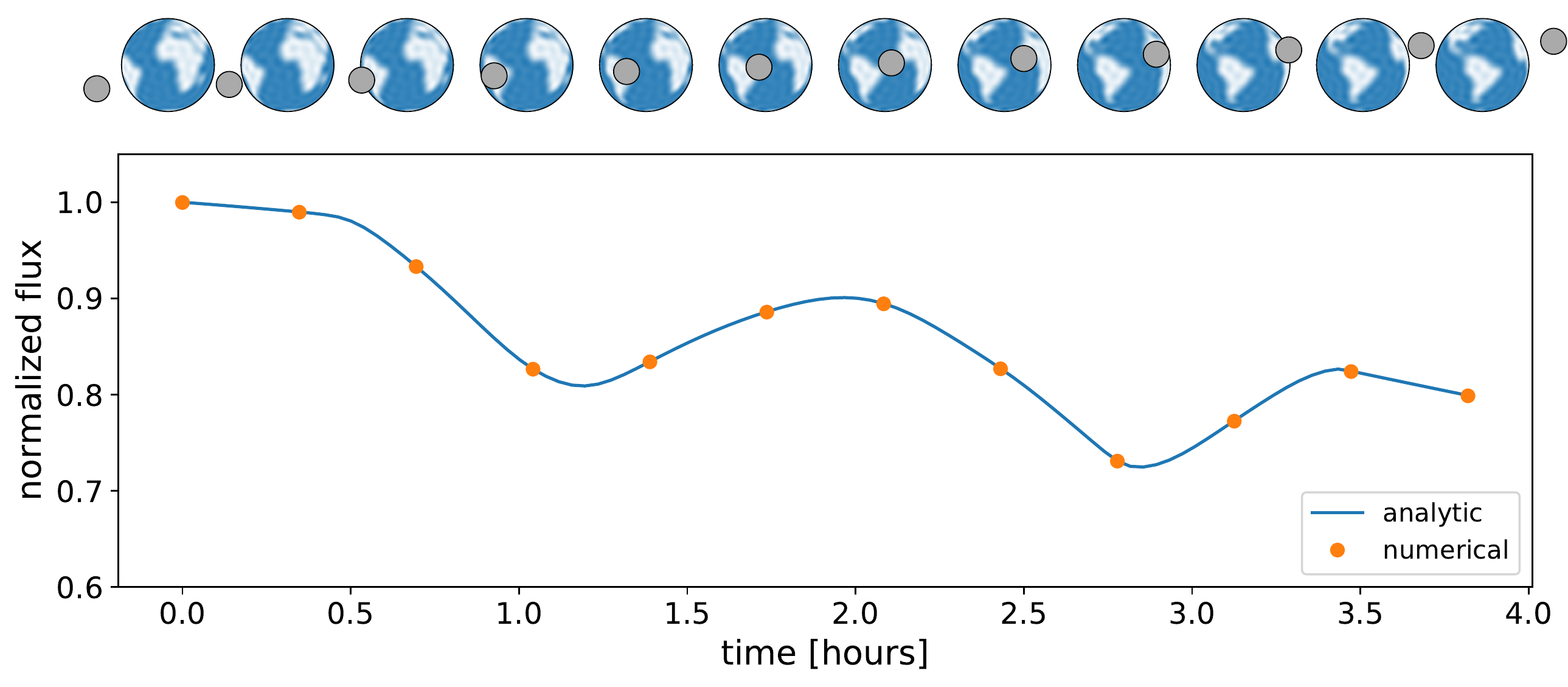}
        \includegraphics[width=\linewidth]{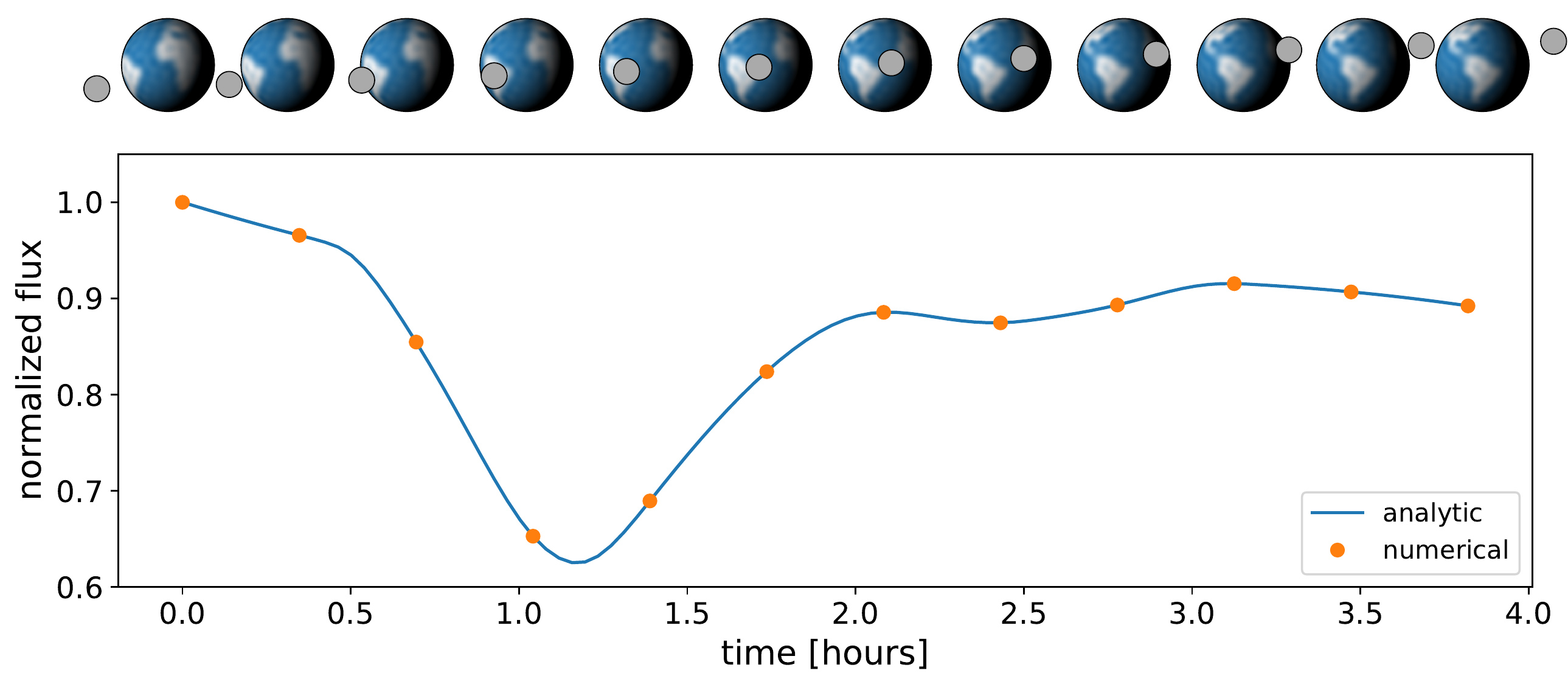}
        \oscaption{earthmoon}{%
            Mock light curves of the Moon occulting a rotating,
            cloudless Earth expanded to spherical harmonic degree $l = 25$.
            Black curves show the analytic solution; orange dots correspond
            to brute force numerical integration on a grid of
            ${\sim}10^5$ points.
            The top panel shows the light curve in emitted light and is
            the same as in Figure~7 in \citet{Luger2019}.
            The bottom panel (this work) shows the same light curve in
            reflected light during northern summer.
            \label{fig:earthmoon}
        }
    \end{centering}
\end{figure}

Figure~\ref{fig:earthmoon} shows another light curve of the rotating
Earth, but this time taken during an occultation by the Moon. The map
of the Earth is the same as before, but the observer is now along the
equatorial plane of the Earth.
The top panel is a reproduction of
Figure~7 in \citet{Luger2019} for the case of thermal light,
where the Moon is seen to travel across the
disk of the Earth from southwest to northeast, progressively occulting
South America (dip), the Atlantic (peak), and Africa (dip).
As before,
the blue curve is the analytic solution and the orange dots correspond
to the numerical solution.

The bottom panel of the figure shows a light curve for the
same occultation geometry, but seen instead in reflected light, with the
Sun to the top left and slightly out of the page, corresponding to
some point during northern summer. Note the same dip-peak-dip pattern,
albeit with significantly different amplitudes. In particular, the
transit across South America is deeper, since it occurs close to local
noon, when the illumination is highest; conversely, the transit across
Africa occurs close to local dusk, when the illumination is close to zero.
As before, the light curves computed using \starry agree to within the
numerical precision of the brute force solution.

\begin{figure}[p!]
    \begin{centering}
        \includegraphics[width=\linewidth]{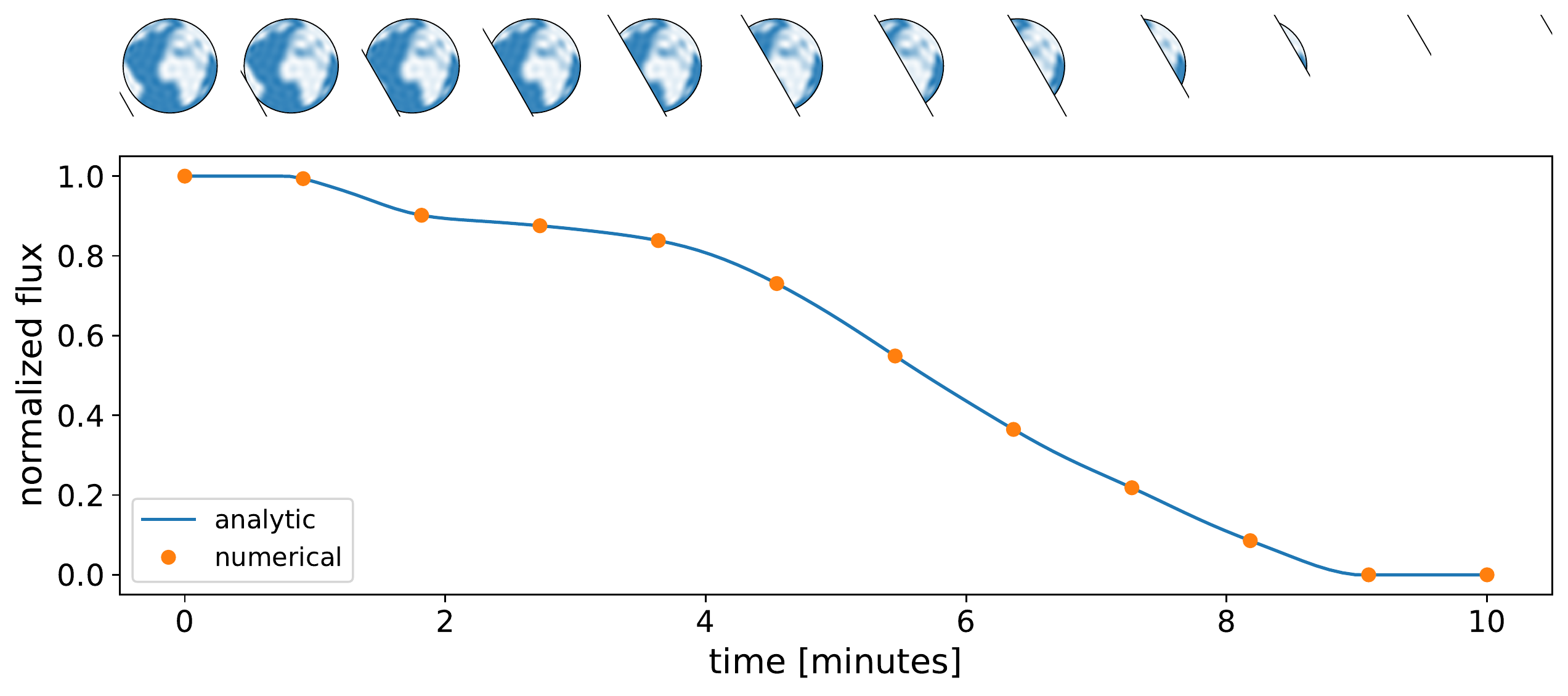}
        \includegraphics[width=\linewidth]{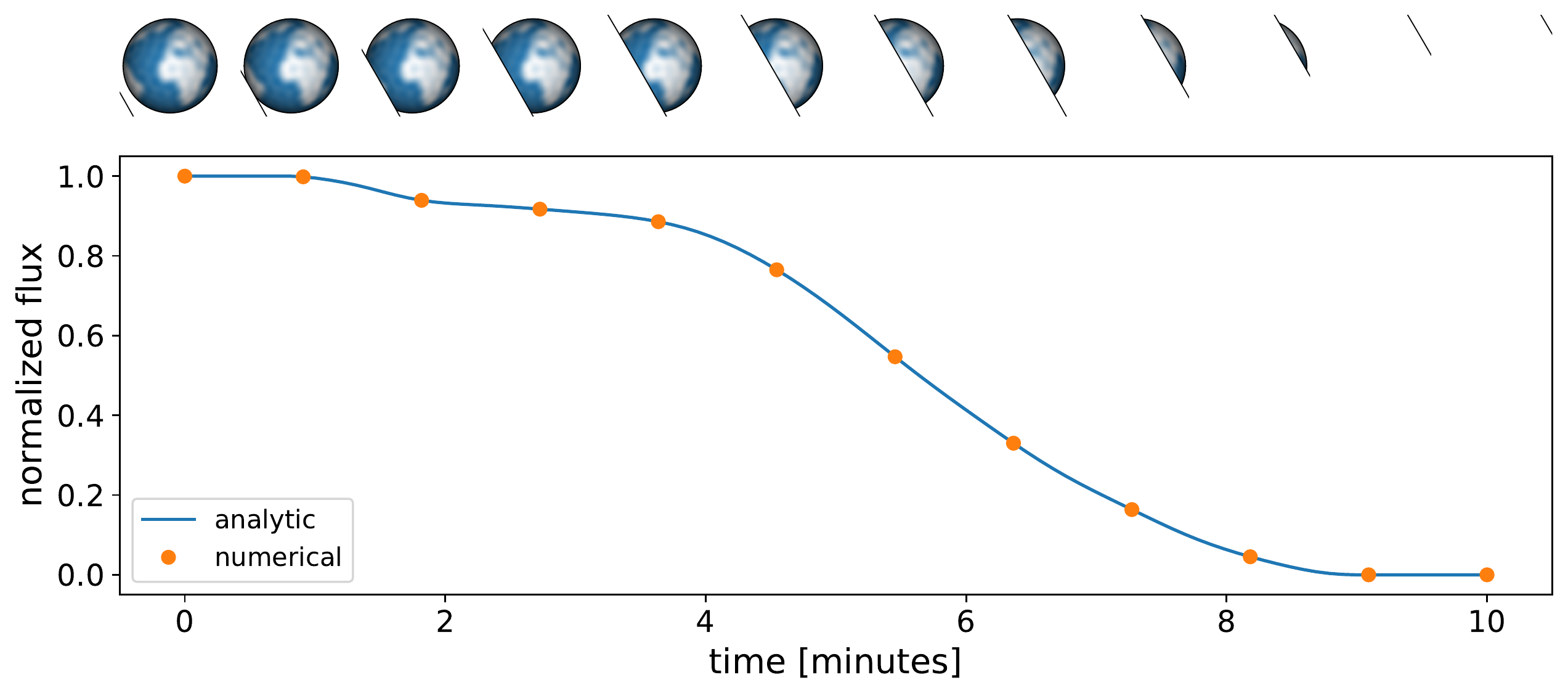}
        \includegraphics[width=\linewidth]{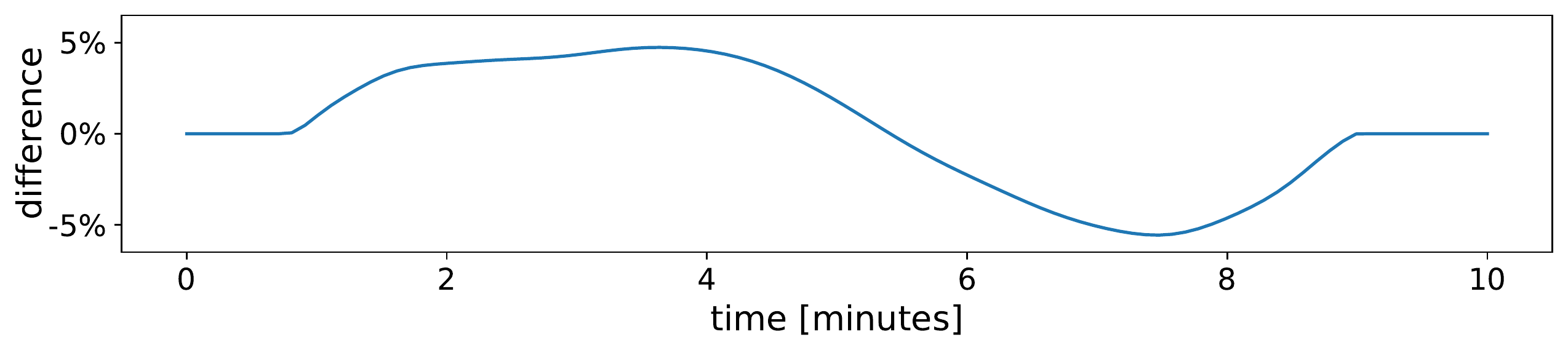}
        \oscaption{earthsun}{%
            Mock secondary eclipse ingress light curves of the
            cloudless Earth expanded to spherical harmonic degree $l = 25$,
            viewed from an orientation where the Earth is occulted behind
            a solar latitude of $30^\circ$.
            Black curves show the analytic solution; orange dots correspond
            to brute force numerical integration on a grid of
            ${\sim}10^5$ points.
            The top panel shows the light curve in emitted light and is
            similar to Figure~13 in \citet{Luger2019}.
            The middle panel (this work) shows the same light curve in
            reflected light.
            The bottom panel shows the difference between the
            normalized reflected and emitted light curves.
            \label{fig:earthsun}
        }
    \end{centering}
\end{figure}

Our last sample light curve is Figure~\ref{fig:earthsun}, which shows
a secondary eclipse light curve of the Earth as it is occulted by the
Sun. The model for the Earth is the same as above, and the observer is
now close to the ecliptic, but slightly misaligned so that the
Earth is occulted behind a solar latitude of $30^\circ$ (i.e., at
a solar impact parameter of $0.5$). The observation takes place at the
June solstice, so the Earth is tilted by $23.5^\circ$ out of the page.
As before, the top panel shows the light curve in thermal light; this
is similar to the top panel of Figure~13 in \citet{Luger2019}. The orange
dots again correspond to the numerical solution.
The center panel shows the same light curve in reflected light.
Because the observation occurs very close to full phase, the normalized
light curves look very similar to each other.
The bottom panel shows the difference between the two
(reflected minus thermal), which is only on the order of a few percent.
In fact, because
the illumination profile is proportional to the cosine of the viewing
angle, $\upmu$, and the reflection is assumed to be isotropic, the
(normalized) secondary eclipse light curve in reflected light is to
good approximation equal
to a limb-darkened thermal occultation light curve with
linear limb darkening coefficient $u_1 = 1$. As we will see later, for very
close-in planets this approximation breaks down, since the illumination
phases at secondary eclipse ingress and egress are sufficiently different
from full phase.

\subsection{Performance}
\label{sec:performance}

As we discuss in the Appendix, the model for phase curves and
occultation light curves in reflected light may be expressed
analytically in terms of purely algebraic and trigonometric functions
and in some cases incomplete elliptic integrals of the first, second,
and third kinds. We have derived efficient and numerically stable
recursion relations to compute the relevant expressions and their
derivatives. At times, these involve the evaluation of certain
expressions numerically, especially when doing so
leads to either a speed-up or a significant gain in numerical precision.
In particular, as we discuss in Appendix~\ref{sec:which-case}, the
integration boundaries during an occultation sometimes depend on the
solution to a quartic equation. While this can be solved in closed form,
the analytic solution can often be very unstable. We therefore solve
the quartic numerically, attaining a precision for the roots within a
few orders of magnitude of machine (double) precision.

\begin{figure}[p!]
    \begin{centering}
        \includegraphics[width=\linewidth]{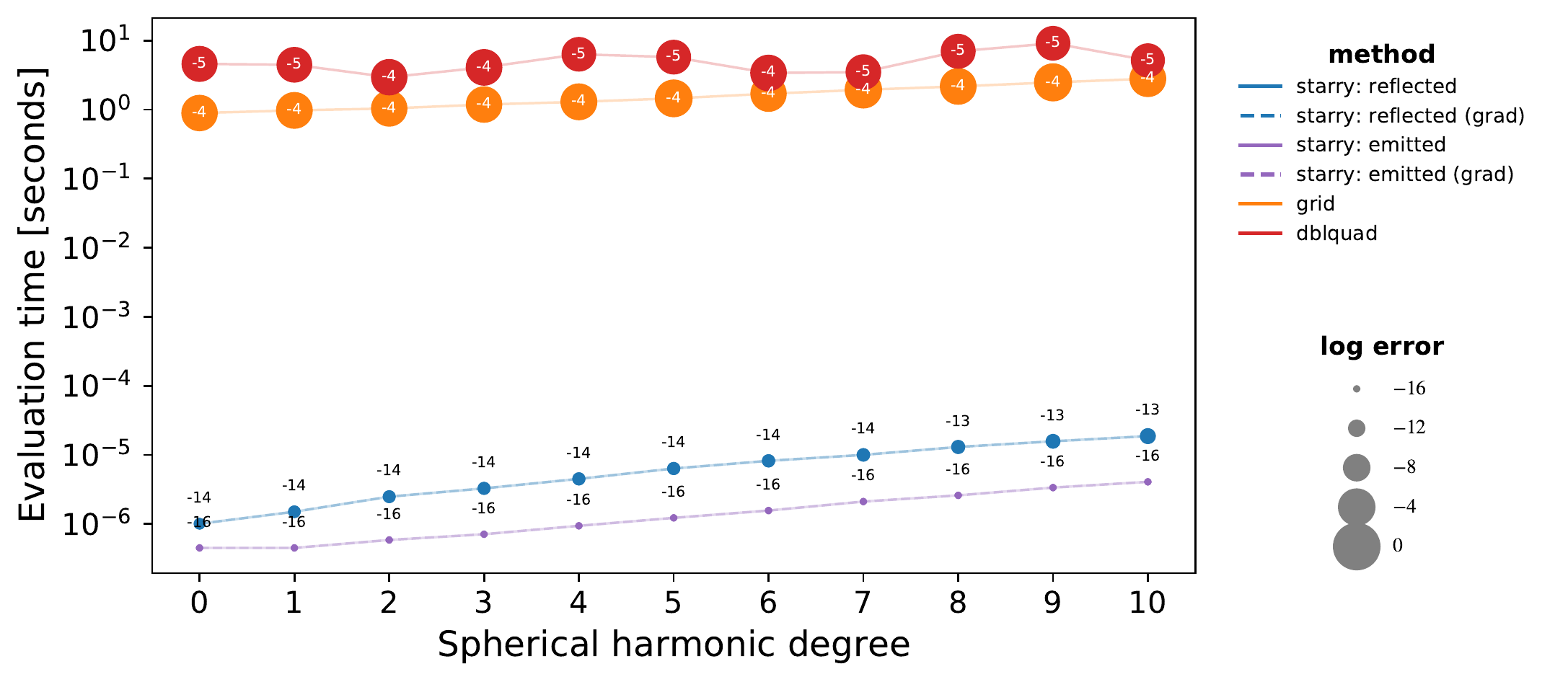}
        \oscaption{speed}{%
            Evaluation time (vertical axis)
            and numerical precision (point size)
            for a single flux evaluation in the absence of an occultor
            as a function of spherical
            harmonic degree for different methods.
            In purple we show results for the emitted light
            \starry algorithm from \citet{Luger2019} (solid: no gradient,
            dashed: with gradient), and in blue we show results for
            the reflected
            light algorithm from this paper (solid: no gradient,
            dashed: with gradient). For comparison, in we also show
            results for discrete integration on a grid (orange) and for
            numerical integration using two-dimensional Gaussian
            quadrature (red); neither of these include gradient
            evaluations. The reflected light algorithm is
            comparable in efficiency and precision to the emitted
            light algorithm. It is ${\sim}5$ orders of magnitude faster
            and ${\sim}10$ orders of magnitude more precise than numerical
            integration.
            \label{fig:speed_no_occ}
        }
    \end{centering}
\end{figure}

\begin{figure}[p!]
    \begin{centering}
        \includegraphics[width=\linewidth]{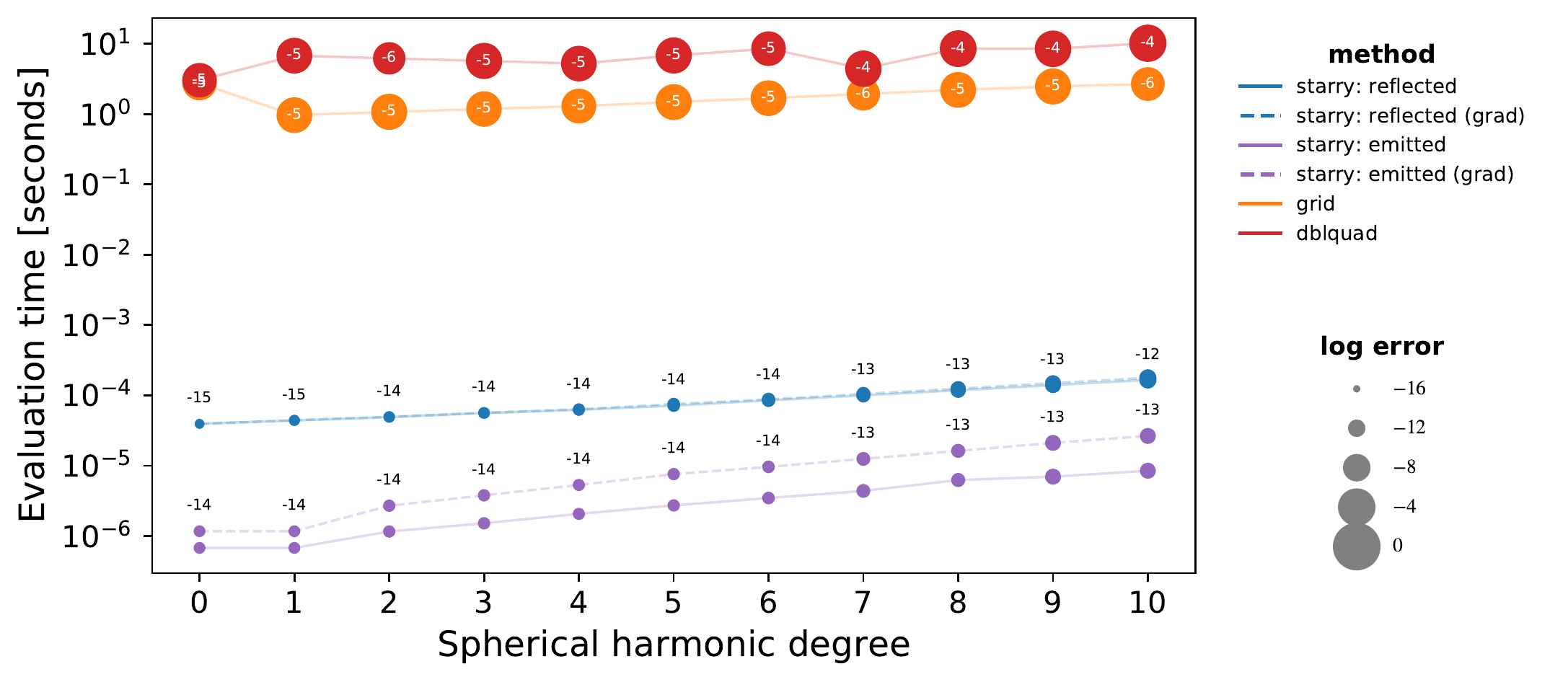}
        \oscaption{speed}{%
            Same as Figure~\ref{fig:speed_no_occ}, but for an occultation
            evaluation in which the occultor intersects the terminator
            (case 6 in Appendix~\ref{sec:solution-occ}).
            The reflected light algorithm is around one
            order of magnitude slower and comparably precise to the emitted
            light algorithm. It is ${\sim}4$ orders of magnitude faster
            and ${\sim}10$ orders of magnitude more precise than numerical
            integration.
            \label{fig:speed}
        }
    \end{centering}
\end{figure}

Figures~\ref{fig:speed_no_occ} and \ref{fig:speed} summarize
the precision and computation time of the \starry algorithm for
two typical scenarios: a phase curve evaluation
(Figure~\ref{fig:speed_no_occ}) and an occultation evaluation
(Figure~\ref{fig:speed}).
Blue points correspond to the reflected light algorithm developed
in this paper, while purple points correspond to the thermal light
algorithm from \citet{Luger2019} for the same occultation geometry,
but without an illumination source.
Solid and dashed lines correspond to evaluations without and
with gradient propagation, respectively (see \S\ref{sec:usage}
for details).
The orange and red dots
correspond to numerical evaluation of the flux:
brute force integration by summation on a grid of ${\sim}10^6$ points
(orange) and two-dimensional adaptive Gaussian quadrature using the
\textsf{dblquad} function in \textsf{scipy} \citep{scipy} with
both absolute and relative error tolerances set to $10^{-3}$ (red).
In both figures, the vertical axis corresponds to the evaluation
time in seconds for a single flux computation, while the size of the
points is proportional to the base-10 log of the relative error. For the
\starry solutions, the latter is estimated as the max-min difference
in the flux over one thousand evaluations in which the input parameters
are perturbed within an order of magnitude of machine epsilon; this is
therefore a probe of the condition number of the starry algorithm and
captures only error due to numerical instabilities.
It is worth emphasizing that this is a measurement of the precision of
the algorithm, rather than the accuracy, because it would be computationally
intractable to compute a solution more accurate than this using a different
algorithm.
We argue that this measurement can be interpreted to mean that the algorithm
is also accurate, but detailed quantification of this difference is beyond the
scope of this paper.
For the numerical
solutions, the error is estimated as the difference between the numerical
flux and the \starry flux.

For both phase curves and occultations, the \starry reflected
light algorithm is 1--2 orders of
magnitude slower than the emitted light algorithm, owing
primarily to the increased complexity of the reflected light model.
For phase curves (Figure~\ref{fig:speed_no_occ}),
the thermal solution vector $\rTe$ (Equation~\ref{eq:rTA1Ry}) is a constant
that can be pre-computed, while the analogous vector in the reflected
light case, $\rT$ (Equation~\ref{eq:rTsoln}), must be evaluated
recursively each time. For occultations (Figure~\ref{fig:speed}),
the slower evaluation in the reflected light case is primarily due to
the time spent solving the quartic equation for the points of intersection
between the occultor and the day/night terminator of the illuminated
body (Appendix~\ref{sec:solution-occ}). This contributes the
same overhead at all map degrees $l$, resulting in a gentler scaling
in $l$ than for the thermal case; for large $l$, the evaluation time
for the two algorithms is within a factor of 2--3.
In terms of precision, the algorithms are comparable, particularly
for occultations. For both phase curves and occultations, the numerical error up
to $l=10$ is less than one part per trillion ($10^{-12}$) for both thermal
and reflected light curves.

Compared to either numerical evaluation method, the \starry reflected light
solutions are
\textbf{4--5 orders of magnitude faster} and about
\textbf{10 orders of magnitude more precise}. While different
grid sizes and different error settings for the numerical integration change
the balance slightly between these numbers, the \starry solution is always
many orders of magnitude faster and more precise than either method.
In particular, because of the complicated integration boundaries
(see, for example, Figure~\ref{fig:cases}), two-dimensional Gaussian quadrature
struggles to reach adequate accuracy in a reasonable amount of time,
while integration on a grid fails to capture the curvature of the
integration boundaries. Moreover, neither method yields the gradient of the
solution with respect to the input parameters, which can be extremely useful
for optimization and inference problems (see \S\ref{sec:usage}) below.

Note, importantly, that as we mentioned above, the reported error of the \starry
solution is only the \emph{numerical} error of the algorithm: it does not
capture any systematic error due to, say, an error in the derivation of
the method. To this end, we rely on the Jupyter notebooks containing
derivations and validations of the main equations in the Appendix, whose
links \prooficon\, appear next to the equation labels.
We have also developed an extensive suite of
\href{https://github.com/rodluger/starry/tree/master/tests}{unit tests}
comparing the \starry solution to the numerical solution over a large
grid of input parameter values, and verified that the solutions agree
to within the precision of the numerical method.
That said, there are specific cases in which the algorithm presented in the
Appendix suffers from numerical instabilities. These generally happen due
to division by small numbers or catastrophic cancellation in the recursion,
and often occur near configurations involving grazing occultations,
near-total occultations, terminator semi-minor axis $b \approx 0$ or
$b \approx 1$, etc. To mitigate these, we introduce various tolerance
parameters in the code to either nudge the inputs away from these singular
points or switch to a different evaluation method. These parameters are
outlined in Table~\ref{tab:tolerance} at the end. In the vicinity of the cases
described in that table, the precision of the \starry algorithm will be
reduced to (roughly) the value of the tolerance parameter, which in
extremely rare cases can be as high as $10^{-5}$.

\subsection{Implementation and usage}
\label{sec:usage}

The algorithm presented in this paper has been implemented in the \Python
package \starry, which can be installed from
\href{https://github.com/rodluger/starry}{GitHub} or via the Python package
manager \textsf{pip}. The algorithm is coded in a mixture of \textsf{C++}
with forward automatic differentiation using the \textsf{Eigen} library
\citep{eigen} and \Python with backward differentiation using
just-in-time compiled \textsf{theano}
operations \citep{theano}. The user interface, however, is purely in
\Python. The \textsf{theano} backend facilitates integration with the
\textsf{exoplanet} modeling package \citep{exoplanet} and in particular
with \textsf{pymc3} \citep{pymc3} for inference with gradient-based
Markov Chain Monte Carlo (MCMC) schemes
such as Hamiltonian Monte Carlo \citep[HMC;][]{Duane1987}
and No-U-Turn Sampling \citep[NUTS;][]{Hoffman2011}.
Complete \href{https://starry.readthedocs.io}{documentation} and an
extensive library of tutorials is available online.
The links next to each
of the figures \codeicon\, point to the \Python scripts used to generate
them and may also help in learning how to use \starry.

\section{Extensions}
\label{sec:extensions}

The algorithm discussed above and derived in the Appendix computes light
curves in the limit that (1) the body is illuminated by a point source
and (2) the body is an ideal Lambertian scatterer. Both of these
assumptions can be relaxed within \starry, and below we discuss modifications
to the code to allow for this.

\pagebreak

\subsection{Extended illumination source}
\label{sec:extended}

\begin{figure}[t!]
    \begin{centering}
        \includegraphics[width=\linewidth]{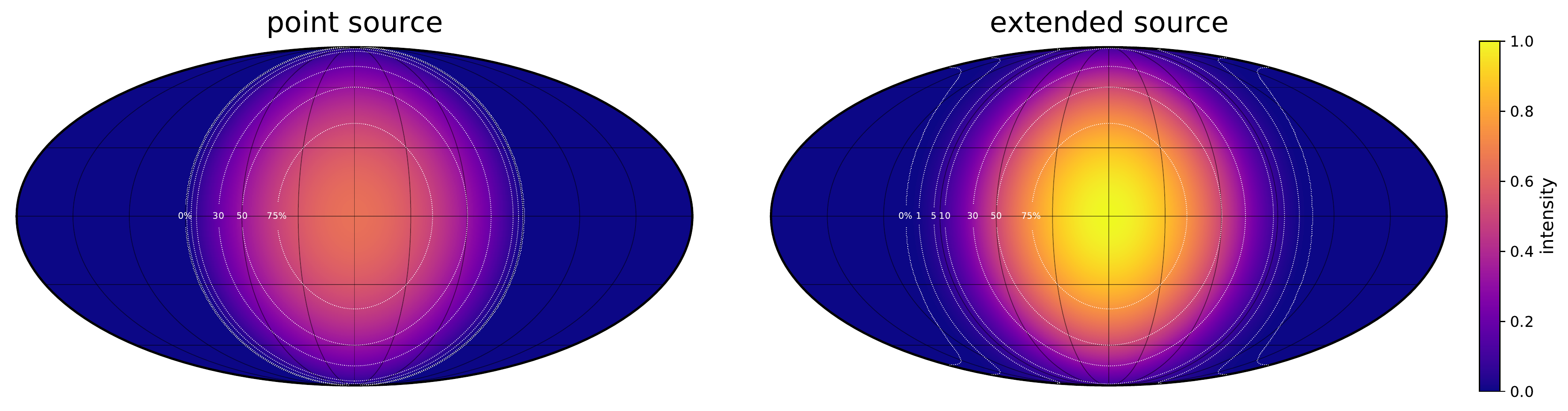}
        \oscaption{extended}{%
            Normalized surface intensity on Kelt-9b viewed in a
            Mollweide projection assuming a point illumination source
            (left) and accounting for the finite extent of the star
            (right). The day/night terminator extends about $17^\circ$
            past where it is in the point source case. The sub-stellar
            intensity is higher in the extended source case because the
            sub-planet point on the star is
            closer to the planet than in the case where the star is a point
            source located at the center of the star.
            \label{fig:extended}
        }
    \end{centering}
\end{figure}

In the limit that the angular size of the star as seen from the planet is
small, the illumination profile on the surface of the planet will
decrease as the cosine of the angle between the surface normal and the star,
reaching zero at the day/night terminator, an angle $\nicefrac{\pi}{2}$
away from the sub-stellar point. However, if the star is sufficiently large
and the planet is sufficiently close-in, rays originating from near the limb
of the star will reach points on the planet surface beyond this angle.
If the stellar radius $R_\star$
is larger than the planet radius $R_\mathrm{p}$, the angular extent of the true
day/night terminator past $\nicefrac{\pi}{2}$ is given by
\begin{proof}{tau}
    \label{eq:tau}
    \tau &= \arcsin\left( \frac{1 - \nicefrac{R_\mathrm{p}}{R_\star}}{\nicefrac{a}{R_\star}} \right)
\end{proof}
where $a$ the semi-major axis of the orbit (where we implictly assume
the eccentricity is zero).
For planets like the Earth, this quantity is only about $0.26^\circ$,
resulting in a negligible effect on the planet's light curve. However,
for very close-in planets, the effect can be significant. For instance,
the hot Jupiter Kelt-9b has $\nicefrac{R_\mathrm{p}}{R_\star} = 0.083$ and
$\nicefrac{a}{R_\star} = 3.16$
planet \citep{Wong2019}. Assuming zero eccentricity and ignoring any
stellar oblateness
\citep[see][]{Ahlers2020},
the day/night terminator extends $\tau \approx 17^\circ$ past the limb
of that planet. Figure~\ref{fig:extended} shows the illumination profile
of this planet in a Mollweide projection, with the sub-stellar point at
the center, for the point source approximation (left) and accounting for
the finite size of the star (right). In addition to the
displaced day/night terminator, the main difference between
the two profiles is the sub-stellar intensity, which is significantly higher
in the extended source case. This is due to the simple fact that in the point
source case the illumination source is placed at the center of the star,
which is one stellar radius farther from the planet than the point closest
to the planet (the sub-planetary point) in the extended source case.
Once accounting for this difference, the fractional change in the intensity
on the planet away from the sub-stellar point is similar in both cases,
and the intensity anywhere beyond $\nicefrac{\pi}{2}$ is less than one-tenth
the peak value.

The illumination profile in the extended source case may be computed as the
two-dimensional integral of the point source illumination profile over
the visible portion of the stellar disk. While this integral may in theory be
computed analytically
\citep[see, for instance,][who derived series solutions to this problem]{Kopal1954},
the resulting profile on the planet surface
will not in general be exactly expressible in terms of spherical harmonics, a
necessary condition for the \starry algorithm.
For simplicity, we therefore compute the illumination profile for
extended sources by averaging the contribution of \textsf{source\_npts} point
sources uniformly distributed across the projected disk corresponding to
the portion of the stellar surface
visible from the planet, where \textsf{source\_npts} is a user-supplied
value. In the limit $R_\mathrm{p} \ll R_\star$, this is a spherical cap centered at
the sub-planet point with radius $R_\star \cos\tau$. In
Figure~\ref{fig:extended} we set
\textsf{source\_npts} = 300, but in practice we find that ${\sim}30$
points are sufficient for even the most extreme cases such as Kelt-9b.
Note, importantly, that while this method allows one to account for the
effect of stellar limb darkening on the illumination profile of the planet,
this has not been implemented in \starry.

\begin{figure}[p!]
    \begin{centering}
        \includegraphics[width=\linewidth]{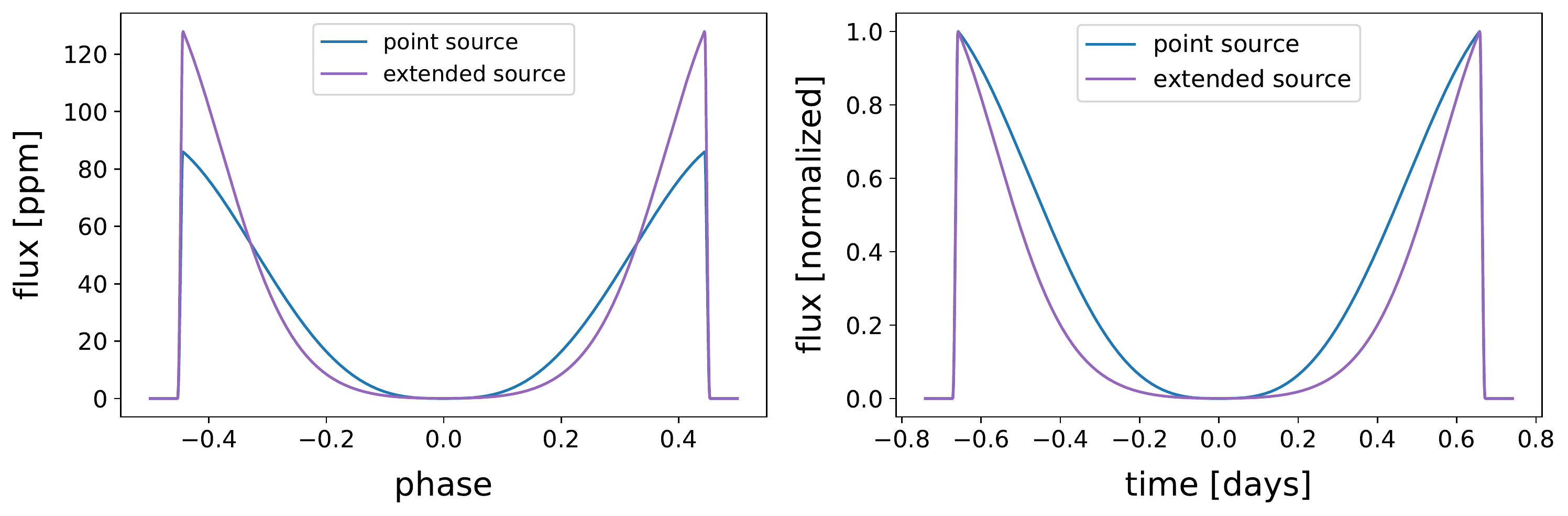}
        \oscaption{kelt9b}{%
            Reflected light phase curve model for Kelt-9b, assuming
            a spherical albedo $A = 0.2$. The transit (phase zero)
            is not included. Blue is the \starry model
            assuming a point source illumination; purple accounts for
            the finite size of the star.
            The left panel shows the two models in parts per million
            of the stellar flux; the right panel shows the models
            normalized so their maximum value is unity.
            The primary effect of the extended source size is to
            increase the planet flux near full phase, since the stellar
            surface is on average slightly closer to the planet, and
            to change the overall curvature of the phase curve.
            The shape of secondary eclipse, however, is relatively
            insensitive to the point source approximation
            (see Figure~\ref{fig:kelt9b_eclipse}).
            \label{fig:kelt9b}
        }
    \end{centering}
\end{figure}
\begin{figure}[p!]
    \begin{centering}
        \includegraphics[width=\linewidth]{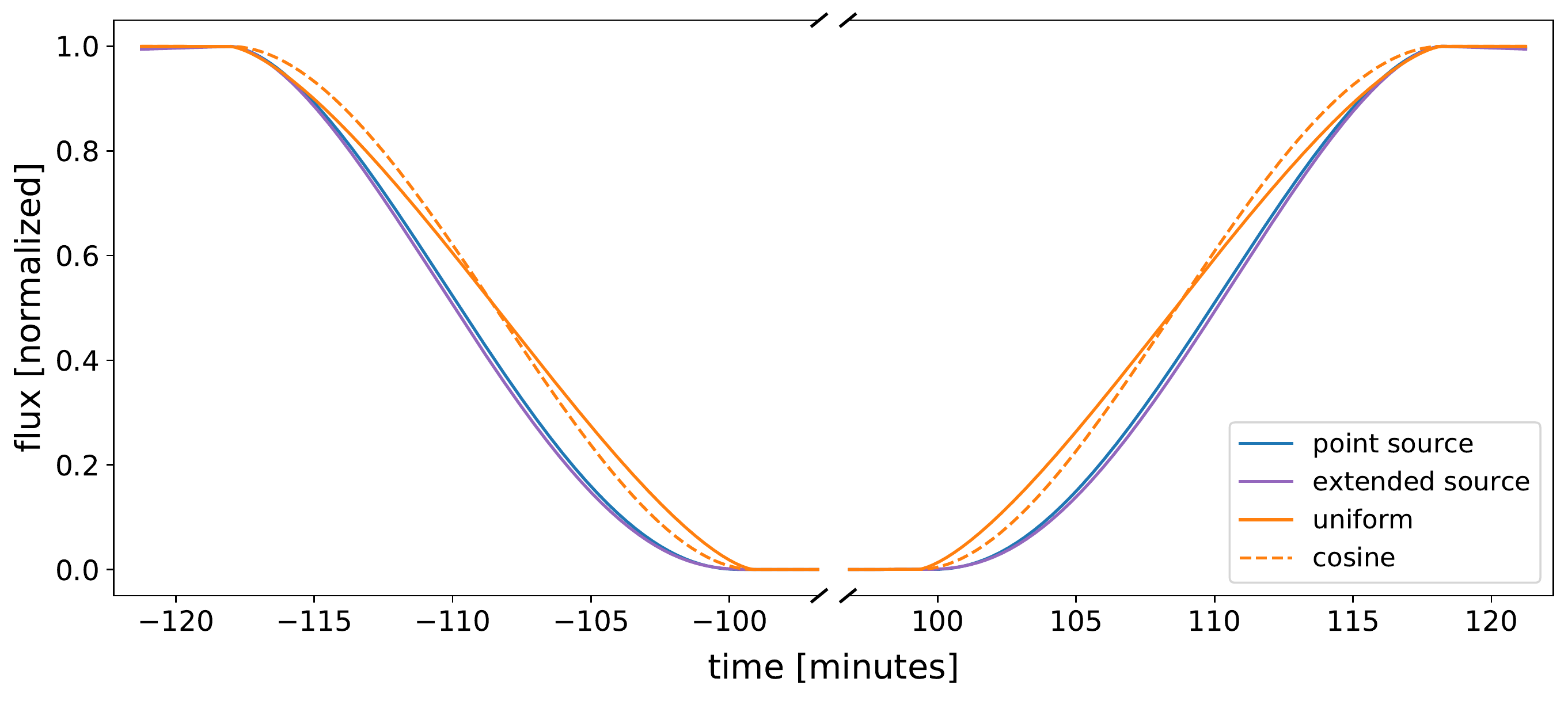}
        \oscaption{kelt9b}{%
            Normalized reflected light secondary eclipse model for Kelt-9b,
            assuming a spherical albedo $A = 0.2$. Blue is the \starry model
            assuming a point source illumination; purple accounts for
            the finite size of the star. The orange curves show
            approximate models computed using the \citet{MandelAgol2002}
            model: a sphere of uniform intensity (solid orange)
            and a sphere whose intensity falls as the cosine of the
            viewing angle from the center of the planet disk
            (dashed orange).
            Because Kelt-9b is so close to its host star, the illumination
            phase changes significantly from ingress to egress, and neither
            approximation accurately captures the behavior of the light curve.
            On the other hand, the point source approximation agrees well with
            the extended source solution, modulo the difference in
            the depth (see Figure~\ref{fig:kelt9b}).
            \label{fig:kelt9b_eclipse}
        }
    \end{centering}
\end{figure}

Figure~\ref{fig:kelt9b} shows the practical implications of the finite
stellar size for the phase curve of Kelt-9b. The left panel shows the
reflected light phase curve of the planet in parts per million, assuming
a spherical albedo of 0.2, for the point source approximation
(blue) and including the effect of the extended source with
\textsf{source\_npts} = 300 (purple). The increased
illumination at the sub-stellar point results in a ${\sim}30$ ppm increase
in the value of the phase curve close to full phase ($\pm\nicefrac{1}{2}$).
Close to a phase of zero, the extended source results in decreased
flux, since the portion of the star illuminating the planet (the region
close to the limb) is slightly
\emph{farther} away, by a factor of
$\sqrt{1 + (\nicefrac{a}{R_\star})^{-2}}$. This results in a steeper
phase curve, which can be seen in the right panel, where the light curves
have been normalized to their maximum value.

While modeling the extended size of the star is essential to
getting the shape of the phase curve correct, the same is not true
for secondary eclipse. Figure~\ref{fig:kelt9b_eclipse} shows
the normalized secondary eclipse model for Kelt-9b under the point
source approximation (blue) and the extended source model (purple).
Neglecting the fact that the \emph{depth} of secondary eclipse is
significantly different between the two models (see the left panel
of Figure~\ref{fig:kelt9b}), the difference in \emph{shape} between
the two curves is almost negligible.
For reference, the figure shows
two additional models one might consider using to fit a secondary
eclipse light curve: a uniform (unilluminated) disk (solid orange)
and a disk whose intensity falls as $\upmu$, the cosine of the
viewing angle (dashed orange). Both can be computed using the
classical \citet{MandelAgol2002} transit model; the latter corresponds
to a linearly limb-darkened sphere and is functionally equivalent to
a Lambertian sphere seen at full phase.
However, neither approximation is particularly good, since Kelt-9b
changes illumination phase significantly from ingress to egress
owing to its proximity to the star.

\pagebreak

\subsection{Non-Lambertian scatterers}
\label{sec:nonlambertian}

The second assumption we now seek to relax is that of Lambertian
scattering. A perfect Lambert sphere reflects light isotropically, so
the measured intensity at a point on the surface is strictly
proportional to the product of the cosine of the angle of incidence
and the cosine of the viewing angle. While this is convenient
from a modeling standpoint, it is hardly ever true in practice.
For planets and moons in particular, there is often a strong phase
dependence in the scattering. Rayleigh scattering in planetary
atmospheres is preferentially in the forward/backward direction,
while clouds and oceans can contribute strong specular reflection.
Moreover, rough surfaces can have complex scattering behavior due
to changes in the orientation of the surface normal on small scales
and effects such as multiple reflections and self-shading.

\begin{figure}[t!]
    \begin{centering}
        \includegraphics[width=0.5\linewidth]{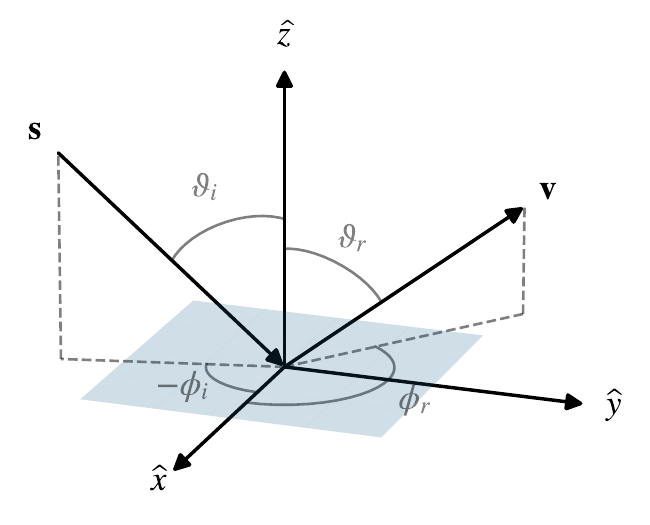}
        \oscaption{scattering}{%
            Scattering geometry for non-Lambertian reflection. Based on
            Figure~3 of \citet{OrenNayar1994}. The incident radiation
            is labeled $\mathbf{s}$ and the outgoing radiation is
            labeled $\mathbf{v}$. The shaded region is a small patch
            of surface, oriented so that the normal vector points along
            $\hat{z}$. The four angles relevant to the computation of
            the emergent intensity are also indicated.
            \label{fig:scattering}
        }
    \end{centering}
\end{figure}

In principle, any of these processes can be accounted for in
the \starry algorithm by modifying the linear operator
$\mathbf{I}$ (Equation~\ref{eq:Imat}), which in the Lambertian
case simply weights the spherical harmonic expansion of the albedo
by the cosine-like illumination profile to obtain the observed
intensity at a point on the surface
(see Equations~\ref{eq:sTA2IA1RRy} and \ref{eq:rTIA1RRy}).
For non-Lambertian scattering, this matrix must also account for
the phase dependence of the reflection: in particular, it will depend
not only on the angle between the surface normal and the incident
radiation, $\vartheta_\mathrm{i}$, but also on the angle between the surface
normal and the reflected radiation (i.e., the direction toward the
observer), $\vartheta_\mathrm{r}$. It may also depend on the azimuthal angles
of the incident and reflected rays, $\phi_\mathrm{i}$ and $\phi_\mathrm{r}$, respectively.
These four angles are shown in Figure~\ref{fig:scattering}, showing
the incoming radiation (source) vector $\mathbf{s}$ and the outgoing
radiation (viewer) vector $\mathbf{v}$ in a frame in which the $z$-axis
points along the surface normal.

Treatment of a generalized, flexible scattering model is beyond the scope
of this paper; see \citet{Heng2021} for recent results on this front.
However, as an example of how a scattering model may be
incorporated
into the \starry algorithm, we consider in detail the case of the rough surface
scattering model of \citet{OrenNayar1994}, commonly used in computer
graphics applications and solar system body modeling
\citep[e.g.,][]{Morgado2019}. In this model, the surface is treated as a
collection of a large number of Lambertian facets oriented at random
angles relative to the average surface normal, whose net contribution to
the total intensity can depart significantly from the Lambertian case.
While the general model accounts for interreflections, shadowing, and
an arbitrary distribution of facet orientations, in its simplest form
the intensity observed at a point $(x, y)$ on the (projected) surface of a
body of unit spherical albedo may be approximated as
\citep[c.f. Equation~30 in][]{OrenNayar1994}
\begin{proof}{OrenNayar}
    \label{eq:OrenNayar}
    \mathcal{I} & =
    \mathcal{I}_\text{Lamb}
    \bigg\{ c_0 + c_1 \, \text{max}\Big( 0, \, \cos(\phi_\mathrm{r} - \phi_\mathrm{i}) \Big)
    \sin \alpha \tan \beta
    \bigg\}
\end{proof}
where%
\\
\begin{minipage}{.5\linewidth}
    \begin{align}
        c_0 & = 1 - 0.5 \left(\frac{\sigma^2}{\sigma^2 + 0.33}\right)
        \nonumber                                                     \\
        c_1 & = 0.45 \left(\frac{\sigma^2}{\sigma^2 + 0.09}\right)
        \nonumber
    \end{align}
\end{minipage}%
\begin{minipage}{.49\linewidth}
    \begin{proof}{OrenNayar}
        \alpha & = \text{max}\Big( \vartheta_\mathrm{r}, \, \vartheta_\mathrm{i} \Big)
        \nonumber                                                    \\
        \beta  & = \text{min}\Big( \vartheta_\mathrm{r}, \, \vartheta_\mathrm{i} \Big)
    \end{proof}
\end{minipage}
\\[1em]
and the angles $\vartheta_\mathrm{i}$, $\vartheta_\mathrm{r}$, $\phi_\mathrm{i}$, and $\phi_\mathrm{r}$ are all
implicit functions of $x$, $y$, and the illumination source position.
The term $\mathcal{I}_\text{Lamb}$ is the Lambertian illumination profile,
given by Equation~(\ref{eq:LambertsLaw}).
At a given point on the surface, and for a given source position,
the intensity $\mathcal{I}$ is therefore a function of a single
parameter, $\sigma$, defined as the standard deviation in radians of the
distribution of facet angles (which is assumed to be a zero-mean Gaussian).

\begin{figure}[t!]
    \begin{centering}
        \includegraphics[width=\linewidth]{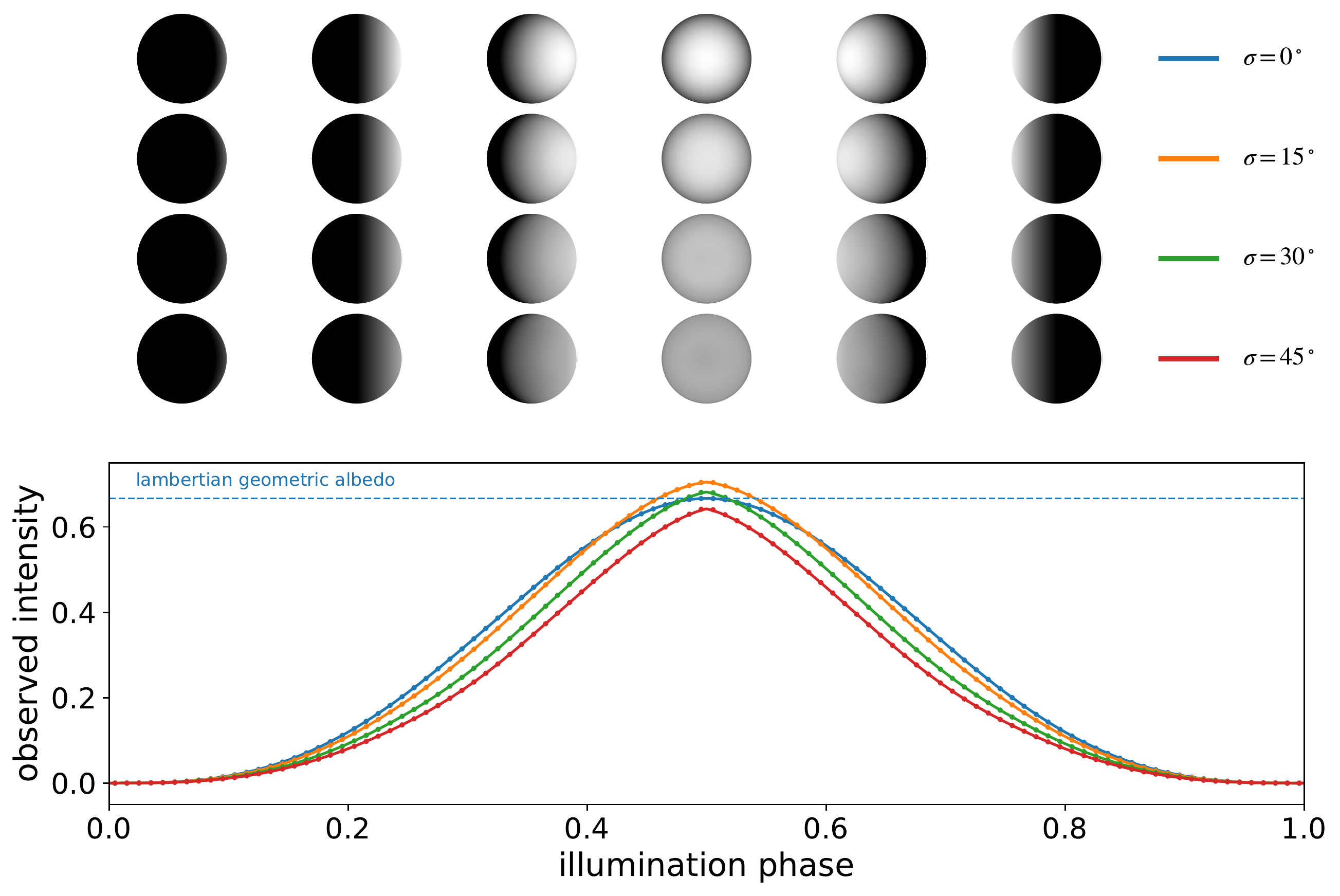}
        \oscaption{oren_nayar}{%
            Intensity measured from a sphere at varying illumination
            phase under the \citet{OrenNayar1994} scattering model.
            The top panel shows spheres rendered with different surface
            roughness coefficients ranging from $\sigma = 0^\circ$
            (the Lambertian case) to $\sigma = 45^\circ$. The bottom
            panel shows the corresponding phase curves for a sphere
            of unit spherical albedo illuminated by a point source, computed
            analytically from a degree \STARRYORENNAYARDEG expansion
            of the scattering law. Dots
            correspond to the intensity computed numerically directly
            from Equation~(30) in \citet{OrenNayar1994}.
            \label{fig:oren_nayar}
        }
    \end{centering}
\end{figure}

In order to incorporate this scattering model into \starry, we must weight
the spherical harmonic expansion of the albedo, $\mathbf{y}$, by
Equation~(\ref{eq:OrenNayar}) instead of Equation~(\ref{eq:LambertsLaw}).
Weighting by Equation~(\ref{eq:LambertsLaw}) is (relatively) straightforward,
since $\mathcal{I}_\text{Lamb}$ is a piecewise function of the $l=1$ spherical
harmonics (see Appendix~\ref{sec:adapting-starry}).
The function we must integrate when computing fluxes is therefore
exactly expressible in terms of spherical harmonics and thus
\starry-integrable. However, Equation~(\ref{eq:OrenNayar}) cannot be expressed
exactly in terms of spherical harmonics, so we must instead approximate it.
To this end, we evaluate Equation~(\ref{eq:OrenNayar}) on a
grid of $x$ and $y$ spanning the unit disk, as well as the illumination phase,
parametrized by $b$, the semi-minor axis of the elliptical segment defining
the day/night terminator (see Appendix~\ref{sec:adapting-starry}).
We then fit to this a polynomial of total degree \STARRYORENNAYARDEG
in $x$, $y$, and $z \equiv \sqrt{1 - x^2 - y^2}$ and degree
\STARRYORENNAYARNB in $b$ and degree \STARRYORENNAYARNBC in
$b_\mathrm{c} \equiv \sqrt{1 - b^2}$.
Then, for a given value of $b$ and $b_\mathrm{c}$, we construct the operator
$\mathbf{I}$ out of the polynomial coefficients in the same way as we
constructed the Lambertian operator in Appendix~\ref{sec:adapting-starry}.
More details about our approximation can be found in the Jupyter notebook
accompanying Equation~(\ref{eq:OrenNayar}).

Figure~\ref{fig:oren_nayar} shows spheres of unit albedo
with different surface roughness
coefficients $\sigma$ and their corresponding phase curves.
The sphere in the top row is perfectly Lambertian; its phase curve (blue)
peaks at a value of $\nicefrac{2}{3}$, equal to the geometric albedo
of a Lambert sphere. Increasing the surface roughness results in a greater
relative contribution of flux from the limb of the object near full
phase, since there now exist facets reflecting light directly back toward the
observer (remaining rows and curves in the phase curve plot).
Conversely, less light is scattered
back to the observer at the
sub-illumination point. These competing effects lead to phase curves that
peak at a super-Lambertian value for small roughness
coefficients ($\sigma = 15^\circ$, orange) and at a sub-Lambertian value for
large roughness coefficients ($\sigma = 45^\circ$, red).
We validate our calculations by computing the phase curves by numerically
integrating the \citet{OrenNayar1994} model over the visible disk; these
are shown as the small dots in the figure, which agree to within
$350$ ppm of the body's flux for $\sigma = 45^\circ$.

\begin{figure}[t!]
    \begin{centering}
        \includegraphics[width=\linewidth]{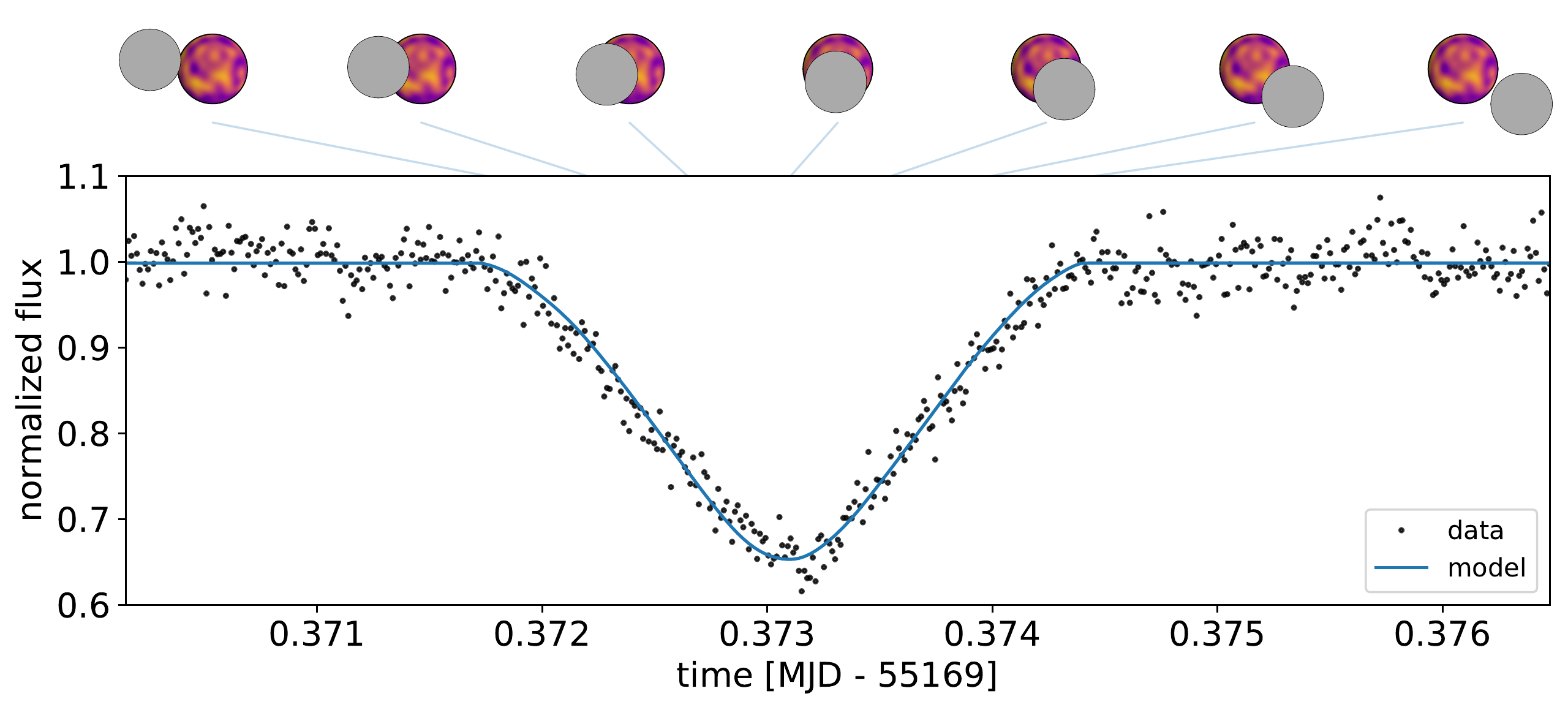}
        \oscaption{io_europa}{%
            Visible-light occultation of Io by Europa observed
            on 04 Dec 2009 by the PHEMU09 campaign \citep{Arlot2014}.
            The blue line is the \starry model, based of an
            $l=15$ spherical harmonic fit to the
            Galileo global color mosaic of Io \citep{Becker2005},
            an $l=\STARRYORENNAYARDEG$ expansion of the
            \citet{OrenNayar1994} scattering law, and orbital
            information from the JPL Horizons database.
            See text for details.
            \label{fig:io_europa}
        }
    \end{centering}
\end{figure}

Our implementation of the scattering model extends just as easily
to occultations and to cases where the surface does not have uniform albedo.
Figure~\ref{fig:io_europa} shows a visible-light observation of the
occultation of Io by Europa on 04 Dec 2009 taken by the PHEMU09
campaign \citep{Arlot2014}. The trajectory of Europa relative to Io,
computed from the \href{https://ssd.jpl.nasa.gov/horizons.cgi}{JPL Horizons database}
using ephemerides from \citet{Folkner2014}
is shown at the top. We fit to this data a \starry occultation model
with the scattering law discussed above.
For simplicity, we set the surface map equal too
an $l=15$ spherical harmonic expansion of the Galileo global color mosaic
of Io \citep{Becker2005} and use ephemerides from the JPL Horizons
database, allowing for a small static $x-y$ offset between Europa and Io as
in \citep{Arlot2014} due to the uncertainty in the database. In total,
we fit for five parameters: the two Cartesian offset terms,
the flux contribution from Europa, the average
albedo of Io, and the average surface roughness of Io, $\sigma$. The model
is displayed in blue and closely matches the data.
Note, however, that this is meant simply as a demonstration of the \starry
algorithm, as our model for the surface is approximate at best, given
differences in the wavelength band between the Galileo observations and
those of the PHEMU09 campaign, changes in the albedo of Io since the Galileo
measurements, and the fact that the orientation and extent of any shadows
due to volcanoes on the surface are likely different between the Galileo
and PHEMU09 observations. Furthermore,
proper modeling would entail the
joint analysis of all light curves of Io taken in a given season, for which
we can afford to simultaneously fit for the surface map without risk of
overfitting \citep{Bartolic2021}.

\section{Discussion}
\label{sec:discussion}

\subsection{Linearity}
\label{sec:linearity}

In the Appendix we derive closed-form expressions for the flux
as a function of the spherical harmonic expansion of the albedo,
$\mathbf{y}$: Equation~(\ref{eq:sTA2IA1RRy}) for occultations and
Equation~(\ref{eq:rTA1Ry}) for phase curves. Inspection of those
equations reveals that they are both \emph{linear} in $\mathbf{y}$:
the flux is simply the dot product of several matrices and the
vector of spherical harmonic coefficients. We may therefore write
both expressions in the form
\begin{align}
    f = \mathbf{x}^\top \mathbf{y}
\end{align}
where $f$ is a scalar representing the model for the flux at a particular
point in time and $\mathbf{x}^\top$ is a row vector equal to
$\sT\mathbf{A_2}\mathbf{I}\mathbf{A_1}\mathbf{R'}\mathbf{R}$
(in the case of an occultation)
or
$\rT\mathbf{I}\mathbf{A_1}\mathbf{R''}\mathbf{R}$
(in the case of a phase curve; see Appendix~\ref{sec:adapting-starry}
for details on what each of the terms represent). Now, if we let $\mathbf{f}$
be the vector of values of $f$ for each point in the timeseries and
construct the matrix $\mathbf{X}$ out of the stacked row vectors
$\mathbf{x}^\top$, we may write our model for the entire timeseries as
the dot product
\begin{align}
    \mathbf{f} = \mathbf{X} \mathbf{y}
    \quad.
\end{align}
The linearity of the \starry model is useful in several ways. For one,
it can be exploited to cheaply compute the same model for different
input vectors $\mathbf{y}$. This is useful for multi-band light curves,
where the same matrix $\mathbf{X}$ dots into several vectors $\mathbf{y}$,
one for each observation band, or for time-dependent models, in which
the model for the flux might be the Taylor series
\begin{align}
    \mathbf{f} =
    \mathbf{X}
    \bigg(
    \mathbf{y}(t)\bigg|_{t=t_0} +
    \frac{\dd\mathbf{y'}(t)}{\dd t}\bigg|_{t=t_0} t +
    \frac{1}{2}\frac{\dd^2\mathbf{y'}(t)}{\dd t^2}\bigg|_{t=t_0} t^2 +
    \cdots
    \bigg)
    \quad,
\end{align}
where $\mathbf{y}(t)$ is a time-dependent representation of the surface
map, which we expand about $t=t_0$ \citep[see][]{Luger2019b}.
But perhaps even more importantly, linear
models are particularly useful for inference, since under Gaussian
noise properties the posterior is analytic. In particular, if
our light curve measurements are given by the data vector $\mathbf{d}$
whose noise model is specified by the covariance matrix $\pmb{\Sigma}$,
and we place a Gaussian prior on $\mathbf{y}$ with mean $\pmb{\mu}$ and
covariance $\pmb{\Lambda}$, the posterior mean may be written
\begin{align}
    \label{eq:posterior_mu}
    \hat{\bvec{y}} & =
    \bvec{C}
    \left(
    \mathbf{X}^\top
    \pmb{\Sigma}^{-1}
    \bvec{d}
    +
    \pmb{\Lambda}^{-1}
    \pmb{\mu}
    \right)
    \quad,
\end{align}
where $\bvec{C}$ is the posterior covariance, given by
\begin{align}
    \label{eq:posterior_cov}
    \bvec{C} & =
    \left(
    \mathbf{X}^\top
    \pmb{\Sigma}^{-1}
    \mathbf{X}
    +
    \pmb{\Lambda}^{-1}
    \right)^{-1}
    \quad.
\end{align}
Because of this linearity, and the analyticity of the \starry
model, inference on datasets comprising thousands of points and
spherical harmonic degree $l \leq 20$ takes a
\emph{fraction of a second} on a typical computer.
\textbf{Full posterior inference with \starry can thus be faster than
    the numerical evaluation of a single forward model}
(c.f. Figure~\ref{fig:speed}).

It is important to note, however, that the \starry model is linear
only in the map coefficients $\mathbf{y}$. In any real application,
there will be uncertainty in the inputs of $\mathbf{X}$, such
as the orbital parameters, the occultor parameters, the scattering
law, etc. These parameters must typically be sampled over, since the model
is a nonlinear function of them. However, the analyticity---and in particular,
the differentiability---of the \starry model makes sampling via
gradient-based MCMC easy. Moreover, the linearity of the model
with respect to $\mathbf{y}$ allows one to efficiently marginalize over
those parameters when sampling over the nonlinear parameters.
Tutorials on how to do this can be found in the
\starry \href{https://starry.readthedocs.io}{documentation}.

\subsection{The information content of reflected light curves}
\label{sec:information}

One of the fundamental difficulties with the mapping problem is that
the process of inferring a two-dimensional map from a light curve
is almost always ill-posed. This has been known since at least
the turn of the last century, when \citet{Russell1906} discussed
how, because of symmetry,
all odd harmonics above $l = 1$ are in the null space for thermal
phase curves of spherical bodies, meaning those terms do not contribute
at all to the disk-integrated flux. As discussed in \citet{Luger2019}
and \citet{PaperI},
the problem is even more ill-posed than that: for any even
degree $l > 0$, there are $2l + 1$ modes on the \emph{surface}
(one for each value of $m$), but only $2$ Fourier modes in the
\emph{light curve} (i.e., a sine and a cosine). Thus, for every
mode that can be constrained from the light curve, there are far
more modes that cannot, a problem that only gets worse as $l$
increases.

\begin{figure}[t!]
    \begin{centering}
        \includegraphics[width=\linewidth]{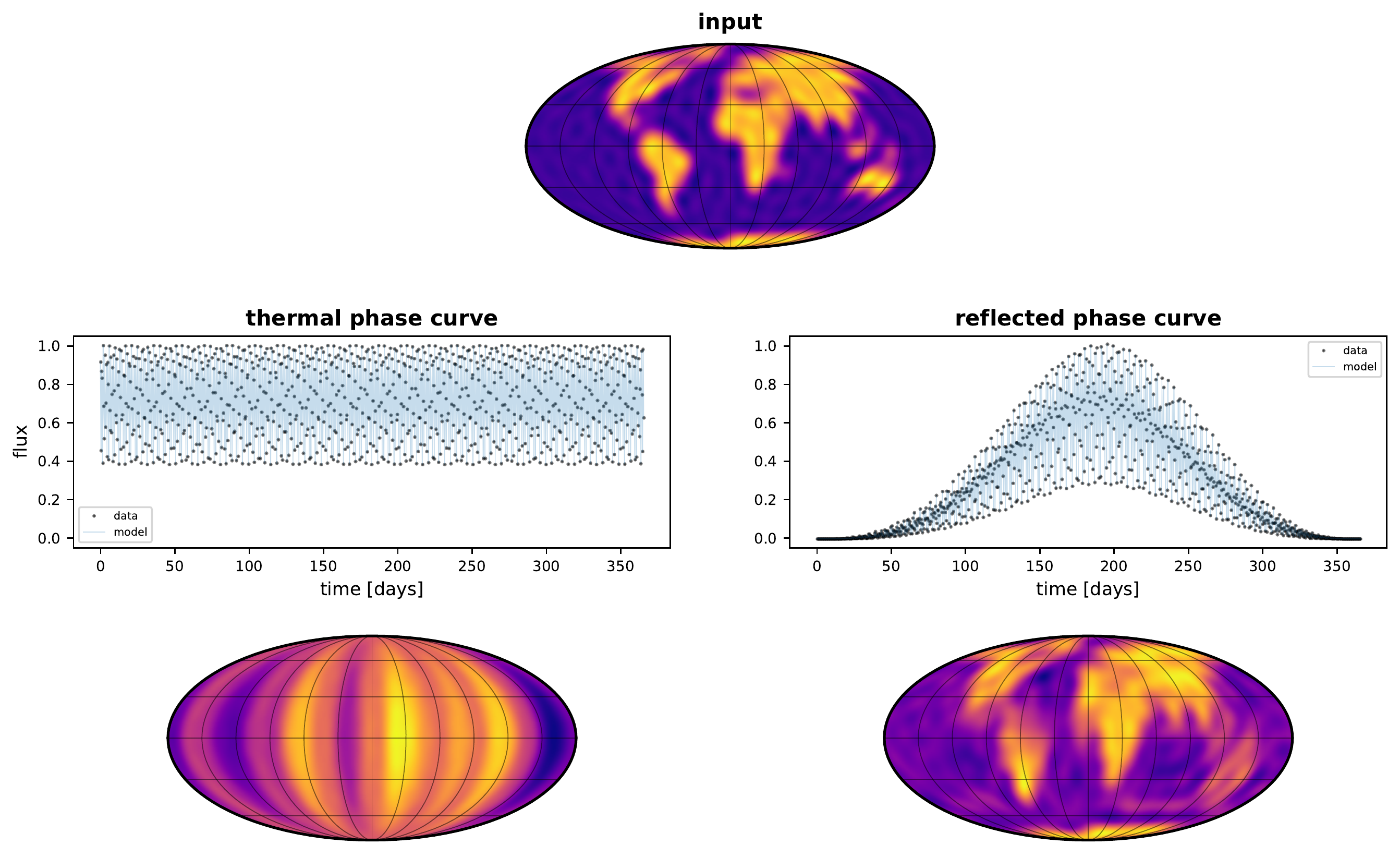}
        \oscaption{inference}{%
            Example of an inference problem for a thermal phase curve
            (center left) and a reflected phase curve (center right).
            In both cases,
            a mock phase curve is generated from an $l=20$ expansion
            of the cloudless Earth (top) with 1,000 evenly spaced points
            over the course of one year and an extremely small
            photometric uncertainty of 1 ppm. The observer sits along the
            ecliptic and the obliquity of the Earth is set to $23.5^\circ$.
            Data is shown as the black points, and the maximum likelihood
            \starry model is shown in blue. While both models fit the
            data equally well, the same is not true of the inferred
            surface maps (bottom row): only in the reflected case are the
            continental outlines recovered.
            The thermal phase curve problem is
            extremely ill-conditioned, but the analogous problem in reflected
            light is much better posed.
            \label{fig:inference}
        }
    \end{centering}
\end{figure}

The left panel of Figure~\ref{fig:inference} shows this issue in practice.
We generate a mock thermal light curve (center left) from an $l = 20$
expansion of the Earth (top) with 1,000 points over the course of one year
with an exquisite photometric precision of 1 ppm. The Earth is given
an obliquity of $23.5^\circ$ on the plane of the sky but is viewed
along the ecliptic, rotating edge-on with an inclination of $90^\circ$.
The data are shown in black, and in blue is the posterior mean model
(Equation~\ref{eq:posterior_mu}),
in which we assume a prior variance of $10^{-3}$ for all spherical
harmonic coefficients (and zero covariance). The corresponding surface
map is shown at the bottom. As expected, this looks nothing like the
true map of the Earth. For a body seen rotating edge-on, the information
content of the light curve is strictly longitudinal. While the
inferred map captures the average brightness of the Earth at each
longitude fairly well, it is missing all latitudinal information.
This is independent of the signal-to-noise or the cadence of the
dataset---it is a fundamental limitation of phase curves in thermal light.

The same is not true for the case of reflected light phase curves. In
the right panel of Figure~\ref{fig:inference} we show the same
mock light curve and perform the same inference step,
but this time for observations in reflected light; the light curve
is similar to that in Figure~\ref{fig:earthphase}. Because of the
presence of a day-night terminator beyond which features on the surface
contribute zero flux, none of the symmetry arguments above apply. In
particular, the facts that (1) the Earth is seen at different phases and
(2) the terminator is inclined relative to the rotational axis mean
that the region of the surface contributing to the phase curve is
always changing, resulting in a complex light curve that encodes
significantly more information than its thermal counterpart.
The result is an inferred map that is largely faithful to the
true map (bottom). While the continental outlines are somewhat fuzzy and some
artifacts are present at high latitudes, it is clear that the
inference problem is much less ill-posed in this case.

Unlike thermal light curves, reflected light phase curves have the
potential to robustly constrain two-dimensional maps of exoplanets.
This result is
not new, and has been discussed at length in the literature
\citep[e.g.,][]{Fujii2012,Berdyugina2017,Luger2019b,Aizawa2020,Kawahara2020}.
In particular, \citet{Kawahara2010} demonstrated the uniqueness of their
inferred map from mock reflected light curves of the Earth. As we argued
above, for specific geometrical configurations, the mapping problem
in reflected light can actually be well-posed, meaning it has no
null space up to a certain degree $l$. Reflected light phase curves of
terrestrial planets with JWST and future direct imaging missions thus
have the potential to reveal detailed information about their surfaces.

\begin{figure}[p!]
    \begin{centering}
        \includegraphics[width=\linewidth]{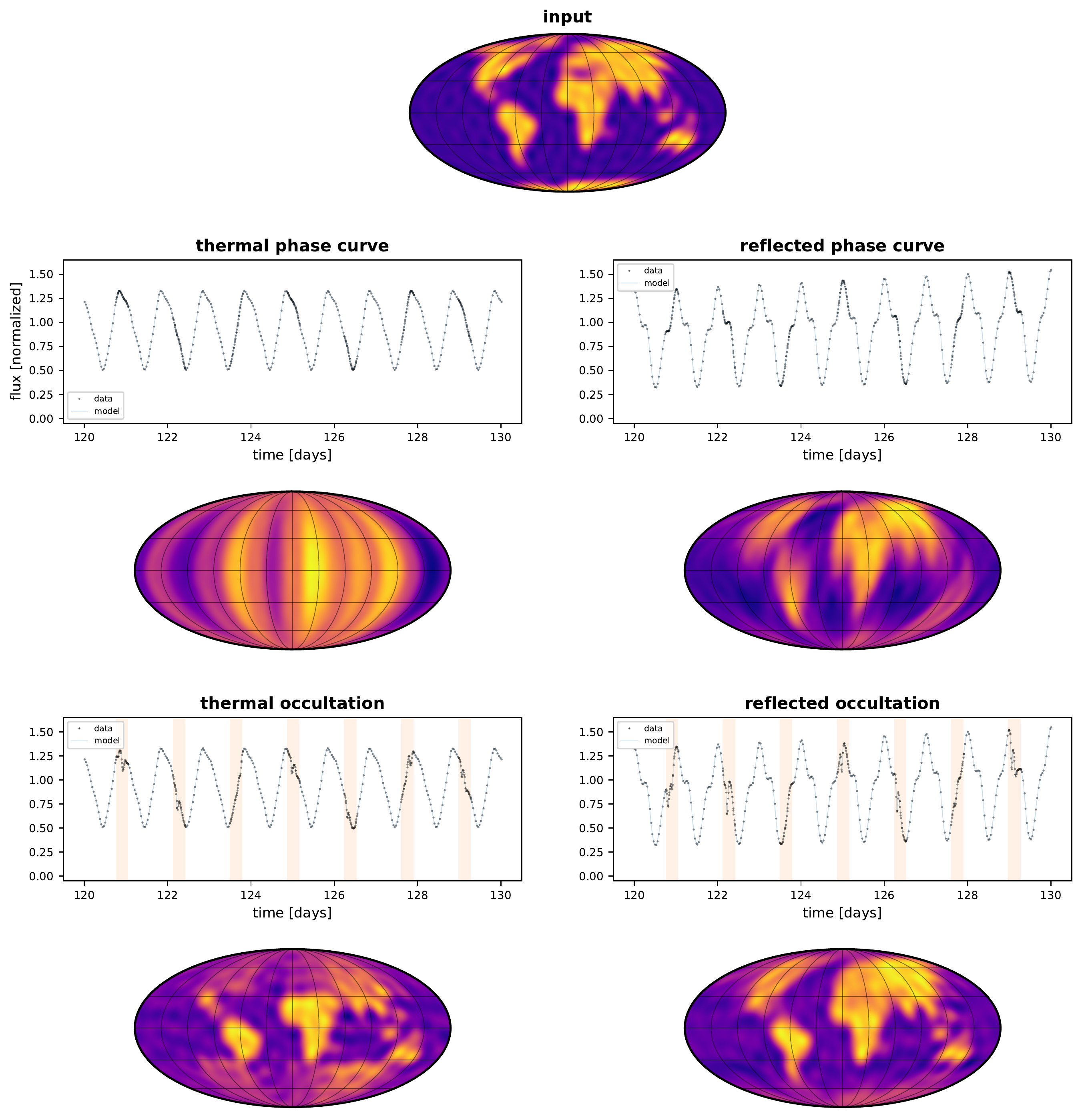}
        \oscaption{inference_occ}{%
            Similar to Figure~\ref{fig:inference}, but for 500 high
            signal to noise observations taken
            over ten days near quadrature.
            At the top we show the thermal and reflected
            phase curves and the corresponding inferred maps.
            At the bottom, we show the same light curves but this
            time including 7 equatorial occultations by
            a (very) short period moon one-quarter the size of the planet.
            The presence of the occultations significantly increases the
            fidelity of the recovered maps near the equator (the path
            of the occultor). While in the thermal case there is significant
            overfitting at high latitudes, in the reflected case the map
            accurately recovers features across the entire planet.
            \label{fig:inference_occ}
        }
    \end{centering}
\end{figure}

Nevertheless, at finite signal to noise and for limited observation
duration or cadence, there may still be significant degeneracies in
the reflected phase curve problem. In \citet{Luger2019} we argued that
occultations can be used to break many of these degeneracies, since they
directly probe the surface at scales inaccessible to phase curves.
The same is true in reflected light. Figure~\ref{fig:inference_occ}
demonstrates this for mock observations of the same Earth-like
planet as in Figure~\ref{fig:inference}, but this time taken
over ten days near quadrature, when the disk of the planet is seen at
half phase. The top rows show the thermal and reflected phase
curves and the inferred maps, which look similar to those in
the previous figure.
At the bottom we show the same light curve, but this time including
seven equatorial occultations by a moon one-quarter the size of the planet.
The moon's period and occultation duration are unrealistically short, but
the inferred maps at the bottom show the exquisite constraining power of these
occultations.
In the thermal case, the presence of the moon allows us to infer the
two-dimensional distribution of surface features along its equatorial
occultation path; however, there is little information
in the light curve about features at higher latitudes, and the \starry model
overfits. Conversely, in the reflected light case, there is information
about all latitudes and longitudes, and the inferred map is
largely faithful to the true map.

It should be kept in mind that the kinds of observations
mentioned above will be very challenging for exoplanets, even with
next-generation observatories such as HabEx or LUVOIR. Observations of
occultations of planets by moons, in particular, are likely several decades
away at least. There is some hope that planet-planet occultations may
be detectable in the near future for specific planetary
systems such as TRAPPIST-1
\citep{Luger2019b}, but at extremely limited signal-to-noise.
Even phase curve observations will be difficult
because of their limited signal-to-noise, and in practice many
degeneracies will likely remain. Several studies have found
that color information can greatly help in the interpretation of
reflected phase curves \citep{Cowan2009,Kawahara2011,LustigYaeger2018,Kawahara2020}, while others
have explored in detail the best kinds of priors to assume
\citep[e.g.,][]{Aizawa2020}. Future maps of exoplanets---particularly
terrestrial ones in the habitable zone---will require every tool
in the toolbox.

\subsection{Limitations}
\label{sec:limitations}

There are a few limitations to the \starry model that are important
to bear in mind. The primary limitation concerns the maximum spherical
harmonic degree of the model. While the expressions derived here are
valid at arbitrary degree, we find that their numerical stability quickly
degrades above $l \sim 20-25$ for occultations and $l \sim 35-40$ for phase
curves. The same is true for the model in thermal light \citep{Luger2019},
and is due to (1) the large condition number of the change-of-basis matrix
from spherical harmonics to polynomials and (2) instabilities in the
many recursion relations used to evaluate the solution vectors
$\sT$ and $\rT$.
In principle, one could improve the numerical stability by evaluating
all expressions at higher floating-point precision, but in practice
the computational cost of this becomes quickly prohibitive.
However, it is important to keep in mind that our current best image
of an exoplanet is the $l=1$ map of HD189733b
\citep{Knutson2007,Majeau2012,deWit2012}. Next-generation facilities
such as JWST may allow us to probe surface modes as small as $l=5$
for some planets (Luger et al., \emph{in prep}), but even with future
telescopes such as LUVOIR it is extremely unlikely we will do better than
$l=20$. If cases arise requiring a resolution smaller than about
$180^\circ / 20 = 9^\circ$ on the surface, the \starry algorithm
will have to be revisited.

The second limitation concerns the flexibility of the \starry model.
While we presented ways to capture non-Lambertian scattering
in \starry (\S\ref{sec:nonlambertian}), there are certain aspects
of light curves in reflected light that cannot be captured by the model.
One example of this is shadowing. Craters on the moon or
volcanoes on Io can cast large shadows visible from space, particularly
if viewed near crescent phase. Unfortunately, there is no way to model
this within the \starry framework. Another example is multiple scattering,
as in optically thick atmospheres, for which a proper radiative transfer model
must be used. There is also the case of specular reflection,
or ``glint'', which is a pronounced signal for the Earth due its
oceans \citep{Robinson2014} and on Titan due to
its hydrocarbon lakes \citep{Barnes2011}. In principle, glint could be
modeled in the same way as non-Lambertian sccattering, by constructing the
linear operator $\mathbf{I}$ in such a way as to downweight portions of the
projected disk where $\vartheta_\mathrm{i} \neq \vartheta_\mathrm{r}$. In practice, however,
if the size of the glint spot is small (which is typically the case),
an expansion at extremely high $l$ ($l \sim 360$ in the case of the Earth)
would be required, which would not work for the reasons above.
Instead, it may be possible to combine the \starry algorithm with the
formalism of \citet{Haggard2018}, who derived analytic expressions for
phase curves of delta function maps ($\delta$-map), to model glint.
As discussed in \citet{LustigYaeger2018}, glint mapping is an extremely
powerful way to not only map terrestrial planets but also to confirm
their habitability via the presence of an ocean.  Lastly, the presence
of time-variable features, such as clouds, dust storms, or seasonal
variations of vegetation are not accounted for in the model, although
the \href{https://starry.readthedocs.io}{documentation} discusses
how one may approach the modeling of temporal features.

Finally, we would like to emphasize that while spherical harmonics are
a convenient basis for the purpose of computing light curves, they
have a significant drawback when it comes to modeling real planetary
surfaces: it can be difficult to strictly enforce physical
values of the albedo everywhere on the surface map when doing inference.
That is because there is no analytic way to determine whether a
spherical harmonic representation is positive-valued
(or restricted to a given range) everywhere on the sphere.
Instead, this must be checked numerically, by evaluating the function on a
discrete grid. This makes it somewhat cumbersome to implement positivity as a prior
when doing inference; in particular, this prior cannot be expressed as
a Gaussian, so the analytic expression for the posterior
discussed in \S\ref{sec:linearity} will generally have nonzero support for
negative albedo values. This is particularly problematic when the data is
not very constraining and the posterior is prior-dominated, as
positivity can be an \emph{extremely} informative prior
\citep[e.g.,][]{Fienup1978}.
We therefore recommend that in such cases
users of the \starry algorithm use HMC/MCMC to do inference, either
(1) sampling over the spherical
harmonic coefficients $\mathbf{y}$ and imposing a uniform prior in the range
$[0, 1]$ on the albedo values $\mathbf{a}$
evaluated on a discrete grid on the sphere or
(2) sampling over the albedo values $\mathbf{a}$ with the same uniform prior,
but using $\mathbf{y}$ to compute the actual light curve model
\citep{Bartolic2021}. In both
cases, there exists a linear operator that transforms between
$\mathbf{y}$ and $\mathbf{a}$:
\begin{align}
    \label{eq:Py}
    \mathbf{a} = \mathbf{P} \mathbf{y}
\end{align}
and
\begin{align}
    \label{eq:PInvy}
    \mathbf{y} = \mathbf{P}^+ \mathbf{a}
    \quad,
\end{align}
where
\begin{align}
    \label{eq:PInv}
    \mathbf{P}^+ =
    (\mathbf{P}^\top \mathbf{P} + \lambda \boldsymbol{\mathsf{I}})^{-1} \mathbf{P}^\top
\end{align}
is the pseudoinverse of $\mathbf{P}$, with (small) regularization parameter
$\lambda$ and where $\boldsymbol{\mathsf{I}}$ is the identity matrix.
Each row of the matrix $\mathbf{P}$ is constructed from the value
of each of the spherical harmonics at the corresponding point on the
grid; both $\mathbf{P}$ and its inverse may be pre-computed for
efficiency. In both cases, the grid should be fine enough to ensure
positivity over most of the sphere but not so fine as to throttle the
computation; as a rule of thumb, we find that grids with ${\sim}4$ times
as many pixels
as spherical harmonic coefficients are sufficient.
The \href{https://starry.readthedocs.io}{documentation}
includes tutorials on how to implement this in practice.

\section{Conclusions}
\label{sec:conclusions}

We have presented an efficient, numerically stable,
closed-form algorithm for computing phase curves
and occultation light curves of spherical bodies in reflected
(scattered) light.
This algorithm is an extension of the algorithm presented
in \citet{Luger2019} for light curves in thermal light and
is generally applicable to exoplanetary phase
curves, secondary eclipses, and occultations by moons and other planets,
as well as to light curves of planets and moons in our solar system.
We derive the solution for the case of a Lambert sphere illuminated by a
point source, but extend it to the case of an extended illumination
source and non-Lambertian scattering parametrized by a surface roughness
coefficient.
The algorithm is ${\sim}4-5$ orders of magnitude faster
and ${\sim}10$ orders of magnitude more precise than other numerical
approaches for computing these light curves. The algorithm is
also differentiable, enabling inference with efficient gradient-based samplers
such as Hamiltonian Monte Carlo (HMC), and linear in the spherical
harmonic coefficients describing the surface albedo, enabling fast,
closed-form solutions for the albedo posterior distribution under a
Gaussian noise model.
We implement the algorithm within the \starry software, an open-source \Python
package for inferring surface maps of unresolved celestial bodies. The
algorithm is coded in a combination of \cpp and \Python compiled using
the \textsf{theano} package \citep{theano}. The interface was designed specifically for
compatibility with the \textsf{exoplanet} modeling package
\citep{exoplanet} and the \textsf{pymc3} inference suite \citep{pymc3}.

Upcoming telescopes will enable measurements of exoplanet phase curves
and secondary eclipses at unprecendented precision. While the
James Webb Space Telescope (JWST) will be primarily sensitive to thermal
emission from exoplanets (which can currently be modeled with \starry),
next-generation direct imaging facilities such as the
Large UV/Optical/IR Surveyor (LUVOIR)
will enable measurements in reflected light, in particular for terrestrial
planets in the habitable zone. Because of the changing illumination
pattern over the course of an orbit of the planet, phase curves and
occultation light curves in reflected
light contain vastly more information about the two-dimension albedo
distribution of the body than their thermal counterparts.
With careful modeling, light curves in reflected light are likely to give us
the first images of potentially habitable exoplanets, enabling the
detection of clouds, continents, oceans, and perhaps even life.

\vspace{1em}

The software presented in this work is open source under the MIT License and
is available at \url{https://github.com/rodluger/starry}, with documentation
and tutorials hosted at \url{https://starry.readthedocs.io}.
The code used to generate the figures in this paper is hosted
at \url{https://github.com/rodluger/starrynight}.

\software{
    astroquery \citep{astroquery1,astroquery2},
    Eigen v3 \citep{eigen},
    exoplanet \citep{exoplanet},
    pybind11 \citep{pybind11},
    pymc3 \citep{pymc3},
    scipy \citep{scipy},
    starry \citep{starry},
    theano \citep{theano}.
}

\vspace{1em}

We would like to thank Nicolas Cowan, Christina Hedges, and the
Astronomical Data Group at the Center for Computational Astrophysics for
many thought-provoking discussions that helped shape this paper.

\bibliography{bib}

\appendix

\section{The Problem}
\label{sec:the-problem}
This paper closely follows the
notation and formalism introduced in \citet{Luger2019}. While we
include all of the relevant equations and definitions below,
the reader is encouraged to
review \citet{Luger2019} before proceeding.
To improve the readability of this paper,
Tables~\ref{tab:symbols}--\ref{tab:matrices} at the end list the principal
symbols and quantities used throughout the text, with links to the
sections and equations in which they are defined.
Because of the large number of symbols used in this paper, we adopt
the following conventions: scalars are represented by regular lowercase or
occasionally uppercase letters (i.e., $x$ or $X$), vectors are
represented by boldface lowercase letters
($\mathbf{x}$), and matrices and other linear operators are represented
by boldface capital letters ($\mathbf{X}$).
With a few exceptions,
Greek letters are reserved for angular quantities and may be either
scalars ($\alpha$) or vectors ($\pmb{\alpha}$).
Script font is typically used to denote curves or frames of reference
($\mathcal{F}$). Primes are used to
distinguish between frames of reference ($x$ and $x'$ are used
to denote the same quantity, but in frames $\mathcal{F}$ and $\mathcal{F}'$,
respectively). Tildes are used to denote basis vectors ($\by$). Finally,
blackboard vectors ($\mathbbb{x}$) correspond to solutions
to the various ``primitive'' integrals that arise in the occultation
problem.

\subsection{Review of the \starry algorithm in emitted light}
\label{sec:starry-review}
Without loss of generality, assume the body whose flux we wish to compute
has radius unity and sits at the origin of a right-handed Cartesian coordinate
system in some frame $\mathcal{F}_0$. In this frame,
the surface (emitted) intensity field of the body is described by a
vector $\mathbf{y}$ of coefficients in the spherical harmonic
basis $\by$:
\begin{align}
    \label{eq:by}
    \by(x, y) & =
    \begin{pmatrix}
        Y_{0, 0}  &
        Y_{1, -1} & Y_{1, 0}  & Y_{1, 1} &
        Y_{2, -2} & Y_{2, -1} & Y_{2, 0} & Y_{2, 1} & Y_{2, 2} &
        \cdot\cdot\cdot
    \end{pmatrix}^\top
    \quad,
\end{align}
where the component at index $n$ is the spherical harmonic $Y_{l,m}(x, y)$ with
\begin{align}
    \label{eq:l-m}
    l & = \floor*{\sqrt{n}}
    \nonumber               \\
    m & = n - l^2 - l
    \quad.
\end{align}
The spherical harmonics are traditionally expressed in spherical coordinates,
but for our purposes
it is more conventient to express them in Cartesian coordinates on
the sky-projected disk, in which case they are simply polynomials in
$x$, $y$, and $z$ \citep[see Appendix~A in][]{Luger2019}.

An observer views the body from a large distance in the sky frame
$\mathcal{F}$, in which the $x$-axis points to the right,
the $y$-axis points up, and the $z$-axis points out of the sky
toward the observer. Following \citet{Luger2019}, if
an occultor of radius $r_\mathrm{o}$ is located at sky position $(x_\mathrm{o}, y_\mathrm{o})$,
we compute the visible thermal flux $f_\mathrm{T}$ from
\begin{align}
    \label{eq:sTARRy}
    f_\mathrm{T} = \sTe \mathbf{A} \mathbf{R}' \mathbf{R} \mathbf{y}
    \quad,
\end{align}
where, from right to left, $\mathbf{R} = \mathbf{R}(\text{I}, \Lambda, \Theta)$
is a Wigner rotation matrix that rotates $\bvec{y}$ from $\mathcal{F}_0$
to the sky frame $\mathcal{F}$
given the body's inclination $\text{I}$, obliquity
$\Lambda$, and rotational phase $\Theta$
\citep[Appendix C in][]{Luger2019},
$\mathbf{R}' = \mathbf{R}'(x_\mathrm{o}, y_\mathrm{o})$ rotates the body on the plane
of the sky into the integration frame $\mathcal{F}'$, in which the
occultor lies along the $+y'$-axis,
$\mathbf{A}$
\citep[Equation~B13 in][]{Luger2019}
is the change-of-basis matrix from $\by$
to the \emph{Green's basis} $\bg$ in which the integrals are computed,
whose component at index $n$ is
\begin{align}
    \label{eq:bg}
    \tilde{g}_{n}(x, y) & =
    \begin{dcases}
        \frac{\mu+2}{2}x^\frac{\mu}{2} y^\frac{\nu}{2}
         & \qquad \mu, \nu \, \text{even}
        \\[1em]
        z(x, y)
         & \qquad \mu = \nu = 1
        \\[1em]
        3x^{l-2}yz(x, y)
         & \qquad \nu \, \text{odd}, \,
        \mu = 1, \,
        \frac{\mu + \nu}{2} \, \text{even}
        \\[1em]
        z(x, y)
        \bigg(
        -x^{l-3} + x^{l-1} + 4x^{l-3}y^2
        \bigg)
         & \qquad \nu \, \text{odd}, \,
        \mu = 1, \,
        \, \text{odd}
        \\[1em]
        z(x, y)
        \bigg(
        \frac{\mu-3}{2} x^\frac{\mu-5}{2} y^\frac{\nu-1}{2}
        \ - \
        \frac{\mu-3}{2} x^\frac{\mu-5}{2} y^\frac{\nu+3}{2}
        \\
        \qquad\qquad \ - \
        \frac{\mu+3}{2} x^\frac{\mu-1}{2} y^\frac{\nu-1}{2}
        \bigg)
         & \qquad \text{otherwise}
        \quad,
    \end{dcases}
\end{align}
with
\begin{align}
    \label{eq:mu-nu}
    \mu & \equiv l - m
    \nonumber          \\
    \nu & \equiv l + m
\end{align}
and
\begin{align}
    \label{eq:z}
    z(x, y) \equiv \sqrt{1 - x^2 - y^2}
    \quad,
\end{align}
and $\sTe = \sTe(b_\mathrm{o}, r_\mathrm{o})$ is the vector of solutions to the integral over
the projected visible disk of the body for each term in $\bg$
\citep[Equation~26 in][]{Luger2019}, with $b_\mathrm{o} = \sqrt{x_\mathrm{o}^2 + y_\mathrm{o}^2}$.

If instead no occultor is present, we compute the total
visible thermal flux $f_\mathrm{T_0}$ from this body as
\begin{align}
    \label{eq:rTA1Ry}
    f_\mathrm{T_0} = \rTe \mathbf{A_1} \mathbf{R}'' \mathbf{R} \mathbf{y}
    \quad,
\end{align}
where, as before, $\mathbf{R} = \mathbf{R}(\text{I}, \Lambda, \Theta)$
rotates the body from $\mathcal{F}_0$
to the sky frame $\mathcal{F}$,
$\mathbf{R}''$ rotates the body on the plane
of the sky into the integration frame
$\mathcal{F}''$%
\footnote{%
    In \citet{Luger2019}, $\mathcal{F}'' = \mathcal{F}'$, so
    this rotation is trivial: $\mathbf{R}''$ is just the identity matrix.
},
$\mathbf{A_1}$
\citep[Equation~B11 in][]{Luger2019}
is the change-of-basis matrix from the spherical harmonic
basis $\by$ to the \emph{polynomial basis} $\bp$ in which the integrals
are computed, whose component at index $n$ is
\begin{align}
    \label{eq:bp}
    \tilde{p}_n(x, y) & =
    \begin{dcases}
        x^\frac{\mu}{2} y^\frac{\nu}{2}
         & \qquad \mu, \nu \, \text{even}
        \\[1em]
        x^\frac{\mu-1}{2} y^\frac{\nu-1}{2} z(x, y)
         & \qquad \text{otherwise}
        \quad,
    \end{dcases}
\end{align}
and $\rTe$ is the vector of solutions to the integral over
the projected visible disk of the body for each term in $\bp$
\citep[Equation~19 in][]{Luger2019}.

\subsection{Adapting the algorithm to the reflected light case}
\label{sec:adapting-starry}
In order to compute light curves in reflected light, we must make two
modifications to the \starry algorithm. First,
the expressions above assume that the coefficient vector
$\mathbf{y}$ describes the \emph{emissivity} of the body, which (in the
absence of limb darkening) is assumed to be Lambertian, i.e., all points on the
surface emit equally in all directions.
Here, we wish to derive the solution for the flux in the case of Lambertian
reflectance, in which case the vector $\mathbf{y}$ is taken to describe the
spherical albedo of the surface, $A$.

Second, we must explicitly model the illumination of the body. We assume the
body is illuminated by a point-like source whose flux measured by the observer
is unity. In this case, the observed intensity at any
point on the surface is proportional to the cosine of the angle $\vartheta_\mathrm{i}$ between
the incident light and the surface normal. Points for which
$\vartheta_\mathrm{i} \ge \nicefrac{\pi}{2}$ are unilluminated and
therefore have an intensity of zero.
If the point-like illumination source is placed at sky coordinates
$(x_\mathrm{s}, y_\mathrm{s}, z_\mathrm{s})$ in units of the radius of the illuminated body,
the day/night terminator on the body is a half-ellipse
of semi-major axis unity that is fully described by its (signed) semi-minor
axis,
\begin{proof}{illumination}
    \label{eq:b}
    b = -\frac{z_\mathrm{s}}{r_\mathrm{s}}
    \quad,
\end{proof}
where $r_\mathrm{s} = \sqrt{x_\mathrm{s}^2 + y_\mathrm{s}^2 + z_\mathrm{s}^2}$ is the distance to the source,
and the angle by which its semi-major axis is rotated away from the
$+x$-axis,
\begin{proof}{illumination}
    \label{eq:theta}
    \theta = -\atantwo(x_\mathrm{s}, y_\mathrm{s})
    \quad,
\end{proof}
where $\atantwo(a, b)$ is the quadrant-aware arctangent of
$\nicefrac{a}{b}$.
Given this formulation, and assuming that
$r_\mathrm{s} \gg 1$,
it is straightforward to show that the illumination
$\mathcal{I}$ at a point $(x, y)$ on the projected disk of the body is given
by the function
\begin{proof}{illumination}
    \label{eq:illum}
    \mathcal{I}(b, \theta, r_\mathrm{s}; x, y)&=
    \text{max}\bigg( 0, I(b, \theta, r_\mathrm{s}; x, y) \bigg)
\end{proof}
where
\begin{proof}{illumination}
    \label{eq:illum_poly}
    I(b, \theta, r_\mathrm{s}; x, y) &=
    \frac{1}{\pi r_\mathrm{s}^2}
    \cos \vartheta_\mathrm{i}
    \nonumber \\
    &= \frac{1}{\pi r_\mathrm{s}^2}
    \bigg(
    -b_\mathrm{c}\sin\theta x + b_\mathrm{c}\cos\theta y - bz(x, y)
    \bigg)
\end{proof}
with $b_\mathrm{c} \equiv \sqrt{1 - b^2}$ and $z(x, y) = \sqrt{1 - x^2 - y^2}$.
The illumination $\mathcal{I}$ is a unitless quantity, normalized such that
the integral of $A\mathcal{I}$ over the unit
disk is equal to the flux measured
by the observer as a fraction of the flux of the illumination source.
In particular, if we place the illumination source along the $+z$-axis at
$(0, 0, 1)$, the body is seen at full phase, so $b = -1$, $b_\mathrm{c} = 0$, and
\begin{proof}{illumination}
    \mathcal{I}_\text{full}(x, y) = \frac{\sqrt{1 - x^2 - y^2}}{\pi}
    \quad.
\end{proof}
Multiplying this by the albedo and integrating over the unit disk,
we obtain the reflected flux measured
by the observer in units of the flux of the illumination source:
\begin{proof}{illumination}
    \mathcal{f}_\text{full} &=
    \int_{-1}^{1}
    \int_{-\sqrt{1 - x^2}}^{\sqrt{1 - x^2}}
    A
    \frac{\sqrt{1 - x^2 - y^2}}{\pi}
    \,
    \dd y
    \,
    \dd x
    \nonumber \\[0.5em]
    &= \frac{2}{3}A
    \quad,
\end{proof}
which is precisely the geometric albedo of a Lambert
sphere of spherical albedo $A$ \citep[see, e.g.][]{Seager2010}.

In principle, our task is now straightforward: weight each of the
terms in the Green's basis (Equation~\ref{eq:bg}) and integrate them
over the visible portion of the body's disk to obtain the reflected
light solution vector, $\sT$. Unfortunately, the piecewise nature
of Equation~(\ref{eq:illum}) makes direct evaluation of these integrals
extremely difficult in practice.
We find that it is more tractable to weight our basis terms by
the function $I$ (Equation~\ref{eq:illum_poly}) and to modify the limits
of integration to exclude the nightside of the body, where $I$ is
(unphysically) negative.
In particular, since $I$ is just a polynomial in $x$, $y$, and $z(x, y)$, we
can express it as a vector $\mathbf{i}(b, \theta)$ in the polynomial basis $\bp$.
Recalling the structure of the basis (Equation~\ref{eq:bp}),
we may write
\begin{proof}{illumination}
    \label{eq:ivec}
    \mathbf{i}(b, \theta, r_\mathrm{s}) & =
    \frac{1}{\pi r_\mathrm{s}^2}
    \begin{pmatrix}
        0                       \\
        -b_\mathrm{c}\sin\theta \\
        -b                      \\
        b_\mathrm{c}\cos\theta
    \end{pmatrix}
    \quad.
\end{proof}
This fact allows us to construct a linear operator $\mathbf{I}$ to weight a map
vector in the polynomial basis by the illumination profile.
If we think about how each of the terms in $\bp$ transforms under $\mathbf{I}$,
\\[1em]
\begin{minipage}{0.22\linewidth}
    \begin{align}
        \begin{pmatrix}
            1 \\
            0 \\
            0 \\
            0
        \end{pmatrix}
         & \pmb{\rightarrow}
        \begin{pmatrix}
            \bvec{i}_0 \\ 
            \bvec{i}_1 \\ 
            \bvec{i}_2 \\ 
            \bvec{i}_3 \\ 
            0          \\ 
            0          \\ 
            0          \\ 
            0          \\ 
            0             
        \end{pmatrix}
        \nonumber
    \end{align}
\end{minipage}
\begin{minipage}{0.22\linewidth}
    \begin{align}
        \begin{pmatrix}
            0 \\
            1 \\
            0 \\
            0
        \end{pmatrix}
         & \pmb{\rightarrow}
        \begin{pmatrix}
            0          \\ 
            \bvec{i}_0 \\ 
            0          \\ 
            0          \\ 
            \bvec{i}_1 \\ 
            \bvec{i}_2 \\ 
            \bvec{i}_3 \\ 
            0          \\ 
            0    
        \end{pmatrix}
        \nonumber
    \end{align}
\end{minipage}
\begin{minipage}{0.22\linewidth}
    \begin{align}
        \begin{pmatrix}
            0 \\
            0 \\
            1 \\
            0
        \end{pmatrix}
         & \pmb{\rightarrow}
        \begin{pmatrix}
            \bvec{i}_2  \\ 
            0           \\ 
            \bvec{i}_0  \\ 
            0           \\ 
            -\bvec{i}_2 \\ 
            \bvec{i}_1  \\ 
            0           \\ 
            \bvec{i}_3  \\ 
            -\bvec{i}_2    
        \end{pmatrix}
        \nonumber
    \end{align}
\end{minipage}
\begin{minipage}{0.22\linewidth}
    \begin{align}
        \begin{pmatrix}
            0 \\
            0 \\
            0 \\
            1
        \end{pmatrix}
         & \pmb{\rightarrow}
        \begin{pmatrix}
            0          \\ 
            0          \\ 
            0          \\ 
            \bvec{i}_0 \\ 
            0          \\ 
            0          \\ 
            \bvec{i}_1 \\ 
            \bvec{i}_2 \\ 
            \bvec{i}_3    
        \end{pmatrix}
        \nonumber
    \end{align}
\end{minipage}
\begin{minipage}{0.05\linewidth}
    \begin{align}
    \end{align}
\end{minipage}
\\[1em]
we can compose $\mathbf{I}$ out of these column vectors:
\begin{proof}{illumination}
    \label{eq:Imat}
    \mathbf{I}(b, \theta, r_\mathrm{s}) & =
    \frac{1}{\pi r_\mathrm{s}^2}
    \begin{pmatrix}
        0                       & 0                       & -b                      & 0                       & \cdots \\
        -b_\mathrm{c}\sin\theta & 0                       & 0                       & 0                       & \cdots \\
        -b                      & 0                       & 0                       & 0                       & \cdots \\
        b_\mathrm{c}\cos\theta  & 0                       & 0                       & 0                       & \cdots \\
        0                       & -b_\mathrm{c}\sin\theta & b                       & 0                       & \cdots \\
        0                       & -b                      & -b_\mathrm{c}\sin\theta & 0                       & \cdots \\
        0                       & b_\mathrm{c}\cos\theta  & 0                       & -b_\mathrm{c}\sin\theta & \cdots \\
        0                       & 0                       & b_\mathrm{c}\cos\theta  & -b                      & \cdots \\
        0                       & 0                       & b                       & b_\mathrm{c}\cos\theta  & \cdots \\
        \vdots                  & \vdots                  & \vdots                  & \vdots                  & \ddots
    \end{pmatrix}
\end{proof}
where the dimensions of the matrix are $\big((l + 2)^2, (l + 1)^2\big)$, where
$l$ is the spherical harmonic degree of the map (this operator raises the
degree of the map by one).
Note, again, that this weighting is valid only on the dayside
hemisphere (see Equation~\ref{eq:illum}), as the operator $\mathbf{I}$ weights
points on the nightside by a \emph{negative} amount, which is clearly
unphysical. As we will see momentarily, we account for this by excluding the
nightside from the integration region in our flux integrals.

\begin{figure}[t!]
    \begin{centering}
        \includegraphics[width=\linewidth]{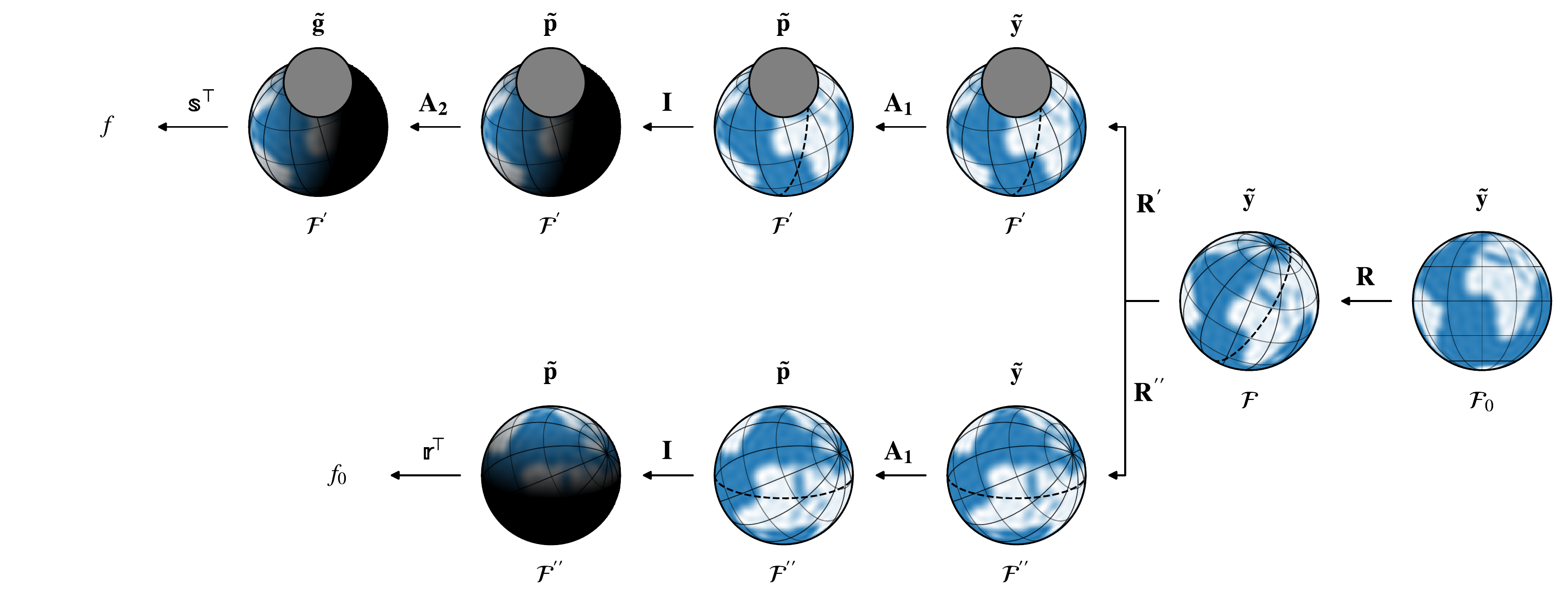}
        \oscaption{frames}{%
            How \starry computes the flux from a body in reflected light,
            tracking each of the linear transformations from the input map
            (far right) to the output (far left). The label below each map
            denotes the reference frame, while the label above
            each map denotes the basis in which the map is represented.
            Arrows indicate linear operations and are labeled accordingly.
            The upper branch corresponds to the occulted case
            (Equation~\ref{eq:sTA2IA1RRy}), while the
            lower branch corresponds to the case where the body
            is unocculted (Equation~\ref{eq:rTIA1RRy}).
            See text for details.
            \label{fig:frames}
        }
    \end{centering}
\end{figure}

We may now re-write Equations~(\ref{eq:sTARRy}) and (\ref{eq:rTA1Ry}) to
account for this illumination transformation. The flux during an occultation
is now given by
\begin{align}
    \label{eq:sTA2IA1RRy}
    f & =
    \sT(b, \theta', b_\mathrm{o}, r_\mathrm{o})
    \mathbf{A_2}
    \mathbf{I}(b, \theta', r_\mathrm{s})
    \mathbf{A_1}
    \mathbf{R}'(x_\mathrm{o}, y_\mathrm{o})
    \mathbf{R}(\text{I}, \Lambda, \Theta)
    \mathbf{y}
    \quad,
\end{align}
where
\begin{align}
    \label{eq:theta'}
    \theta' = \atantwo(x_\mathrm{o}, y_\mathrm{o}) - \atantwo(x_\mathrm{s}, y_\mathrm{s})
\end{align}
is the angle of the terminator in the frame $\mathcal{F}'$.
Note that we made use of the fact that
$\mathbf{A} = \mathbf{A_2} \mathbf{A_1}$
\citep[Equation~14 in][]{Luger2019}, where $\mathbf{A_1}$ transforms from
the spherical harmonic basis $\by$ to the polynomial basis $\bp$, and
$\mathbf{A_2}$ transforms from $\bp$ to the Green's basis $\bg$.

Similarly, the flux when there is no occultation is now given by
\begin{align}
    \label{eq:rTIA1RRy}
    f_0 & =
    \rT(b)
    \mathbf{I}(b, \theta'', r_\mathrm{s})
    \mathbf{A_1}
    \mathbf{R}''(x_\mathrm{s}, y_\mathrm{s})
    \mathbf{R}(\text{I}, \Lambda, \Theta)
    \mathbf{y}
    \quad,
\end{align}
where
\begin{align}
    \label{eq:theta''}
    \theta'' = 0
\end{align}
is the angle of the terminator in the frame $\mathcal{F}''$, by
construction. The transformation
$\mathbf{R}'' = \mathbf{R}''(x_\mathrm{s}, y_\mathrm{s})$ rotates
the body
through an angle $\atantwo(x_\mathrm{s}, y_\mathrm{s})$
so the semi-major axis of the terminator is aligned
with the $x''$-axis; as will become clear in \S\ref{sec:solution-no-occ} below,
this greatly simplifies the integration step.

Note that in both equations we replaced the integral vectors
$\rTe$ and $\sTe(b_\mathrm{o}, r_\mathrm{o})$
with the vectors
$\rT(b)$ and $\sT(b, \theta', b_\mathrm{o}, r_\mathrm{o})$,
respectively.
As we mentioned above, we must modify the integration limits to exclude the
nightside, where the weighting by $\mathbf{I}$ is unphysical.
The vectors $\rT$ and $\sT$ correspond to these
modified integrals, which we devote the rest of this paper to
computing.

Figure~\ref{fig:frames} summarizes the transformations involved in the two
equations above. Starting on the right with a map vector $\mathbf{y}$ in
the spherical harmonic basis $\by$, defined in some observer-independent frame
$\mathcal{F}_0$, we first rotate it via $\mathbf{R}$ to the sky frame
$\mathcal{F}$, in which the body is viewed by the
observer. If an occultor is present (upper branch of the figure),
we rotate the map from $\mathcal{F}$ via $\mathbf{R}'$ to the frame
$\mathcal{F}'$, in which the occultor lies along the
$+y'$-axis. We then apply $\mathbf{A_1}$ to change basis to $\bp$ and $\mathbf{I}$
to weight the map by the illumination. Finally, we change basis
via $\mathbf{A_2}$ to the Green's basis, in which we compute and dot the
integrals $\sT$.
If, on the other hand, there is no occultation (lower branch of the figure),
we instead rotate the map via $\mathbf{R}''$ to the integration frame
$\mathcal{F}''$, in which the terminator is parallel to the
$x''$-axis. We then apply $\mathbf{A_1}$ to change basis to $\bp$, apply the
illumination transform $\mathbf{I}$, and finally dot in the solutions to the
surface integrals $\rT$.

\section{The Solution: No Occultation}
\label{sec:solution-no-occ}
Before we tackle configurations involving occultations, we must address the
simpler problem of computing the total visible flux from an unocculted
body in reflected light (Equation~\ref{eq:rTIA1RRy}). This problem was
originally solved by \citet{Haggard2018} and subsequently by
\citet{Luger2019b}, but for completeness we present the detailed
derivation in the \starry formalism here.

As we discussed above, we perform the integration in a frame
$\mathcal{F}''$
in which the semi-major axis of the terminator is aligned with the
$x''$-axis, with the illumination source at $y'' \ge 0$.
The solution vector may then be computed from
\begin{align}
    \label{eq:rT}
    \rT(b) & =
    \int_{-1}^{1}
    \int_{b\sqrt{1 - x''^2}}^{\sqrt{1 - x''^2}}
    \bp(x'', y'')
    \ \dd y'' \ \dd x''
    \quad,
\end{align}
which is identical to Equation~(20) in \citet{Luger2019} except for the
lower integration limit of the inner integral. The lower limit is now
the equation describing the terminator, which ensures we always exclude the
nightside from the integration region.
Equation~(\ref{eq:rT}) may be solved analytically in terms of purely
trigonometric and algebraic functions of $b$.
The component of $\rT$
at index $n$ is given by
\begin{proof}{PhaseCurve}
    \label{eq:rTsoln}
    \mathbb{r}_n(b) & =
    \begin{cases}
        \frac{\left(1 - b^{\frac{\nu + 2}{2}}\right)}{2}
        \mathbb{L}_{\frac{\mu}{2}, \frac{\nu}{2}}
         &
        \qquad
        \frac{\mu}{2} \ \text{even}
        \\[1em]
        \mathbb{k}_{\frac{\nu - 1}{2}}(b) \mathbb{M}_{\frac{\mu - 1}{2}, \frac{\nu - 1}{2}}
         &
        \qquad
        \frac{\mu - 1}{2} \ \text{even}
        \\[1em]
        0
         &
        \qquad
        \text{otherwise}
    \end{cases}
\end{proof}
where the components of $\mathbbb{k}$, $\mathbbb{L}$, and $\mathbbb{M}$
are given by
\begin{proof}{PhaseCurve}
    \label{eq:HJK}
    \mathbb{k}_{j}(b) &= \int_b^1 a^j \sqrt{1 - a^2} \dd a
    \nonumber \\
    \mathbb{L}_{i,j} &=
    \frac{
        \Gamma\left(\frac{i + 1}{2}\right)
        \Gamma\left(\frac{j + 1}{2}\right)
    }
    {
        \Gamma\left(\frac{i + j + 4}{2}\right)
    }
    \nonumber \\
    \mathbb{M}_{i,j} &=
    \frac{
        \Gamma\left(\frac{i + 1}{2}\right)
        \Gamma\left(\frac{j + 4}{2}\right)
    }
    {
        \Gamma\left(\frac{i + j + 5}{2}\right)
    }
    \quad,
\end{proof}
where $\Gamma$ is the gamma function.
Given initial conditions
\\[1em]
\begin{minipage}{.33\linewidth}
    \begin{align}
        \mathbb{k}_{0}(b) & = \frac{\arccos(b) - bb_c}{2}
        \nonumber                                         \\
        \mathbb{k}_{1}(b) & = \frac{b_\mathrm{c}^3}{3}
        \nonumber
    \end{align}
\end{minipage}%
\begin{minipage}{.32\linewidth}
    \begin{align}
        \mathbb{L}_{0,0} & = \pi
        \nonumber                        \\
        \mathbb{L}_{0,1} & = \frac{4}{3}
        \nonumber
    \end{align}
\end{minipage}%
\begin{minipage}{.33\linewidth}
    \begin{proof}{PhaseCurve}
        \label{eq:IJK0}
        \mathbb{M}_{0,0} &= \frac{4}{3}
        \nonumber \\
        \mathbb{M}_{0,1} &= \frac{3\pi}{8}
    \end{proof}
\end{minipage}
\\[1em]
we may compute all the required higher order terms from the recurrence relations
\begin{proof}{PhaseCurve}
    \label{eq:IJKrec}
    \mathbb{k}_{j}(b) &= \frac{b^{j-1} b_\mathrm{c}^3 + (j - 1) \mathbb{k}_{j - 2}(b)}{j + 2}
    \nonumber \\
    \mathbb{L}_{0,j} &= \left(\frac{j - 1}{j + 2}\right) \mathbb{L}_{0,j-2}
    \nonumber \\
    \mathbb{M}_{0,j} &= \left(\frac{j + 2}{j + 3}\right) \mathbb{M}_{0,j-2}
    \nonumber \\
    \mathbb{L}_{i,j} &= \left(\frac{i - 1}{i + j + 2}\right) \mathbb{L}_{i-2,j}
    \nonumber \\
    \mathbb{M}_{i,j} &= \left(\frac{i - 1}{i + j + 3}\right) \mathbb{M}_{i-2,j}
    \quad.
\end{proof}
Once $\rT$ is known,
the observed total flux in reflected light
is computed from (c.f. Equation~\ref{eq:rTIA1RRy})
\begin{align}
    \label{eq:f0}
    f_0 & =
    \rT(b)
    \mathbf{I}(b, \theta'', r_\mathrm{s})
    \mathbf{A_1}
    \mathbf{R}''(x_\mathrm{s}, y_\mathrm{s})
    \mathbf{R}(\text{I}, \Lambda, \Theta)
    \mathbf{y}
    \quad.
\end{align}
Finally, for future reference, we can also compute what we will call the
\emph{complement} of case 0:
\begin{align}
    \label{eq:f0hat}
    \hat{f}_0 & =
    \Big(
    \mathbf{r}^\top
    -
    \rT(b)
    \Big)
    \mathbf{I}(b, \theta'', r_\mathrm{s})
    \mathbf{A_1}
    \mathbf{R}''(x_\mathrm{s}, y_\mathrm{s})
    \mathbf{R}(\text{I}, \Lambda, \Theta)
    \mathbf{y}
    \quad.
\end{align}
This is the flux contribution from the unphysical night side
(if we were to integrate over it), where
our polynomial illumination (Equation~\ref{eq:illum_poly}) function
yields \emph{negative} intensities. This quantity
will be useful in negating the unphysical contribution
in the integrals of the following section.

\section{The Solution: Occultation}
\label{sec:solution-occ}
\begin{figure}[t!]
    \begin{centering}
        \includegraphics[width=\linewidth]{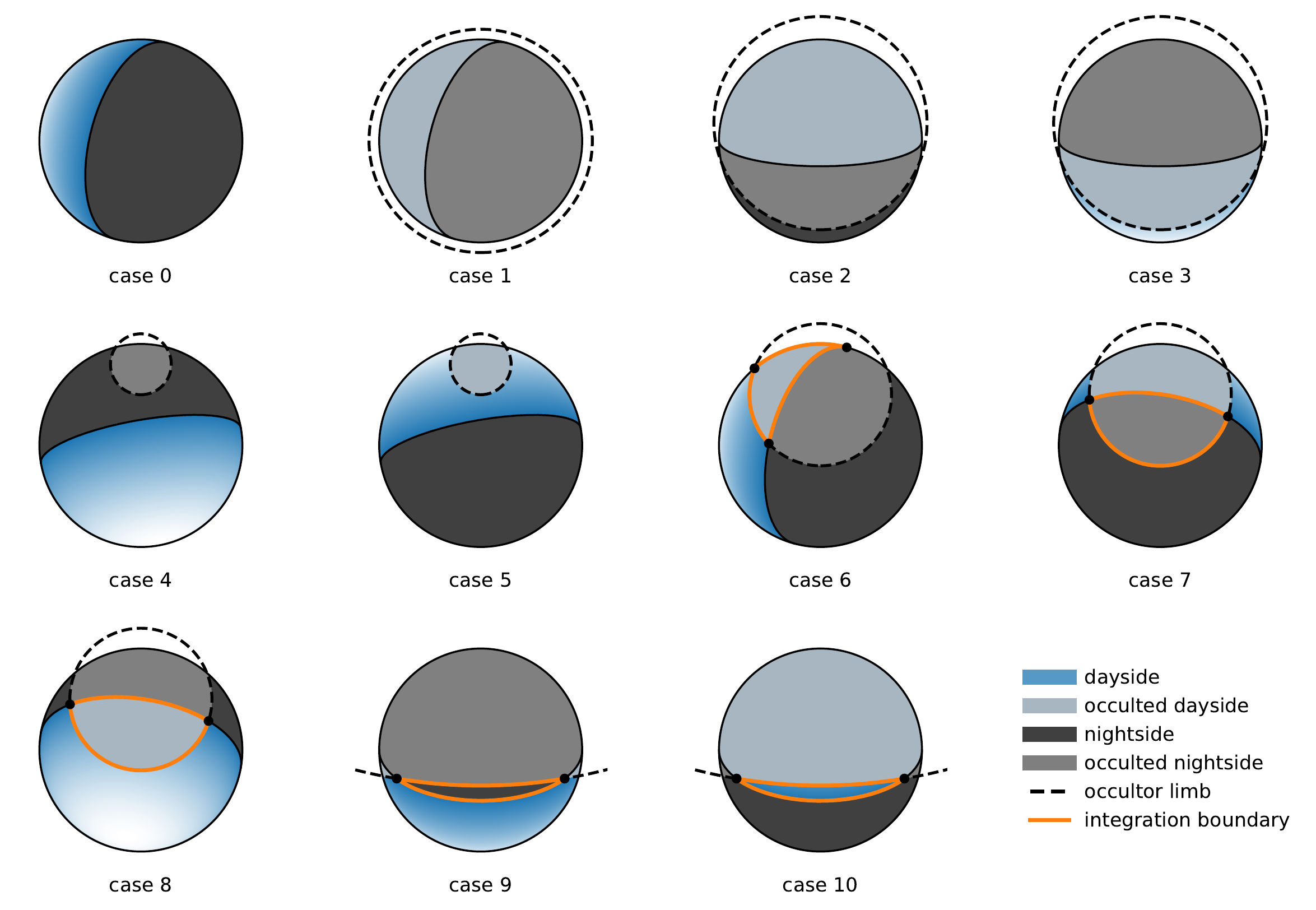}
        \oscaption{cases}{%
            The 10 principal families of cases of occultations in
            reflected light.
            In these figures, the body with the solid outline
            is the one whose flux we are interested in, and the body with the
            dashed outline is the occultor.
            The nightside of the occulted body is colored
            black (dark grey if occulted), and the dayside is colored blue
            (bluish-grey if occulted).
            Case 0 is the unocculted case (\S\ref{sec:solution-no-occ}),
            while cases 1--5 involve
            configurations in which the limb of the occultor does not intersect
            with the terminator at any point, so the visible flux may be
            computed in terms of classical \starry integrals. The remaining
            cases require integration along the orange boundary (the curves
            $\mathcal{P}$, $\mathcal{T}$, and $\mathcal{Q}$ of
            \S\ref{sec:cases-hard}), which
            include the terminator. These involve the evaluation of incomplete
            elliptic integrals and are derived below.
            Note, finally, that there are additional subcases not shown above.
            For instance, cases 4, 5, 7, and 8 also encompass configurations
            in which the occultor does not intersect the limb of the occulted
            body. However, as this distinction does not affect the procedure
            for computing the flux in these cases (see text), we omit these
            subcases from the figure.
            \label{fig:cases}
        }
    \end{centering}
\end{figure}

The integration in the unocculted case presented above is relatively
straightforward, since
the boundaries of integration are always the half-ellipse defining the
terminator and the half-circle defining the upper limb of the body
(Equation~\ref{eq:rT}). When an occultor is present, however, the
integration boundaries are far less trivial, since they may or may not
include sections of the terminator, sections of the limb of the body,
and sections of the limb of the occultor. The integration regions may also
be disjoint; for instance, in case 7 of Figure~\ref{fig:cases}, the
portion of the dayside that is unocculted consists of two separate
regions.

Whereas in \citet{Luger2019} we compute the observed flux by always
integrating over the unocculted portion of the disk, here we find that
it is often easier and more computationally efficient to compute the
integral of the intensity over the
\emph{simplest} region -- meaning the one with the fewest boundaries --
and combine it with the formalism from \S\ref{sec:solution-no-occ} to
compute the visible flux. These integrals may be over the unocculted dayside,
the occulted dayside, the unocculted nightside, or the occulted nightside;
in the case of the latter three, a bit of algebra
(\S\ref{sec:cases-hard}--\S\ref{sec:cases-pathological}) is needed
to relate these to the observed flux. Additionally, in some cases we can
avoid computing new integrals entirely, as the solution can be obtained from
a combination of the classical \starry solution vector $\sT$ and the
formalism from the unocculted case (\S\ref{sec:solution-no-occ}).

After exhaustive experimentation, we identified in total 14 families of
geometrical configurations for the occultation problem,
each defined by a distinct combination of integration
boundaries; these are shown in Figures~\ref{fig:cases} and
\ref{fig:pathological}.
Together, these cases encompass all possible
occultation configurations, for any illumination angle, occultor size, and
occultor position.

Before we discuss how to compute the occultation integrals, we must first
develop a procedure to identify the relevant case given the occultor
impact parameter $b_\mathrm{o} = \sqrt{x_\mathrm{o}^2 + y_\mathrm{o}^2}$ and radius $r_\mathrm{o}$ and
the terminator semi-minor
axis $b$ and angle $\theta'$ in the frame $\mathcal{F}'$.
Then, once the case is determined, we must
identify the relevant integration boundaries, which depend on the points
of intersection between the limb of the body, the limb of the occultor, and
the terminator. We do so in the following sections.

\subsection{Case determination}
\label{sec:which-case}

The key to identifying the case corresponding to a given configuration is
to determine whether or not the limb of the occultor intersects the terminator
of the body, and if so, the points of intersection.
While we perform the integration in frame $\mathcal{F}'$, finding the points
of intersection with the terminator is easier if we temporarily switch to
the frame $\mathcal{F}''$, in which the terminator
is parallel to the $x''$-axis.
In this frame, the equations defining the terminator and the limb of the
occultor are, respectively,
\begin{proof}{quartic}
    y''_1(x'') & = b \sqrt{1 - x''^2}
    \nonumber                                               \\
    y''_2(x'') & = y''_\mathrm{o} \pm \sqrt{r_\mathrm{o}^2 - (x'' - x''_\mathrm{o})^2}
\end{proof}
where
\begin{proof}{quartic}
    x''_\mathrm{o} = b_\mathrm{o}\sin\theta'
    \nonumber \\
    y''_\mathrm{o} = b_\mathrm{o}\cos\theta'
\end{proof}
are the coordinates of the occultor in $\mathcal{F}''$.
We wish to find the vector of $N$ points
$\mathbf{x''} = \left(x_0, x_1, {\cdot\cdot\cdot}, x_{N-1}\right)^\top$
for which
$y''_1(x_n'') - y''_2(x_n'') = 0$. Following \citet{Luger2017}, we may
express this condition as the quartic equation
\begin{proof}{quartic}
    \label{eq:quartic}
    A {x''}^4 + B {x''}^3 + C {x''}^2 + D {x''} + E = 0
\end{proof}
with coefficients
\begin{proof}{quartic}
    \label{eq:quartic-coeffs}
    A &= (1 - b^2)^2
    \nonumber \\
    B &= -4 x''_\mathrm{o} (1 - b^2)
    \nonumber \\
    C &= -2 \bigg(
    b^4
    + r_\mathrm{o}^2
    - 3 {x''_\mathrm{o}}^2
    - {y''_\mathrm{o}}^2
    - b^2 \big(1 + r_\mathrm{o}^2 - {x''_\mathrm{o}}^2 + {y''_\mathrm{o}}^2\big)
    \bigg)
    \nonumber \\
    D &= -4 x''_\mathrm{o} (b^2 - r_\mathrm{o}^2 + {x''_\mathrm{o}}^2 + {y''_\mathrm{o}}^2)
    \nonumber \\
    E &=
    b^4
    - 2 b^2 \big(r_\mathrm{o}^2 - {x''_\mathrm{o}}^2 + {y''_\mathrm{o}}^2\big)
    + \big(r_\mathrm{o}^2 - {x''_\mathrm{o}}^2 - {y''_\mathrm{o}}^2\big)^2
    \quad.
\end{proof}
Although closed-form solutions to quartic equations exist
\citep[see, e.g.,][who solve for the area of overlap between two ellipses
    analytically]{Hughes2011}, they are prone to significant numerical
instabilities. Instead, we solve for the roots of the quartic
numerically by casting it
as an eigenvalue problem \citep[e.g.,][]{Edelman1995} and polish the
results with a few iterations of Newton's method. We find that this is
reasonably computationally efficient and
yields roots with precision within a couple orders of magnitude of machine
epsilon (see \S\ref{sec:performance}).

In general, the quartic defined by Equation~(\ref{eq:quartic}) has
$N=4$ (potentially degenerate) roots, some of which may be complex, and some
of which correspond to intersections with the wrong half of the
terminator ellipse (i.e., the section of the terminator on the far side
of the body). After excluding the unphysical solutions, we are still left with
anywhere between zero and four roots.

Cases with zero roots (case 1 -- case 5) are treated in \S\ref{sec:cases-easy},
while cases with one or two roots (case 6 -- case 10) are treated in
\S\ref{sec:cases-hard}. Cases with three or four roots
(case 11 -- case 14) are rarely encountered
in practice, but are possible for some pathological configurations; these are
treated in \S\ref{sec:cases-pathological}.

\subsection{Cases 1--5}
\label{sec:cases-easy}
Cases 1--5 (see Figure~\ref{fig:cases}) involve configurations in which the
occultor does not intersect with
the terminator of the occulted body, and are therefore fairly
straightforward to solve. In particular, we can use the original emitted
light solution from \citet{Luger2019}, provided we weight the map by our
polynomial illumination function:
\begin{align}
    \label{eq:fI}
    f_\mathrm{I} & =
    \sTe(b_\mathrm{o}, r_\mathrm{o})
    \mathbf{A_2}
    \mathbf{I}(b, \theta', r_\mathrm{s})
    \mathbf{A_1}
    \mathbf{R}'(x_\mathrm{o}, y_\mathrm{o})
    \mathbf{R}(\text{I}, \Lambda, \Theta)
    \mathbf{y}
    \quad,
\end{align}
where $\sTe(b_\mathrm{o}, r_\mathrm{o})$ is the emitted light solution vector
\citep[Equation~26 in][]{Luger2019}. The flux $f_\mathrm{I}$ is the flux one would
measure from a body whose surface map is weighted by the illumination function
$\mathbf{I}(b, \theta', r_\mathrm{s})$ during an occultation. Note that this is not
necessarily the \emph{observed} flux, since this may include the unphysical
negative contribution from the nightside. We must compute the actual
observed flux on a case-by-case basis.

Case 1 corresponds to any complete occultation of the body
($b_\mathrm{o} \le r_\mathrm{o} - 1$), so the solution for the flux is trivial:
\begin{align}
    \label{eq:f1}
    f_1 = 0
    \quad.
\end{align}
Case 2 corresponds to occultations in which the occultor blocks \emph{all} of
the dayside of the body and \emph{some} of the nightside. In this
configuration, the unocculted part of the disk consists only of nightside, so
the solution is again trivial:
\begin{align}{}
    \label{eq:f2}
    f_2 = 0
    \quad.
\end{align}
Conversely, case 3 corresponds to occultations in which the occultor blocks
\emph{all} of the nightside of the body and \emph{some} of the dayside.
Since the visible portion of the disk consists only of dayside, we can
simply use the weighted solution in emitted light
(Equation~\ref{eq:fI}):
\begin{align}
    \label{eq:f3}
    f_3 = f_\mathrm{I}
\end{align}
Case 4 involves any occultation in which the occultor blocks \emph{only} the
nightside of the body (regardless of whether or not it intersects with the
limb of the body). Since the nightside intensity is zero everywhere, this case
is also trivial, as the flux is equal to the flux in the no occultation case
(Equation~\ref{eq:f0}):
\begin{align}
    \label{eq:f4}
    f_4 = f_0
    \quad.
\end{align}
Finally, case 5 involves any occultation in which the occultor blocks
\emph{only} the dayside of the body (regardless of whether or not it
intersects with the limb). We first compute the illumination-weighted flux
$f_\mathrm{I}$ as above, then negate the unphysical nightside contribution
using Equation~(\ref{eq:f0hat}):
\begin{align}
    \label{eq:f5}
    f_5 = f_\mathrm{I} - \hat{f}_0
    \quad.
\end{align}

\subsection{Cases 6--10}
\label{sec:cases-hard}

\begin{figure}[t!]
    \begin{centering}
        \includegraphics[width=\linewidth]{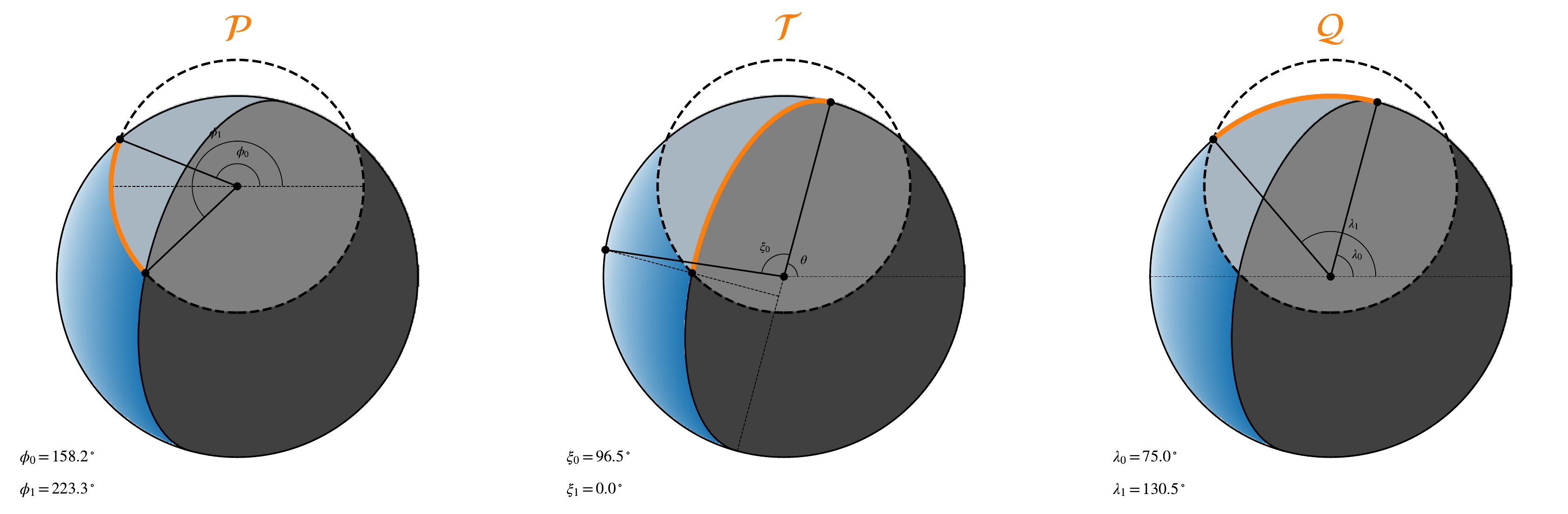}
        \oscaption{geometry}{%
            Geometry of an occultation in reflected light, corresponding
            to case 6 in Figure~\ref{fig:cases}. The surface integral over
            the occulted portion of the dayside (bluish-grey region)
            is computed from the line integrals of the antiderivatives of the
            surface intensity map along the boundary curves
            $\mathcal{P}$, $\mathcal{T}$, and $\mathcal{Q}$. See text for details.
            \label{fig:geometry}
        }
    \end{centering}
\end{figure}

Cases 6--10 (see Figure~\ref{fig:cases}) correspond to configurations in which
the limb of the occultor intersects with the terminator at either one point
(case 6) or two points (cases 7--10). Because of these intersections, we
cannot simply re-weight the emitted light solution, as the integration
boundaries are now different. In general, we may compute the flux by
integrating the components of the Green's basis $\bg$
over the region $S$ bounded by three curves, which we denote
$\mathcal{P}$, $\mathcal{T}$, and $\mathcal{Q}$. These are shown in orange
in Figure~\ref{fig:cases} and presented in more detail in
Figure~\ref{fig:geometry}.
Curve $\mathcal{P}$ is a segment of the limb of
the occultor, parametrized by the angle $\phi \in [\phi_0, \phi_1]$;
curve $\mathcal{T}$ is a segment of the terminator,
parametrized by the angle $\xi \in [\xi_0, \xi_1]$;
and curve $\mathcal{Q}$ is a segment of the limb of the occulted body,
parametrized by the angle $\lambda \in [\lambda_0, \lambda_1]$.
The endpoints $\phi_0, \phi_1, \xi_0, \xi_1, \lambda_0,$ and
$\lambda_1$ are functions of the solutions to the quartic from
\S\ref{sec:which-case} and will be presented in \S\ref{sec:sT}.

Let $\sT$ be the integral of $\bg^\top$ over $S$:
\begin{align}
    \label{eq:sTint}
    \sT(b, \theta', b_\mathrm{o}, r_\mathrm{o}) & =
    \iint\limits_{S(b, \theta', b_\mathrm{o}, r_\mathrm{o})}
    \bg^\top(x', y')
    \ \dd x' \ \dd y'
    \quad,
\end{align}
We defer the solution to Equation~(\ref{eq:sTint}) to \S\ref{sec:sT} below,
as it is quite lengthy. Given $\sT$, the flux $f_\mathrm{S}$ over the integration
region $S$ is computed from Equation~(\ref{eq:sTA2IA1RRy}):
\begin{align}
    \label{eq:fS}
    f_\mathrm{S} & =
    \sT(b, \theta', b_\mathrm{o}, r_\mathrm{o})
    \mathbf{A_2}
    \mathbf{I}(b, \theta', r_\mathrm{s})
    \mathbf{A_1}
    \mathbf{R}'(x_\mathrm{o}, y_\mathrm{o})
    \mathbf{R}(\text{I}, \Lambda, \Theta)
    \mathbf{y}
    \quad.
\end{align}
Note again that this is not necessarily the \emph{observed} flux, which we must
compute on a case-by-case basis below.

Case 6 corresponds to configurations in which the limb of the occultor
intersects the terminator at a single point. The integration region
(see Figures~\ref{fig:cases} and \ref{fig:geometry}) is the occulted
portion of the dayside, which is bounded by all three curves $\mathcal{P}$, $\mathcal{Q}$, and $\mathcal{T}$.
The total flux may
be computed by subtracting the occulted flux $f_\mathrm{S}$ from the total dayside
flux $f_0$:
\begin{align}
    \label{eq:f6}
    f_6 = f_0 - f_\mathrm{S}
    \quad.
\end{align}
Cases 7--10 involve two points of intersection between the occultor limb and
the terminator.
Cases 7 and 8 correspond to occultors that block some of the
nightside and some of the dayside, but \emph{neither} of the extrema of the
terminator ellipse. In case 7 a lens-shaped region is formed by the
intersection of the occultor limb and the terminator on the
\emph{nightside}, while in case 8 this region is formed on the \emph{dayside}.
In case 7, we begin by computing the
flux over the unocculted region, $f_\mathrm{I}$, which includes the spurious
nightside contribution. We then remove this contribution by noting that it
is equal to the total nightside contribution, $\hat{f}_0$, minus the
occulted nightside flux, $f_\mathrm{S}$:
\begin{align}
    \label{eq:f7}
    f_7 = f_\mathrm{I} - (\hat{f}_0 - f_\mathrm{S})
    \quad.
\end{align}
Case 8, on the other hand, is equivalent to case 6, since
the integration region consists of occulted dayside:
\begin{align}
    \label{eq:f8}
    f_8 = f_0 - f_\mathrm{S}
    \quad.
\end{align}
Cases 9 and 10 correspond to occultors that also block some nightside and some
dayside, along with \emph{both} of the extrema of the ellipse; these are
therefore exclusively for large occultors ($r_\mathrm{o} > 1$). Case 9
involves occultations in which only a small lens-shaped region of the
nightside is visible. The total flux is the visible dayside plus unphysical
nightside contribution, $f_\mathrm{I}$, minus the nightside contribution, which we
compute from Equation~(\ref{eq:fS}):
\begin{align}
    \label{eq:f9}
    f_9 = f_\mathrm{I} - f_\mathrm{S}
    \quad.
\end{align}
Conversely, case 10 involves occultations in which only a small lens-shaped
region of the dayside is visible. In this case, we may compute the observed
flux from Equation~(\ref{eq:fS}) directly:
\begin{align}
    \label{eq:f10}
    f_{10} = f_\mathrm{S}
    \quad.
\end{align}

\subsection{Cases 11--14}
\label{sec:cases-pathological}

\begin{figure}[t!]
    \begin{centering}
        \includegraphics[width=\linewidth]{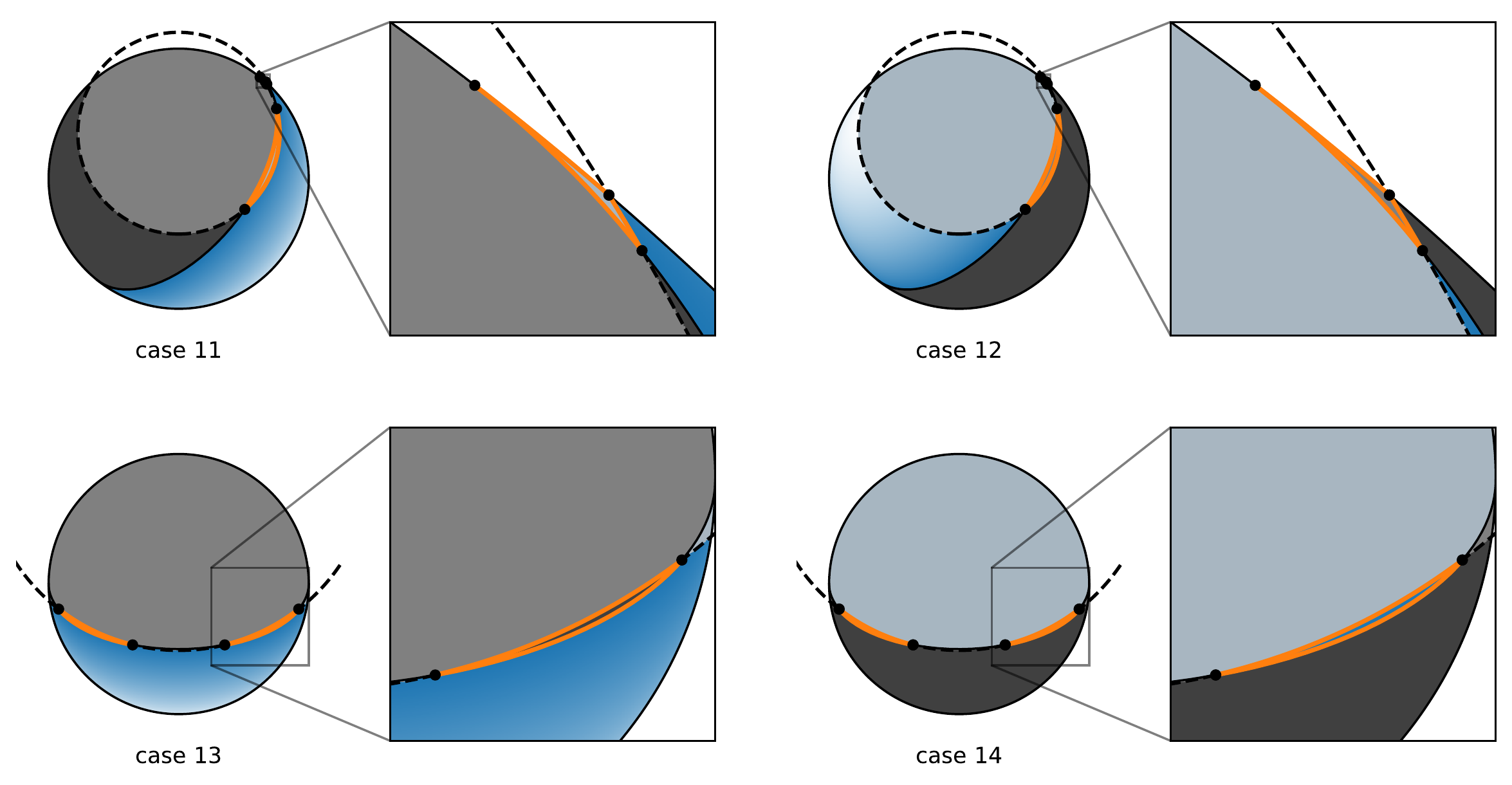}
        \oscaption{pathological}{%
            Four additional families of occultations in reflected light,
            involving rare triple (cases 11 and 12, top) and quadruple
            (cases 13 and 14, bottom) intersections between the limb of the
            occultor and the terminator of the occulted body. All four cases
            involve integration over two disjoint regions (bounded by
            the orange curves in the figure). The insets next to each case
            show a zoomed-in version of four such regions. See text for
            more details.
            \label{fig:pathological}
        }
    \end{centering}
\end{figure}

Cases 11--14 correspond to (rare) configurations involving three
or four roots to Equation~(\ref{eq:quartic}) and are
illustrated in Figure~\ref{fig:pathological}. All four involve integration over
two disjoint regions (see the figure).
Cases 11 and 12 involve three points of intersection between the terminator
and the occultor limb. In case 11, the regions of integration $S_1$ and
$S_2$ are the
occulted portion of the dayside, so the solution is similar to
that of cases 6 and 8:
\begin{align}
    \label{eq:f11}
    f_{11} = f_0 - (f_{\mathrm{S}_1} + f_{\mathrm{S}_2})
    \quad,
\end{align}
where $f_{\mathrm{S}_1}$ and $f_{\mathrm{S}_2}$ are computed from Equation~(\ref{eq:fS}) for
each of the integration regions. Conversely, in case 12 the two regions
are the occulted portion of the nightside, so the solution is similar to that
of case 7:
\begin{align}
    \label{eq:f12}
    f_{12} = f_\mathrm{I} - \big(\hat{f}_0 - (f_{\mathrm{S}_1} + f_{\mathrm{S}_2})\big)
    \quad,
\end{align}
Finally, cases 13 and 14 involve four points of intersection between the
terminator and the occultor limb. The regions of integration in case 13 are
the visible portion of the nightside, so this case is equivalent to case 9:
\begin{align}
    \label{eq:f13}
    f_{13} = f_\mathrm{I} - (f_{\mathrm{S}_1} + f_{\mathrm{S}_2})
    \quad.
\end{align}
Conversely, the regions of integration in case 14 are the
visible portion of the dayside, so this case is equivalent to case 10:
\begin{align}
    \label{eq:f14}
    f_{14} = f_{\mathrm{S}_1} + f_{\mathrm{S}_2}
    \quad.
\end{align}

\subsection{Computing the integrals $\sT$}
\label{sec:sT}
In the previous sections, we discussed how to identify the case
corresponding to a specific configuration of the occultor and the
illumination source. We showed how in some cases
(1--5; \S\ref{sec:cases-easy}) the total flux may be
computed by exploiting the classical \starry integrals (Equation~\ref{eq:fI}).
In all other cases (6--14; \S\ref{sec:cases-hard} and
\S\ref{sec:cases-pathological}), however, the flux computation involves
evaluation of
Equation~(\ref{eq:fS}), where the solution vector $\sT$
is the vector of integrals (in the frame $\mathcal{F}'$)
of each of the terms in the Green's basis $\bg$
over a region $S$ of the
projected disk of the occultor (Equation~\ref{eq:sTint}).
As in \citet{Luger2019}, the approach to computing $\sT$
is to use Green's theorem to transform the surface integrals into line
integrals along the curves $\mathcal{P}$, $\mathcal{T}$, and $\mathcal{Q}$
(see Figure~\ref{fig:geometry}). Specifically, we write%
\footnote{%
    In this section, we deliberately drop the dependence of $\sT$ and
    the primitive
    integrals on the geometrical parameters $b, \theta', b_\mathrm{o}, r_\mathrm{o}$
    for clarity.
}
\begin{align}
    \label{eq:greens}
    \sT
     & =
    \iint\limits_{S}
    \bg^\top(x', y')
    \ \dd x' \ \dd y'
    \nonumber \\[0.5em]
     & =
    \oint \bvec{G}^\top (x', y') \cdot
    \dd \bvec{r} (x', y')
    \quad,
\end{align}
where $\bvec{G} (x', y')$
is a vector of two-dimensional Cartesian vectors chosen such that its
exterior derivative is $\bg$,
\begin{align}
    \label{eq:DGg}
    \frac{\dd \bvec{G}_{y'}(x', y')}{\dd x'}
    - \frac{\dd \bvec{G}_{x'}(x', y')}{\dd y'} = \bg(x', y')
    \quad,
\end{align}
and
\begin{align}
    \dd \mathbf{r} (x', y') & =
    \left(\frac{\dd x'}{\dd \varphi}\right) \dd \varphi \, \xhat' +
    \left(\frac{\dd y'}{\dd \varphi}\right) \dd \varphi \, \yhat'
    \quad,
\end{align}
where $\varphi$ is the parametrized angle along the integration path
and the integral is taken in a counter-clockwise direction relative to
the center of the integration region.
\citet{Luger2019} showed that one possible solution to
Equation~(\ref{eq:DGg}) consists of the vector whose $n^\text{th}$
component is given by
\begin{align}
    \label{eq:G}
    \mathbf{G}_n (x', y') & =
    \begin{dcases}
        x'^{\frac{\mu + 2}{2}}
        y'^{\frac{\nu}{2}}
        \,\yhat'
         & \qquad \mu, \nu \, \text{even}
        \\[1em]
        \frac{1-z(x', y')^3}{3(1-z(x', y')^2)}\bigg(-y' \, \xhat' + x' \, \yhat'\bigg)
         & \qquad \mu = \nu = 1
        \\[1em]
        x'^{l-2}
        z(x', y')^3
        \,\xhat
         & \qquad \nu \, \text{odd}, \,
        \mu = 1, \,
        l \, \text{even}
        \\[1em]
        x'^{l-3}
        y'
        z(x', y')^3
        \,\xhat
         & \qquad \nu \, \text{odd}, \,
        \mu = 1, \,
        l \, \text{odd}
        \\[1em]
        x'^{\frac{\mu-3}{2}}
        y'^{\frac{\nu-1}{2}}
        z(x', y')^3
        \,\yhat
         & \qquad \text{otherwise,}
    \end{dcases}
\end{align}
where the indices $l, m, \mu, \nu$ are given by
Equations~(\ref{eq:l-m}) and (\ref{eq:mu-nu}).

We showed in the previous sections that there are at most three curves
$\mathcal{P}$, $\mathcal{T}$, and $\mathcal{Q}$ bounding a given closed surface of integration
(see Figure~\ref{fig:geometry}). We may therefore express
Equation~(\ref{eq:greens}) as
\begin{proof}{Greens}
    \label{eq:sT}
    \sT & =
    \pT + \tT + \qT
    \quad,
\end{proof}
where we define the primitive integrals%
\footnote{%
    The components of the vectors $\pT$ and
    $\qT$ are analogous to the primitive integrals
    $\mathcal{P}$ and $\mathcal{Q}$
    defined in Equations~(30)--(32) in \citet{Luger2019}, although the
    integration limits of both and the sense of integration of $\mathcal{P}$
    are different.
}
\begin{proof}{Greens}
    \label{eq:pT}
    \pT
    & =
    \int\limits_{\pmb{\phi}}
    \mathbf{G}^\top(x'_\mathrm{p}, y'_\mathrm{p})
    \cdot \dd \mathbf{r}(x'_\mathrm{p}, y'_\mathrm{p})
    \\
    \label{eq:tT}
    \tT
    & =
    \int\limits_{\pmb{\xi}}
    \mathbf{G}^\top(x'_\mathrm{t}, y'_\mathrm{t})
    \cdot \dd \mathbf{r}(x'_\mathrm{t}, y'_\mathrm{t})
    \\
    \label{eq:qT}
    \qT
    & =
    \int\limits_{\pmb{\lambda}}
    \mathbf{G}^\top(x'_\mathrm{q}, y'_\mathrm{q})
    \cdot \dd \mathbf{r}(x'_\mathrm{q}, y'_\mathrm{q})
\end{proof}
to be the line integrals of $\mathbf{G}$ along each of the curves
$\mathcal{P}$, $\mathcal{T}$, and $\mathcal{Q}$, respectively, where the coordinates along each
curve are parametrized
in terms of $\varphi$ as follows:
\\[1em]
\begin{minipage}{0.3\linewidth}
    \begin{align}
        x'_\mathrm{p} & = r_\mathrm{o} \cos\varphi
        \nonumber                                                 \\
        y'_\mathrm{p} & = b_\mathrm{o} + r_\mathrm{o} \sin\varphi
        \nonumber
    \end{align}
\end{minipage}
\begin{minipage}{0.34\linewidth}
    \begin{align}
        x'_\mathrm{t} & = \cos\theta' \cos\varphi - b \sin\theta' \sin\varphi
        \nonumber                                                             \\
        y'_\mathrm{t} & = \sin\theta' \cos\varphi + b \cos\theta' \sin\varphi
        \nonumber
    \end{align}
\end{minipage}
\begin{minipage}{0.3\linewidth}
    \begin{proof}{Greens}
        \label{eq:xy_pqt}
        x'_\mathrm{q} & =\cos\varphi
        \nonumber           \\
        y'_\mathrm{q} & = \sin\varphi
        \quad,
    \end{proof}
\end{minipage}
\\[1em]
and we define
\begin{align}
    \label{eq:vint}
    \int\limits_{\pmb{\varphi}} & \equiv
    \int\limits_{\varphi_{0}}^{\varphi_{1}}
    +
    \int\limits_{\varphi_{2}}^{\varphi_{3}}
    +
    \cdots
    +
    \int\limits_{\varphi_{N - 2}}^{\varphi_{N - 1}}
    \nonumber                            \\
                                & \equiv
    \sum_{i = 0}^{\frac{N}{2} - 1}
    \int\limits_{\varphi_{2i}}^{\varphi_{2i+1}}
\end{align}
to be the sum of definite integrals between pairs of limits $\varphi_i$
arranged in a vector $\pmb{\varphi}$ of length $N$.
For future reference, it will also be useful to define the operator
\begin{align}
    \label{eq:pairdiff}
    \Delta \mathbf{x} \equiv \sum_{i=0}^{\frac{N}{2} - 1}
    \left( x_{2i + 1} - x_{2i} \right)
    \quad,
\end{align}
which sums the difference of successive pairs of values in
a vector
$\mathbf{x} = \left( x_0, x_1, x_2, x_3, {\cdot\cdot\cdot}, x_{N - 1} \right)^\top$.
This will come in handy when computing definite integrals. Specifically,
if $g$ is the antiderivative of some function $f$, we may use the
fundamental theorem of calculus to compute the integral of
$f$ over the interval(s) given
by the vector of limit pairs $\pmb{\varphi}$:
\begin{align}
    \vint{\pmb{\varphi}}{f(\varphi)}
     & = \int\limits_{\varphi_{0}}^{\varphi_{1}} f(\varphi) \dd\varphi
    +
    \int\limits_{\varphi_{2}}^{\varphi_{3}} f(\varphi) \dd\varphi
    +
    {\cdot\cdot\cdot}
    +
    \int\limits_{\varphi_{N - 2}}^{\varphi_{N - 1}} f(\varphi) \dd\varphi
    \nonumber                                                          \\
     & = \Delta \mathbf{g}(\pmb{\varphi})
\end{align}
where $\mathbf{g}$ is the vector given by
\begin{align}
    \mathbf{g}(\pmb{\varphi}) =
    \bigg( g(\varphi_0), g(\varphi_1), g(\varphi_2), g(\varphi_3),
    {\cdot\cdot\cdot}, g(\varphi_{N - 2}), g(\varphi_{N - 1}) \bigg)^\top
    \quad.
\end{align}
Note that most of the cases (1--10) involve integration over a single closed
region, so Equations~(\ref{eq:vint}) and (\ref{eq:pairdiff}) reduce to
\begin{align}
    \int\limits_{\pmb{\varphi}} & \equiv
    \int\limits_{\varphi_{0}}^{\varphi_{1}}
\end{align}
and
\begin{align}
    \Delta \mathbf{x} \equiv x_1 - x_0
    \quad.
\end{align}
For cases 11--14, we must integrate over two disjoint regions, so we
sum over two pairs of limits.
In the next three sections, we derive the solutions to each of the
primitive integrals $\pT$, $\tT$, and $\qT$.

\subsection{The integral along the occultor limb, $\pT$}
\label{sec:pT}
In this section we present a solution to Equation~(\ref{eq:pT}). The first
order of business is to derive expressions for the integration limits
$\pmb{\phi}$. Depending on the integration case, these limits will correspond
to the point of intersection between the limb of the occultor and the
limb of the occulted body and/or the point of intersection between the limb
of the occultor and the terminator of the occulted body.
The former is given by
\citep[c.f. Equation~24 in][]{Luger2019}
\begin{proof}{pT}
    \phi_0 & =
    \frac{\pi}{2} \pm \left(\arcsin\left(\frac{1 - r_\mathrm{o} ^ 2 - b_\mathrm{o} ^ 2}{2 b_\mathrm{o} r_\mathrm{o}}\right) - \frac{\pi}{2}\right)
    \quad,
\end{proof}
where the sign is chosen such that the point
$(r_\mathrm{o}\cos\phi_0, b_\mathrm{o} + r_\mathrm{o}\sin\phi_0)$ is on the dayside of the occulted body,
and the latter (of which there may be multiple) is given by
\begin{proof}{pT}
    \pmb{\phi_1} & =
    \theta' +
    \atantwo
    \left(b\sqrt{1 - {\mathbf{x''}}^2} - y''_\mathrm{o}, \mathbf{x''} - x''_\mathrm{o}\right)
    \quad
\end{proof}
where $\mathbf{x''}$ are the roots of the quartic (Equation~\ref{eq:quartic}).
These angles are then wrapped to the range $[0, 2\pi)$
and sorted into the vector $\pmb{\phi}$ such that
the integration is always performed in a counter-clockwise sense about
the center of the integration region.
The left panel in Figure~\ref{fig:geometry} shows a configuration in which
the lower integration limit $\phi_0 = 158.2^\circ$ corresponds to the
point of intersection between the limbs of the two bodies and the upper
integration limit $\phi_1 = 223.3^\circ$ corresponds to the
limb-terminator intersection. Both angles are measured counter-clockwise
from the line $x' = b_\mathrm{o}$.

In order to evaluate the integral in Equation~(\ref{eq:pT}),
we follow the reparametrization tricks of \S{D.2.3} in \citet{Luger2019}.
The algebra is long and tedious, so we merely present the result
(alongside the usual validation links). The $n^\text{th}$ component
of $\pT$ is
\begin{proof}{pT}
    \label{eq:pTsoln}
    \mathbb{p}_n & =
    \resizebox{.8\hsize}{!}{$
            \begin{cases}
                2(2r_\mathrm{o})^{l+2}
                \begin{cases}
                    V
                    \left(
                    \frac{\mu+4}{4},
                    \frac{\nu}{2},
                    0;
                    \viI
                    \right)
                     & \qquad \qquad \qquad \qquad \qquad
                    \qquad \qquad \qquad \quad
                    \frac{\mu}{2} \, \text{even}
                    \\[0.5em]
                    V
                    \left(
                    \frac{\mu + 2}{4},
                    \frac{\nu}{2},
                    0;
                    \viU
                    \right)
                     & \qquad \qquad \qquad \qquad \qquad
                    \qquad \qquad \qquad \quad
                    \frac{\mu}{2} \, \text{odd}
                \end{cases}
                 & \qquad
                \mu, \nu \, \text{even}
                \\[2em]
                \mathbb{p}_2
                 & \qquad
                \mu = \nu = 1
                \\[1.5em]
                \beta (2r_\mathrm{o})^{l-1}
                \begin{cases}
                    \begin{cases}
                        V
                        \left(
                        \frac{l-2}{2},
                        0,
                        0;
                        \viJ
                        \right) -
                        2
                        V
                        \left(
                        \frac{l-2}{2},
                        0,
                        1;
                        \viJ
                        \right)
                         & \qquad \quad  \enspace
                        l \, \text{even}
                        \\[1em]
                        V
                        \left(
                        \frac{l-3}{2},
                        1,
                        0;
                        \viJ
                        \right) -
                        2V
                        \left(
                        \frac{l-3}{2},
                        1,
                        1;
                        \viJ
                        \right)
                         & \qquad \quad  \enspace
                        l \, \text{odd} \ne 1
                    \end{cases}
                     & \qquad
                    \mu = 1
                    \\[3em]
                    \begin{cases}
                        2
                        V
                        \left(
                        \frac{\mu-1}{4},
                        \frac{\nu-1}{2},
                        0;
                        \viJ
                        \right)
                         & \qquad \qquad \qquad \qquad
                        \frac{\mu - 1}{2} \, \text{even}
                        \\[1em]
                        2
                        V
                        \left(
                        \frac{\mu-1}{4},
                        \frac{\nu-1}{2},
                        0;
                        \viW
                        \right)
                         & \qquad \qquad \qquad \qquad
                        \frac{\mu - 1}{2} \, \text{odd}
                    \end{cases}
                     & \qquad
                    \mu > 1
                \end{cases}
                 & \qquad
                \mu, \nu \, \text{odd}
            \end{cases}
        $}
\end{proof}
where $\beta = \left(1 - (b_\mathrm{o} - r_\mathrm{o})^2\right)^\frac{3}{2}$
and we define the Vieta operator
\begin{proof}{pT}
    \label{eq:V}
    V\left(u, v, w; \mathbf{x}\right) \equiv
    \sum_{i=0}^{u + v}
    \mathcal{A}_{u, v, i}
    x_{u + w + i}
\end{proof}
as the dot product of a vector $\mathbf{x}$ and
the vector of Vieta's theorem coefficients, where
\citep[c.f. Equation~D34 in][]{Luger2019}
\begin{proof}{pT}
    \label{eq:vieta}
    \mathcal{A}_{u,v,i} & =
    \sum_{j=\text{max}(0,u-i)}^{\text{min}(u+v-i,u)}
    \binom{u}{j}
    \binom{v}{u+v-i-j}
    (-1)^{u+j}\left(\frac{b_\mathrm{o}-r_\mathrm{o}}{2r_\mathrm{o}}\right)^{u+v-i-j}
    \quad.
\end{proof}
The vectors $\viI$, $\viJ$,
$\viU$, and $\viW$ are solutions
to specific integrals, which we compute recursively below. As in
\citet{Luger2019} the $n = 2$ term of $\pT$, $\mathbb{p}_2$,
is handled separately; we also compute this below.

Note that several of the cases in Equation~(\ref{eq:pTsoln}) are
identical to those in Equation~(D35) of \citet{Luger2019}, provided
we replace their integrals $\mathcal{I}$ and $\mathcal{J}$ with our
integrals $\viI$ and $\viJ$, respectively. The integrals themselves
are similar, except for a change in the limits of integration, which
are no longer symmetric about zero. As we will see, this leads to the
dependence of these expressions on \emph{incomplete} elliptic integrals.
Note also that the integrals $\viU$ and $\viW$ are new, as certain
cancellations in \citet{Luger2019} resulted in the corresponding cases
contributing zero net flux \citep[last case in Equation~D35 of][]{Luger2019}.

\subsubsection{The vector $\viI$}
\label{sec:I}
The components of the vector $\viI$ are given by the integral
\begin{align}
    \label{eq:I}
    \iI_v(\valpha) & =
    \vint{\valpha}{\sin^{2v}\varphi}
    \quad,
\end{align}
for $v \in [0, \vmax]$,
where we define the helper angle
\begin{align}
    \label{eq:alpha}
    \valpha \equiv \frac{\pmb{\phi}}{2} + \frac{\pi}{4}
    \quad.
\end{align}
The integral in the expression above is the same as that in Equation (D38)
of \citet{Luger2019}, except for a change in the limits of integration.
As in \citet{Luger2019}, we can compute the vector $\viI$ recursively given
a trivial lower boundary condition:
\begin{proof}{I}
    \label{eq:Irec}
    \iI_0(\valpha) &=
    \Delta \valpha
    \nonumber \\
    \iI_v(\valpha) &=
    \frac{1}{2v}
    \bigg(
    (2v - 1) \iI_{v-1}(\valpha) -
    \Delta \left(\sin^{2v - 1}\valpha\cos^{2v -1}\valpha\right)
    \bigg)
\end{proof}
where the last expression is valid for all $v > 0$. We find that this algorithm
is generally stable, except when
$\sin\valpha$ is small.
In that limit, we evaluate $\iI_\vmax(\valpha)$
by numerical integration of
Equation~(\ref{eq:I}) using Gauss-Legendre quadrature with \STARRYQUADPOINTS
points. We then recurse downward by substituting $v \rightarrow v + 1$ in
Equation~(\ref{eq:Irec}) and solving for $\iI_v(\valpha)$.

\subsubsection{The vector $\viJ$}
\label{sec:J}
The components of the vector $\viJ$ are given by the integral
\begin{align}
    \label{eq:J}
    \iJ_v(k^2, \valpha) =
    \vint{\valpha}{
        \sin^{2v}\varphi
        \left(1 - \frac{\sin^2\varphi}{k^2}\right)^\frac{3}{2}
    }
    \quad,
\end{align}
where
\begin{align}
    \label{eq:k2}
    k^2 & \equiv \frac{1 - r_\mathrm{o}^2 - b_\mathrm{o}^2 + 2 b_\mathrm{o} r_\mathrm{o}}{4 b_\mathrm{o} r_\mathrm{o}}
    \quad.
\end{align}
The integral in this expression is again the same as that in Equation (D39)
of \citet{Luger2019}, except for a change in the limits of integration.
In that paper, we computed all terms
$\{ \iJ_0, {\cdot\cdot\cdot}, \iJ_\vmax \}$ from a three-term
recurrence relation and two boundary conditions. In the case of upward
recursion, the boundary conditions $\iJ_0$ and $\iJ_1$ were
computed analytically from the complete elliptic integrals $K(k^2)$
and $E(k^2)$. In cases where upward recursion was not numerically stable, we
evaluated $\iJ_\vmax$ and $\iJ_{\vmax-1}$
via a quickly convergent series expansion and recursed downward.

In order to solve Equation~(\ref{eq:J}), it is possible to
replace the complete elliptic integrals $K(k^2)$ and $E(k^2)$ in the lower
boundary conditions \citep[Equation D46 in ][]{Luger2019} with the
incomplete elliptic integrals
\begin{align}
    \label{eq:F}
    F(\psi \,|\, m) & \equiv \int_0^{\psi} \frac{\dd \varphi}{\sqrt{1 - m \sin^2 \varphi}}
    \\
    \intertext{and}
    \label{eq:E}
    E(\psi \,|\, m) & \equiv \int_0^{\psi} \sqrt{1 - m \sin^2 \varphi} \, \dd \varphi
    \quad,
\end{align}
which we compute from the $el\mathit{2}$ parametrization of
\citet{Bulirsch1965},
then use the same upward
recursion relation to obtain analytic solutions for all $\iJ_v$:
\begin{proof}{J}
    \label{eq:Jrec}
    \iJ_0(k^2, \valpha) &=
    \frac{1}{3} \bigg(
    2 \left(2 - \frac{1}{k^2}\right) \DE +
    \left(\frac{1}{k^2} - 1\right) \DF +
    \Delta \mathbf{z}_0(k^2, \valpha)
    \bigg)
    \nonumber \\
    \iJ_1(k^2, \valpha) &=
    \frac{1}{15} \bigg(
    \left(-3 k^2 + 13 - \frac{8}{k^2}\right) \DE
    \nonumber \\
    &\qquad\quad\quad\quad\quad\quad\quad\quad\quad
    +
    \left(3 k^2 - 7 + \frac{4}{k^2}\right) \DF +
    \Delta \mathbf{z}_1(k^2, \valpha)
    \bigg)
    \nonumber \\
    \iJ_v(k^2, \valpha) &=
    \frac{1}{2v + 3}
    \bigg(
    2 \left( v + 1 + (v - 1) k^2 \right) \iJ_{v - 1}(k^2, \valpha)
    \nonumber \\
    &\qquad\qquad\quad\quad\quad\quad
    -
    (2v - 3) k^2 \iJ_{v - 2}(k^2, \valpha)
    + \Delta \mathbf{z}_v(k^2, \valpha)
    \bigg)
\end{proof}
where the last expression is valid for all $v > 1$ and
\begin{proof}{J}
    \label{eq:Jrec_z}
    \mathbf{z}_0(k^2, \valpha) & =
    \frac{
        \sin\valpha
        \cos\valpha
        \,
        \mathbf{q}(k^2, \valpha)
    }{
        k^2
    }
    \nonumber\\
    \mathbf{z}_1(k^2, \valpha) & =
    \left(3 \sin^2\valpha + 4 - 6k^2\right)
    \mathbf{z}_0(k^2, \valpha)
    \nonumber\\
    \mathbf{z}_v(k^2, \valpha) & =
    k^2
    \sin^{2v - 3}\valpha
    \cos\valpha
    \,
    \mathbf{q}(k^2, \valpha)^5
    \quad,
\end{proof}
with
\begin{align}
    \label{eq:q}
    \mathbf{q}(k^2, \valpha) = \sqrt{1 - \frac{\sin^2\valpha}{k^2}}
    \quad.
\end{align}
Note that when $k^2 < 1$ we use the reciprocal-modulus transformation
to evaluate the elliptic integrals:
\begin{align}
    F\left(\psi \,\Big|\, \frac{1}{k^2}\right) & =
    k \, F(\beta \,|\, k^2)
    \nonumber                                                                               \\
    E\left(\psi \,\Big|\, \frac{1}{k^2}\right) & =
    \frac{E(\beta \,|\, k^2) - (1 - k^2) F(\beta \,|\, k^2)}{k}
    \\
    \intertext{with}
    \beta                                      & = \arcsin\left( \frac{\sin\psi}{k} \right)
    \quad.
\end{align}

In practice, however, we find that this procedure is even more numerically
unstable than it was in \citet{Luger2019}.
To address this, we express the recurrence structure of the problem as
a tridiagonal system with one lower boundary condition $\iJ_0$
and one upper boundary condition $\iJ_\vmax$:
\begin{proof}{J}
    \label{eq:Jtri}
    \begin{pmatrix}
        a_0 & 1   &     &        &         &         \\
        b_1 & a_1 & 1   &        &         &         \\
            & b_2 & a_2 & 1      &         &         \\
            &     & b_0 & a_3    & 1       &         \\
            &     &     & \ddots & \ddots  & \ddots  \\
            &     &     &        & b_\vmax & a_\vmax
    \end{pmatrix}
    \begin{pmatrix}
        \iJ_1           \\
        \iJ_2           \\
        \iJ_3           \\
        \iJ_4           \\
        \cdot\cdot\cdot \\
        \iJ_{\vmax-1}
    \end{pmatrix}
    =
    \begin{pmatrix}
        c_0 - b_0 \iJ_0 \\
        c_1             \\
        c_2             \\
        c_3             \\
        \cdot\cdot\cdot \\
        c_\vmax - \iJ_\vmax
    \end{pmatrix}
\end{proof}
where the recursion coefficients are given by
\begin{proof}{J}
    \label{eq:Jtri_coeffs}
    a_v(k) &= -2\frac{(v + 1) + (v - 1) k^2}{2v + 3} \nonumber \\
    b_v(k) &= \frac{(2v - 3) k^2}{2v + 3} \nonumber \\
    c_v(k^2, \valpha) &= \Delta
    \bigg(
    \frac{
            \mathbf{z}_v(k^2, \valpha)
        }{
            2v + 3
        }
    \bigg)
    \quad.
\end{proof}
Solving this matrix system yields values for all
intermediate $\{ \iJ_1, {\cdot\cdot\cdot}, \iJ_{\vmax - 1} \}$.
While efficient algorithms exist for solving tridiagonal problems, we obtain
far better numerical stability by instead performing traditional LU
decomposition. We find that this algorithm is stable in all the regimes that we
tested.

We evaluate the upper boundary condition $\iJ_{\vmax}$ by numerical
integration of Equation~(\ref{eq:J}) via Gauss-Legendre quadrature with
\STARRYQUADPOINTS points. While the lower boundary condition may be computed
analytically from Equation~(\ref{eq:Jrec}),
in practice we achieve better precision via numerical
integration (as above), with negligible effects on computational performance.

\subsubsection{The vector $\viU$}
\label{sec:U}
The components of the vector $\viU$ are given by the integral
\begin{align}
    \label{eq:U}
    \iU_v(\valpha) =
    \vint{\valpha}{\cos\varphi\sin^{2v + 1}\varphi}
    \quad.
\end{align}
This integral has an analytic solution for all $v$:
\begin{proof}{U}
    \label{eq:Usol}
    \iU_v(\valpha) &= \frac{\Delta \sin^{2v+2}\valpha}{2v + 2}
    \quad.
\end{proof}

\subsubsection{The vector $\viW$}
\label{sec:W}
The components of the vector $\viW$ are given by the integral
\begin{align}
    \label{eq:W}
    \iW_v(k^2, \valpha) =
    \vint{\valpha}{
        \cos\varphi\sin^{2v + 1}\varphi
        \left(1 - \frac{\sin^2\varphi}{k^2}\right)^\frac{3}{2}
    }
    \quad.
\end{align}
We may compute it by either upward or downward recursion. In both cases,
we compute each of the $\iW_v$ from
\begin{proof}{W}
    \iW_v(k^2, \valpha) = \Delta \mathbf{b}_v(k^2, \valpha)
    \quad.
\end{proof}
In the upward case, we start with the lower boundary conditions
\begin{proof}{W}
    \mathbf{b}_0(k^2, \valpha) &=
    \frac{\sin^2\valpha}{5}
    \left(
    \frac{1 - \mathbf{q}(k^2, \valpha)^3}{1 - \mathbf{q}(k^2, \valpha)^2}
    +
    \mathbf{q}(k^2, \valpha)^3
    \right)
    \nonumber \\
    \mathbf{c}_0(k^2, \valpha) & =
    \sin^4\valpha \,
    \frac{\mathbf{q}(k^2, \valpha)^5}{1 - \mathbf{q}(k^2, \valpha)^2}
    \quad,
\end{proof}
and recurse upward in $\mathbf{b}$ and $\mathbf{c}$ simultaneously:
\begin{proof}{W}
    \mathbf{b}_v(k^2, \valpha) & =
    \frac{1}{2v + 5}
    \left(
    \frac{2 v \sin^2\valpha}{1 - \mathbf{q}(k^2, \valpha)^2}
    \mathbf{b}_{v - 1}(k^2, \valpha)
    - \mathbf{c}_{v - 1}(k^2, \valpha)
    \right)
    \nonumber
    \\
    \mathbf{c}_v(k^2, \valpha) & =
    \sin^2\valpha \, \mathbf{c}_{v - 1}(k^2, \valpha)
\end{proof}
for $v > 0$.
In the case of downward recursion, we start with the upper boundary
conditions
\begin{proof}{W}
    \mathbf{b}_\vmax &=
    \frac{\sin^{2\vmax + 2}\valpha}{4\vmax + 10}
    \left(
    \mathbf{f}_\vmax(k^2, \valpha)
    + 2 \mathbf{q}(k^2, \valpha)^3
    \right)
    \nonumber \\
    \mathbf{c}_\vmax &=
    \frac{1}{2}
    \mathbf{q}(k^2, \valpha)^5
    \sin^{2\vmax}\valpha
    \quad,
\end{proof}
where
\begin{proof}{W}
    \mathbf{f}_v(k^2, \valpha)
    &\equiv
    \frac{3}{v+1}
    \,
    {_2\pmb{F}_1}\left(
    -\frac{1}{2},
    v + 1;
    v + 2;
    1 - \mathbf{q}(k^2, \valpha)^2
    \right)
\end{proof}
and ${_2\pmb{F}_1}(a, b; c; \mathbf{z})$ is the Gauss
hypergeometric function, which we compute via its series
definition.
We recurse downward in $\mathbf{b}$ and $\mathbf{c}$ simultaneously:
\begin{proof}{W}
    \mathbf{b}_v(k^2, \valpha) & =
    \frac{1 - \mathbf{q}(k^2, \valpha)^2}{\sin^2\valpha}
    \left(1 + \frac{5}{2v + 2}\right)
    \mathbf{b}_{v + 1}(k^2, \valpha) +
    \frac{\mathbf{c}_{v + 1}(k^2, \valpha)}{v + 1}
    \nonumber
    \\
    \mathbf{c}_v(k^2, \valpha) & =
    \frac{\mathbf{c}_{v + 1}(k^2, \valpha)}{\sin^2\valpha}
    \quad.
\end{proof}

\subsection{The term $\mathbb{p}_2$}
\label{sec:p2}
The final integral we must solve is that corresponding to
$\mathbb{p}_2$ ($\mu = \nu = 1$). As in \citet{Luger2019}, this is
the integral of the linear limb darkening term,
whose solution must be handled
separately due to the fact that the corresponding antiderivative
in Equation~(\ref{eq:G}) is not a polynomial in $x$, $y$, and $z(x, y)$;
also see \citet{Agol2020}.
The integral we must solve is
\begin{align}
    \label{eq:p2}
    \mathbb{p}_2 & =
    \vint{\pmb{\phi}}{
        \frac{1}{3}
        \left(
        \frac{
            1 - z(r_\mathrm{o}\cos\varphi, b_\mathrm{o} + r_\mathrm{o}\sin\varphi)^3
        }{
            1 - z(r_\mathrm{o}\cos\varphi, b_\mathrm{o} + r_\mathrm{o}\sin\varphi)^2
        }
        \right)
        \left(r_\mathrm{o}^2 + b_\mathrm{o} r_\mathrm{o} \sin\varphi\right)
    }
    \quad,
\end{align}
where $z$ is the usual Cartesian coordinate (Equation~\ref{eq:z}).
The solution is tricky, but fortunately a similar integral was solved in
Equation~(34) of \citet{Pal2012}. Adapting their solution to our formalism,
we obtain
\begin{proof}{p2}
    \label{eq:p2_soln}
    \mathbb{p}_2 & =
    \frac{1}{3}
    \Big(
    c_0 +
    c_1 \DF +
    c_2 \DE +
    c_3 \DPi
    \Big)
\end{proof}
where
\begin{proof}{p2}
    c_0 &=
    \Delta
    \Bigg\{
    -\atantwo\left(
    -(b_\mathrm{o} - r_\mathrm{o}) \cos\valpha, (b_\mathrm{o} + r_\mathrm{o}) \sin\valpha
    \right)
    + \valpha
    \nonumber \\
    &\qquad\enspace\enspace
    - \frac{4}{3} b_\mathrm{o} r_\mathrm{o}
    \sin\valpha \cos\valpha
    \sqrt{1 - (b_\mathrm{o} - r_\mathrm{o})^2 - 4 b_\mathrm{o} r_\mathrm{o} \sin^2\alpha}
    \nonumber \\
    &\qquad\enspace\enspace
    + \pmb{\delta}(b_\mathrm{o}, r_\mathrm{o}, \valpha)
    \Bigg\}
    \nonumber \\[0.5em]
    c_1 &=
    \frac{1 + b_\mathrm{o}^4 - b_\mathrm{o}^2(5 + 2 r_\mathrm{o}^2) + r_\mathrm{o}^4 + r_\mathrm{o}^2}{3 \sqrt{1 - (b_\mathrm{o} - r_\mathrm{o})^2}}
    \nonumber \\[0.5em]
    c_2 &= \frac{(b_\mathrm{o}^2+7r_\mathrm{o}^2-4)\sqrt{1 - (b_\mathrm{o} - r_\mathrm{o})^2}}{3}
    \nonumber \\[0.5em]
    c_3 &= \frac{b_\mathrm{o} + r_\mathrm{o}}{(b_\mathrm{o} - r_\mathrm{o})\sqrt{1 - (b_\mathrm{o} - r_\mathrm{o})^2}}
    \quad,
    \\
    \intertext{and}
    \delta(b_\mathrm{o}, r_\mathrm{o}, \alpha) &=
    \begin{cases}
        -2\pi
         &
        \qquad
        \alpha > \frac{3\pi}{2} \,\, \text{and} \,\, b_\mathrm{o} > r_\mathrm{o}
        \\
        +2\pi
         &
        \qquad
        \alpha > \frac{3\pi}{2} \,\, \text{and} \,\, b_\mathrm{o} < r_\mathrm{o}
        \\
        0
         &
        \qquad
        \text{otherwise}
        \quad.
    \end{cases}
\end{proof}
The quantities $\pmb{F}(\valpha \,|\, \nicefrac{1}{k^2})$ and
$\pmb{E}(\valpha \,|\, \nicefrac{1}{k^2})$
are the same incomplete elliptic integrals as those in
\S\ref{sec:J}, while
\begin{align}
    \label{eq:Pi}
    \Pi(n; \psi \,|\, m) & \equiv
    \int_0^{\psi}
    \frac{\dd \varphi}{(1 - n \sin^2\varphi)\sqrt{1 - m \sin^2 \varphi}}
\end{align}
is the incomplete elliptic integral of the third
kind, with
\begin{proof}{p2}
    \label{eq:n}
    n & = -\frac{4 b_\mathrm{o} r_\mathrm{o}}{(r_\mathrm{o} - b_\mathrm{o})^2}
    \quad.
\end{proof}

While stable algorithms exist to evaluate $\Pi(n; \psi \,|\, m)$
\citep[e.g.][]{Bulirsch1969}, we find that the parametrization above has
poor numerical stability, particularly in the vicinity of the singular
points $b_\mathrm{o} = r_\mathrm{o}$ and $b_\mathrm{o} = 1 + r_\mathrm{o}$. In practice, we find that
numerical evaluation of Equation~(\ref{eq:p2}) via Gaussian quadrature
is more numerically stable and just as computationally efficient as
the procedure outlined above.

\subsection{The integral along the terminator, $\tT$}
\label{sec:tT}
In this section we present a solution to Equation~(\ref{eq:tT}), the
line integral along the day/night terminator of the occulted body. As
before, the first thing we must do is derive expressions for the integration
limits $\pmb{\xi}$.
Depending on the integration case, these limits
will corresponds to the point of intersection between the terminator and
the limb of the occultor and/or the point of intersection between the
terminator and the limb of the occulted body. The former is given by
\begin{proof}{tT}
    \pmb{\xi_0} &=
    \atantwo\left(\sqrt{1 - {\mathbf{x''}}^2}, \mathbf{x''} \right)
\end{proof}
where $\mathbf{x''}$ are the roots of the quartic (Equation~\ref{eq:quartic}).
The latter is given by
\begin{proof}{tT}
    \xi_1 &=
    \begin{cases}
        0   & \qquad \qquad (1 - x_\mathrm{o}'')^2 + {y_\mathrm{o}''}^2 < r_\mathrm{o}^2
        \\
        \pi & \qquad \qquad \text{otherwise}
        \quad.
    \end{cases}
\end{proof}
As before, these angles are then
wrapped to the range $[0, 2\pi)$ and
sorted into the vector
$\pmb{\xi}$ such that the integration is performed counter-clockwise
about the center of the integration region.
The middle panel of Figure~\ref{fig:geometry} shows a case where
$\xi_0 = 96.5^\circ$ corresponds to the point of intersection between the occultor
limb and the terminator and $\xi_1 = 0^\circ$ corresponds to the
point where the terminator extends onto the backside of the body.
Note, importantly, that unlike $\pmb{\phi}$, the angle $\pmb{\xi}$ is not
measured between the horizontal and a point on the curve of $\mathcal{T}$.
Recall that $\pmb{\xi}$ is an angular parameter of the ellipse, so it is
measured in the same way as the eccentric anomaly in a Keplerian orbit:
it is the
angle between the semi-major axis of the ellipse and the perpendicular
projection of a point on the ellipse onto the unit circle
(see Figure~\ref{fig:geometry}).

The solution to Equation~(\ref{eq:tT}) involves repeated application
of the binomial theorem.
If we define the quantities
\begin{proof}{tT}
    \label{eq:Zuvjk}
    Z^{u,v}_{j,k}(b, \theta') & =
    \binom{u}{j}
    \binom{v}{k}
    (-1)^{v-k}
    b^{u+v-j-k}
    \sin^{v+j-k}\theta'
    \cos^{u-j+k}\theta'
    \nonumber
    \\
    \intertext{and}
    \delta(b, \xi) & =
    \begin{cases}
        0                        & \qquad 0 \leq \xi < \frac{\pi}{2}     \\
        \pi                      & \qquad \frac{\pi}{2} \leq \xi < \pi   \\
        2 |b| b_\mathrm{c}       & \qquad \pi \leq \xi < \frac{3\pi}{2}  \\
        \pi + 2 |b| b_\mathrm{c} & \qquad \frac{3\pi}{2} \leq \xi < 2\pi
        \quad,
    \end{cases}
\end{proof}
we may express the solution to the $\mathbb{t}_n$ integral as

\clearpage

\begin{proof}{tT}
    \label{eq:tn}
    \mathbb{t}_n & =
    \begin{cases}
        b\cos\theta'
        \sum\limits_{j=0}^{\frac{\nu}{2}}
        \sum\limits_{k=0}^{\frac{\mu + 2}{2}}
        Z^{\frac{\nu}{2}, \frac{\mu+2}{2}}_{j,k}
        (b, \theta') \,
        \iH_{j + k + 1, l + 1 - j - k}(\pmb{\xi})
        \\
        \quad\quad\quad\quad
        - \, \sin\theta'
        \sum\limits_{j=0}^{\frac{\nu}{2}}
        \sum\limits_{k=0}^{\frac{\mu + 2}{2}}
        Z^{\frac{\nu}{2}, \frac{\mu+2}{2}}_{j,k}
        (b, \theta') \,
        \iH_{j + k, l + 2 - j - k}(\pmb{\xi})
         & \qquad
        \mathrel{\raisebox{1.25em}{$\mu, \nu, \, \text{even}$}}
        \\[3em]
        \frac{1}{3}
        \Delta
        \Big\{
        \arctan\left( \frac{|b|\sin\vxi}{\cos\vxi} \right)
        \\
        \qquad\quad\quad
        - \, \sgn\left({\sin\vxi}\right)
        \left(
        \arctan
        \left(
            \frac{
                \left(\frac{\sin\vxi}{1 + \cos\vxi}\right)^2 + 2 b^2 - 1
            }{2 |b| b_\mathrm{c}}
            \right)
        + |b| b_\mathrm{c} \cos\vxi
        \right)
         & \qquad \mu = \nu = 1
        \\
        \qquad\quad\quad
        + \, \pmb{\delta}(b, \vxi)
        \Big\}
        \\[3em]
        -bb_c^3\sin\theta'
        \sum\limits_{j=0}^{l-2}
        Z^{0, l-2}_{0,j}
        (b, \theta') \,
        \iH_{j + 1, l + 1 - j}(\pmb{\xi})
        \\
        \quad\quad\quad\quad
        - \, b_\mathrm{c}^3\cos\theta'
        \sum\limits_{j=0}^{l-2}
        Z^{0, l-2}_{0,j}
        (b, \theta') \,
        \iH_{j, l + 2 - j}(\pmb{\xi})
         & \qquad
        \mathrel{\raisebox{1.25em}{$\nu \, \text{odd}, \mu = 1, l \, \text{even}$}}
        \\[3em]
        -bb_c^3\sin^2\theta'
        \sum\limits_{j=0}^{l-3}
        Z^{0, l-3}_{0,j}
        (b, \theta') \,
        \iH_{j + 2, l - j}(\pmb{\xi})
        \\
        \quad\quad\quad\quad
        - \, b b_\mathrm{c}^3\cos^2\theta'
        \sum\limits_{j=0}^{l-3}
        Z^{0, l-3}_{0,j}
        (b, \theta') \,
        \iH_{j, l + 2 - j}(\pmb{\xi})
         & \qquad
        \mathrel{\raisebox{0em}{$\nu \, \text{odd}, \mu = 1, l \, \text{odd}$}}
        \\
        \quad\quad\quad\quad
        - \, b_\mathrm{c}^5\sin\theta'\cos\theta'
        \sum\limits_{j=0}^{l-3}
        Z^{0, l-3}_{0,k}
        (b, \theta') \,
        \iH_{j + 1, l + 1 - j}(\pmb{\xi})
        \\[3em]
        bb_c^3\cos\theta'
        \sum\limits_{j=0}^{\frac{\nu}{2}}
        \sum\limits_{k=0}^{\frac{\mu + 2}{2}}
        Z^{\frac{\nu-1}{2}, \frac{\mu - 3}{2}}_{j,k}
        (b, \theta') \,
        \iH_{j + k + 1, l + 1 - j - k}(\pmb{\xi})
        \\
        \quad\quad\quad\quad
        - \, b_\mathrm{c}^3\sin\theta'
        \sum\limits_{j=0}^{\frac{\nu}{2}}
        \sum\limits_{k=0}^{\frac{\mu + 2}{2}}
        Z^{\frac{\nu-1}{2}, \frac{\mu - 3}{2}}_{j,k}
        (b, \theta') \,
        \iH_{j + k, l + 2 - j - k}(\pmb{\xi})
         & \qquad
        \mathrel{\raisebox{1.25em}{$\text{otherwise}$}}
    \end{cases}
\end{proof}

\vfill
\clearpage

The solution to Equation~(\ref{eq:tn}) depends on the matrix
$\viH$, whose components are given by the integral
\begin{align}
    \label{eq:H}
    \iH_{u,v}(\vxi) & =
    \vint{\vxi}{
        \cos^u\varphi
        \sin^v\varphi
    }
    \quad.
\end{align}
The $\viH$ integral is the same as that in Equation (D27)
of \citet{Luger2019}, except for a change in the limits of integration.
We can compute this integral recursively given four lower boundary conditions:
\begin{proof}{H}
    \label{eq:Hlower}
    \iH_{0,0}(\vxi) &= \Delta \vxi
    \nonumber \\
    \iH_{1,0}(\vxi) &= \Delta \sin\vxi
    \nonumber \\
    \iH_{0,1}(\vxi) &= -\Delta \cos\vxi
    \nonumber \\
    \iH_{1,1}(\vxi) &= -\frac{\Delta\cos^2\vxi}{2}
    \quad.
\end{proof}
The remaining terms may be computed by upward recursion using the
relations
\begin{proof}{H}
    \label{eq:Hrec1}
    \iH_{u,v}(\vxi) &=
    \frac{
        -\Delta \left(
        \cos^{u + 1} \vxi
        \sin^{v - 1} \vxi
        \right)
        +(v - 1)\iH_{u,v - 2}(\vxi)
    }{u + v}
\end{proof}
for $u < 2, v \ge 2$ and
\begin{proof}{H}
    \label{eq:Hrec2}
    \iH_{u,v}(\vxi) &=
    \frac{
        \Delta \left(
        \cos^{u - 1} \vxi
        \sin^{v + 1} \vxi
        \right)
        + (u - 1)\iH_{u - 2,v}(\vxi)
    }{u + v}
\end{proof}
for all remaining terms.

\subsection{The integral along the occulted body limb, $\qT$}
\label{sec:qT}
The final line integral we must solve is the integral along the boundary
of the occulted body,  Equation~(\ref{eq:qT}). Fortunately, this is
also the easiest of the three.
The limits of integration $\vlambda$
correspond to the point at which the terminator crosses from the dayside
to the night side,
\begin{proof}{qT}
    \lambda_0 &=
    \begin{cases}
        \theta'        & \qquad \qquad \cos^2\theta' + (\sin\theta' - b_\mathrm{o})^2 < r_\mathrm{o}^2
        \\
        \theta ' + \pi & \qquad \qquad \text{otherwise}
        \quad.
    \end{cases}
\end{proof}
and the point of intersection between the limb of the occultor
and the limb of the occulted body,
\begin{proof}{qT}
    \lambda_1 & =
    \frac{\pi}{2} \pm \left(\arcsin\left(\frac{1 - r_\mathrm{o} ^ 2 + b_\mathrm{o} ^ 2}{2 b_\mathrm{o}}\right) - \frac{\pi}{2}\right)
    \quad,
\end{proof}
where the sign is chosen such that the point
$(\cos\lambda_1, \sin\lambda_1)$ is on the dayside of the occulted body.
As before, these angles are
wrapped to the range $[0, 2\pi)$ and
placed in the vector
$\lambda$ such that the line integral is taken in the counter-clockwise
direction about the center of the integration region. The right panel
of Figure~\ref{fig:geometry} shows a case where
$\lambda_0 = 75^\circ$ and $\lambda_1 = 130.5^\circ$. Both angles are
measured counter-clockwise from the $x'$-axis.

Given $\vlambda$, the solution to Equation~(\ref{eq:qT}) is straightforward:
\begin{proof}{qT}
    \mathbb{q}_n &=
    \begin{cases}
        \iH_{\frac{\mu + 4}{2}, \frac{\nu}{2}}(\vlambda)
                                  & \qquad \mu, \nu, \, \text{even}
        \\[1em]
        \frac{1}{3}\Delta\vlambda & \qquad \mu = \nu = 1
        \\[1em]
        0                         & \qquad \text{otherwise}
        \quad,
    \end{cases}
\end{proof}
where the matrix $\viH$ is given by Equation~(\ref{eq:H}).

\section{Caveats}
\label{sec:cases-missing}
The \starry code includes a large suite of unit tests that compare
the flux computations to numerical models and to various benchmarks
for a wide variety of inputs. While we have done our best to
develop tests over the full range of occultation configurations,
there may be edge cases in which the \starry algorithm fails and returns
the wrong flux. This could happen, for instance, if the quartic root
solver (\S\ref{sec:which-case}) fails to find the points of intersection
between the occultor and the day/night terminator, leading to
an incorrect case identification (Figure~\ref{fig:cases}) and thus the
wrong value for the flux. In the development of the algorithm, these cases
would occasionally show up as a single, obvious outlier in a light curve
model. All such cases we encountered
have been fixed by adding consistency
checks in the root solver and switching to alternate evaluation methods
near known singularities. However, it is possible that there may still be
rare cases in which this happens, in which case we ask that users
raise an \href{https://github.com/rodluger/starry/issues}{issue} on
GitHub so that we can provide a fix.

\vfill
\pagebreak
\input{table}

\end{document}

%% file: cortex.tex
\usepackage{url}
\usepackage{amsmath}
\usepackage{mathtools}
\usepackage{amssymb}
\usepackage{natbib}
\usepackage{graphicx}
\usepackage{calc}
\usepackage{etoolbox}
\usepackage{xspace}
\usepackage[T1]{fontenc} 
\usepackage{textcomp}
\usepackage{ifxetex}
\ifxetex
  \usepackage{fontspec}
  \defaultfontfeatures{Extension = .otf}
\fi
\usepackage{fontawesome}
\usepackage{listings}
\usepackage{nicefrac}
\usepackage[bb=boondox]{mathalfa}
\usepackage{booktabs}
\usepackage{longtable}

\newcommand{\starry}{\textsf{starry}\xspace}
\newcommand{\Python}{\textsf{Python}\xspace}
\newcommand{\cpp}{\textsf{C}++\xspace}
\newcommand{\bvec}[1]{{\ensuremath{\mathbf{#1}}}}

\DeclarePairedDelimiter\floor{\lfloor}{\rfloor}

\renewcommand{\eqref}[1]{\ref{eq:#1}}

\definecolor{linkcolor}{rgb}{0.1216,0.4667,0.7059}
\newcommand{\codeicon}{{\color{linkcolor}\faCloudDownload}}
\newcommand{\prooficon}{{\color{linkcolor}\faPencilSquareO}}
\input{gitlinks}

\newtagform{eqtag}[]{(}{)}
\newcommand{\currentlabel}{None}
\newenvironment{proof}[1]{%
  \ifstrempty{#1}{%
    \renewtagform{eqtag}[]{\raisebox{-0.1em}{{\color{red}\faPencilSquareO}}\,(}{)}%
  }{%
    \renewtagform{eqtag}[]{\prooflink{#1}\,(}{)}%
  }%
  \usetagform{eqtag}%
  \renewcommand{\currentlabel}{#1}
  \align%
}{%
  \endalign%
  \renewtagform{eqtag}[]{(}{)}%
  \usetagform{eqtag}%
  \message{<<<\currentlabel: \theequation>>>}%
}

\usepackage[skins]{tcolorbox}
\newtcbox{\figtimebox}{enhanced,nobeforeafter,tcbox raise=-0.8mm,boxrule=0.6pt,
  top=0.5mm,bottom=0mm,right=0mm,left=6mm,arc=1pt,boxsep=2pt,
  before upper={\vphantom{dlg}},colframe=linkcolor,coltext=linkcolor,
  fontupper=\sffamily\bfseries\tiny,colback=white,overlay={\begin{tcbclipinterior}
        \fill[linkcolor] (frame.south west)
        rectangle node[text=white,font=\sffamily\bfseries\tiny,rotate=0]{CPU}
        ([xshift=6mm]frame.north west);\end{tcbclipinterior}}}
\robustify{\figtimebox}
\newcommand{\figtime}[1]{\IfFileExists{figures/#1.py.time}%
  {%
    \cilink{\figtimebox{\input{figures/#1.py.time}\unskip s}}
  }{}}

\newcommand{\oscaption}[2]{\caption{#2 \codelink{#1} \input{figures/#1.py.time}}}

\definecolor{codegreen}{rgb}{0,0.6,0}
\definecolor{codegray}{rgb}{0.5,0.5,0.5}
\definecolor{codepurple}{rgb}{0.58,0,0.82}
\definecolor{backcolour}{rgb}{0.95,0.95,0.95}
\lstdefinestyle{mystyle}{
  backgroundcolor=\color{backcolour},
  commentstyle=\color{codegreen},
  keywordstyle=\color{magenta},
  numberstyle=\tiny\color{codegray},
  stringstyle=\color{codepurple},
  basicstyle=\small\ttfamily,
  breakatwhitespace=false,
  breaklines=true,
  captionpos=b,
  keepspaces=true,
  numbers=left,
  numbersep=5pt,
  showspaces=false,
  showstringspaces=false,
  showtabs=false,
  tabsize=2,
  aboveskip=1em,
  belowskip=1em,
  keywords=[2]{map},
  keywordstyle=[2]{\color{black!80!black}},
  upquote=true
}
\renewcommand\quad{\hskip\fontdimen3\font}

\usepackage{stackengine,scalerel}
\def\lnlam{\ThisStyle{\ensurestackMath{\stackon[-2.4\LMpt]{%
        \SavedStyle\lambda}{\kern-.5pt\kern\LMpt\rule{1\LMex}{.25pt+.15\LMpt}}}}}

%% file: gitlinks.tex
\newcommand{\codelink}[1]{\href{https://github.com/rodluger/starrynight/blob/dd916843cc70822644d0f63612c104154f32cc7d/tex/figures/#1.py}{\codeicon}\,\,}
\newcommand{\animlink}[1]{\href{https://github.com/rodluger/starrynight/blob/dd916843cc70822644d0f63612c104154f32cc7d/tex/figures/#1.gif}{\animicon}\,\,}
\newcommand{\prooflink}[1]{\href{https://github.com/rodluger/starrynight/blob/dd916843cc70822644d0f63612c104154f32cc7d/tex/proofs/#1.ipynb}{\raisebox{-0.1em}{\prooficon}}}
\newcommand{\cilink}[1]{\href{https://dev.azure.com/rodluger/starrynight/_build}{#1}}

%% file: style.tex
\usepackage{xifthen}
\usepackage{stackengine}
\usepackage{array}
\usepackage{upgreek}
\usepackage[bbgreekl]{mathbbol}
\usepackage{afterpage}

\newcommand{\STARRYQUADPOINTS}{100\xspace}
\newcommand{\STARRYORENNAYARDEG}{5\xspace}
\newcommand{\STARRYORENNAYARNB}{5\xspace}
\newcommand{\STARRYORENNAYARNBC}{4\xspace}
\newcommand{\STARRYANGLETOL}{$10^{-13}$}
\newcommand{\STARRYKTWOONETOL}{$10^{-12}$}
\newcommand{\STARRYTTOL}{$10^{-12}$}
\newcommand{\STARRYBOEQUALSROTOL}{$10^{-8}$}
\newcommand{\STARRYCOMPLETEOCCTOL}{$10^{-8}$}
\newcommand{\STARRYMINSINALPHA}{$10^{-8}$}
\newcommand{\STARRYBZEROTOL}{$10^{-8}$}
\newcommand{\STARRYNOOCCTOL}{$10^{-8}$}
\newcommand{\STARRYGRAZINGTOL}{$10^{-7}$}
\newcommand{\STARRYROOTTOLDUP}{$10^{-7}$}
\newcommand{\STARRYBONETOL}{$10^{-6}$}
\newcommand{\STARRYTHETAUNITRADIUSTOL}{$10^{-5}$}

\newcommand{\vint}[2]{\int\limits_{#1}#2\dd\varphi}
\newcommand{\dd}{\ensuremath{\text{d}}}

\newcommand{\sgn}{{\text{sgn}}}
\newcommand{\atantwo}{{\text{arctan2}}}

\newcommand{\bg}{\ensuremath{\tilde{\mathbf{g}}}}
\newcommand{\bp}{\ensuremath{\tilde{\mathbf{p}}}}
\newcommand{\by}{\ensuremath{\tilde{\mathbf{y}}}}

\newcommand{\xhat}{\ensuremath{\pmb{\hat{x}}}\xspace}
\newcommand{\yhat}{\ensuremath{\pmb{\hat{y}}}\xspace}

\newcommand{\sTe}{\ensuremath{\mathbf{s}^\top}}
\newcommand{\rTe}{\ensuremath{\mathbf{r}^\top}}
\newcommand{\sT}{\ensuremath{\mathbbb{s}^\top}}
\newcommand{\rT}{\ensuremath{\mathbbb{r}^\top}}
\newcommand{\pT}{\ensuremath{\mathbbb{p}^\top}}
\newcommand{\tT}{\ensuremath{\mathbbb{t}^\top}}
\newcommand{\qT}{\ensuremath{\mathbbb{q}^\top}}

\newcommand{\iI}{\ensuremath{\mathbb{i}}}
\newcommand{\iJ}{\ensuremath{\mathbb{j}}}
\newcommand{\iU}{\ensuremath{\mathbb{u}}}
\newcommand{\iW}{\ensuremath{\mathbb{w}}}
\newcommand{\iH}{\ensuremath{\mathbb{H}}}
\newcommand{\viI}{\ensuremath{\mathbbb{i}}}
\newcommand{\viJ}{\ensuremath{\mathbbb{j}}}
\newcommand{\viU}{\ensuremath{\mathbbb{u}}}
\newcommand{\viW}{\ensuremath{\mathbbb{w}}}
\newcommand{\viH}{\ensuremath{\mathbbb{H}}}

\newcommand{\DE}{\Delta \pmb{E}\left(\valpha \,\Big|\, \frac{1}{k^2}\right)}
\newcommand{\DF}{\Delta \pmb{F}\left(\valpha \,\Big|\, \frac{1}{k^2}\right)}
\newcommand{\DPi}{\Delta \pmb{\Pi}\left(n; \valpha \,\Big|\, \frac{1}{k^2}\right)}

\newcommand{\valpha}{\pmb{\alpha}}
\newcommand{\vlambda}{\pmb{\lambda}}
\newcommand{\vxi}{\pmb{\xi}}

\newcommand{\vmax}{{v_\text{max}}}

%% file: table.tex
\begin{center}
    \begin{longtable}{W{c}{1cm} W{l}{9cm} W{l}{2cm}}
        \caption{%
            List of common symbols used in this paper.
        }
        \label{tab:symbols}
        \\
        \toprule
        \multicolumn{1}{c}{\textbf{Symbol}}
         &
        \multicolumn{1}{c}{\textbf{Description}}
         &
        \multicolumn{1}{c}{\textbf{Reference}}
        \\
        \midrule
        \endfirsthead
        \multicolumn{3}{c}%
        {{\bfseries \tablename\ \thetable{}}. (continued from previous page)}
        \\[0.5em]
        \toprule
        \multicolumn{1}{c}{\textbf{Symbol}}
         &
        \multicolumn{1}{c}{\textbf{Definition}}
         &
        \multicolumn{1}{c}{\textbf{Reference}}
        \\
        \midrule
        \endhead
        \bottomrule
        \endfoot
        \endlastfoot
        \midrule
        \multicolumn{3}{c}{\emph{Frames of reference}}
        \\
        \midrule
        $\mathcal{F}_0$
         & frame in which surface map is specified
         & \S\ref{sec:starry-review}
        \\
        $\mathcal{F}$
         & observer (sky) frame
         & \S\ref{sec:starry-review}
        \\
        $\mathcal{F}'$
         & integration frame (occultor present)
         & \S\ref{sec:solution-occ}
        \\
        $\mathcal{F}''$
         & integration frame (no occultor)
         & \S\ref{sec:solution-no-occ}
        \\
        \midrule
        \multicolumn{3}{c}{\emph{Lines \& surfaces}}
        \\
        \midrule
        $S$
         & region of integration enclosed by $\mathcal{PQT}$
         & \S\ref{sec:solution-occ}
        \\
        $\mathcal{P}$
         & integration path along occultor limb
         & \S\ref{sec:solution-occ}
        \\
        $\mathcal{Q}$
         & integration path along body limb
         & \S\ref{sec:solution-occ}
        \\
        $\mathcal{T}$
         & integration path along terminator
         & \S\ref{sec:solution-occ}
        \\
        \midrule
        \multicolumn{3}{c}{\emph{Special functions}}
        \\
        \midrule
        $\atantwo$
         & quadrant-aware arctangent
         & (\ref{eq:theta})
        \\
        $F$
         & incomplete elliptic integral of the first kind
         & (\ref{eq:F})
        \\
        ${_2}F_1$
         & Gauss hypergeometric function
         & \S\ref{sec:W}
        \\
        $E$
         & incomplete elliptic integral of the second kind
         & (\ref{eq:E})
        \\
        $\Gamma$
         & gamma function
         & \S\ref{sec:solution-no-occ}
        \\
        $\Pi$
         & incomplete elliptic integral of the third kind
         & (\ref{eq:Pi})
        \\
        \midrule
        \multicolumn{3}{c}{\emph{Operators \& symbols}}
        \\
        \midrule
        $V$
         & Vieta summation operator
         & (\ref{eq:V})
        \\
        $\Delta$
         & pairwise difference operator
         & (\ref{eq:pairdiff})
        \\
        $\int_{\pmb{\phi}}$
         & vectorized integral
         & (\ref{eq:vint})
        \\
    \end{longtable}
\end{center}


\begin{center}
    \begin{longtable}{W{c}{1cm} W{l}{9cm} W{l}{2cm}}
        \caption{%
            List of common scalar quantities used in this paper.
        }
        \label{tab:scalars}
        \\
        \toprule
        \multicolumn{1}{c}{\textbf{Symbol}}
         &
        \multicolumn{1}{c}{\textbf{Description}}
         &
        \multicolumn{1}{c}{\textbf{Reference}}
        \\
        \midrule
        \endfirsthead
        \multicolumn{3}{c}%
        {{\bfseries \tablename\ \thetable{}}. (continued from previous page)}
        \\[0.5em]
        \toprule
        \multicolumn{1}{c}{\textbf{Symbol}}
         &
        \multicolumn{1}{c}{\textbf{Definition}}
         &
        \multicolumn{1}{c}{\textbf{Reference}}
        \\
        \midrule
        \endhead
        \bottomrule
        \endfoot
        %
        %
        %
        %
        %
        \midrule
        \multicolumn{3}{c}{\emph{Integers}}
        \\
        \midrule
        $l$
         & spherical harmonic degree
         & (\ref{eq:l-m})
        \\
        $m$
         & spherical harmonic order
         & (\ref{eq:l-m})
        \\
        $n$
         & vector index
         & ---
        \\
        $\mu$
         & spherical harmonic index, $\mu = l - m$
         & (\ref{eq:mu-nu})
        \\
        $\nu$
         & spherical harmonic index, $\nu = l + m$
         & (\ref{eq:mu-nu})
        \\
        \midrule
        \multicolumn{3}{c}{\emph{Coordinates}}
        \\
        \midrule
        $x$
         & Cartesian $x$ coordinate on the plane of the sky
         & ---
        \\
        $y$
         & Cartesian $y$ coordinate on the plane of the sky
         & ---
        \\
        $z$
         & Cartesian $z$ coordinate, $z = \sqrt{1 - x^2 - y^2}$
         & (\ref{eq:z})
        \\
        \pagebreak 
        \midrule
        \multicolumn{3}{c}{\emph{Geometrical parameters}}
        \\
        \midrule
        $b$
         & semi-minor axis of terminator ellipse
         & (\ref{eq:b})
        \\
        $b_\mathrm{c}$
         & complement of $b$, $b_\mathrm{c} \equiv \sqrt{1 - b^2}$
         & \S\ref{sec:adapting-starry}
        \\
        $b_\mathrm{o}$
         & occultor impact parameter, $b_\mathrm{o}= \sqrt{x_\mathrm{o}^2 + y_\mathrm{o}^2}$
         & \S\ref{sec:starry-review}
        \\
        $k^2$
         & elliptic parameter
         & (\ref{eq:k2})
        \\
        $\text{I}$
         & body inclination
         & \S\ref{sec:starry-review}
        \\
        $n$
         & elliptic characteristic
         & (\ref{eq:n})
        \\
        $r_\mathrm{o}$
         & occultor radius
         & \S\ref{sec:starry-review}
        \\
        $r_\mathrm{s}$
         & distance to illumination source, $r_\mathrm{s} = \sqrt{x_\mathrm{s}^2 + y_\mathrm{s}^2 + z_\mathrm{s}^2}$
         & \S\ref{sec:adapting-starry}
        \\
        $x_\mathrm{o}$
         & occultor $x$ position
         & \S\ref{sec:starry-review}
        \\
        $x_\mathrm{s}$
         & illumination source $x$ position
         & \S\ref{sec:adapting-starry}
        \\
        $y_\mathrm{o}$
         & occultor $y$ position
         & \S\ref{sec:starry-review}
        \\
        $y_\mathrm{s}$
         & illumination source $y$ position
         & \S\ref{sec:adapting-starry}
        \\
        $z_\mathrm{s}$
         & illumination source $z$ position
         & \S\ref{sec:adapting-starry}
        \\
        $\Lambda$
         & body obliquity
         & \S\ref{sec:starry-review}
        \\
        $\theta$
         & angle of rotation of the terminator ellipse
         & (\ref{eq:theta'}), (\ref{eq:theta''})
        \\
        $\Theta$
         & body rotational phasse
         & \S\ref{sec:starry-review}
        \\
        \midrule
        \multicolumn{3}{c}{\emph{Intensities \& fluxes}}
        \\
        \midrule
        $a$
         & orbital semi-major axis
         & \S\ref{sec:extended}
        \\
        $A$
         & albedo (spherical)
         & (\ref{eq:albedo})
        \\
        $f$
         & reflected flux during an occultation
         & \S\ref{sec:solution-occ}
        \\
        $f_0$
         & reflected flux outside of an occultation
         & (\ref{eq:f0})
        \\
        $\hat{f}_0$
         & complement of reflected flux outside of an occultation
         & (\ref{eq:f0hat})
        \\
        $f_1$--$f_{14}$
         & case-dependent reflected flux during occultation
         & \S\ref{sec:solution-occ}
        \\
        $f_\mathrm{I}$
         & intensity-weighted flux during an occultation
         & (\ref{eq:fI})
        \\
        $f_\mathrm{T}$
         & thermal flux during an occultation
         & (\ref{eq:sTARRy})
        \\
        $f_{\mathrm{T}_0}$
         & thermal flux outside of an occultation
         & (\ref{eq:rTA1Ry})
        \\
        $f_\mathrm{S}$
         & reflected flux over integration region $S$
         & (\ref{eq:fS})
        \\
        $I$
         & polynomial intensity at a point on the surface
         & (\ref{eq:illum_poly})
        \\
        $\mathcal{I}$
         & true intensity at a point on the surface
         & (\ref{eq:illum})
        \\
        $R_\mathrm{p}$
         & planet radius
         & \S\ref{sec:extended}
        \\
        $R_\star$
         & stellar radius
         & \S\ref{sec:extended}
        \\
        $\vartheta_\mathrm{i}$
         & polar angle of incidence
         & Figure~\ref{fig:scattering}
        \\
        $\vartheta_\mathrm{r}$
         & polar angle of reflection
         & Figure~\ref{fig:scattering}
        \\
        $\sigma$
         & Oren-Nayar surface roughness coefficient
         & \S\ref{sec:nonlambertian}
        \\
        $\tau$
         & Angular extent of terminator past $\nicefrac{\pi}{2}$
         & (\ref{eq:tau})
        \\
        $\phi_\mathrm{i}$
         & azimuthal angle of incidence
         & Figure~\ref{fig:scattering}
        \\
        $\phi_\mathrm{r}$
         & azimuthal angle of reflection
         & Figure~\ref{fig:scattering}
        \\
    \end{longtable}
\end{center}

\vfill
\pagebreak

\begin{center}
    \begin{longtable}{W{c}{1cm} W{l}{9cm} W{l}{2cm}}
        \caption{%
            List of common vector quantities used in this paper.
        }
        \label{tab:vectors}
        \\
        \toprule
        \multicolumn{1}{c}{\textbf{Symbol}}
         &
        \multicolumn{1}{c}{\textbf{Description}}
         &
        \multicolumn{1}{c}{\textbf{Reference}}
        \\
        \midrule
        \endfirsthead
        \multicolumn{3}{c}%
        {{\bfseries \tablename\ \thetable{}}. (continued from previous page)}
        \\[0.5em]
        \toprule
        \multicolumn{1}{c}{\textbf{Symbol}}
         &
        \multicolumn{1}{c}{\textbf{Definition}}
         &
        \multicolumn{1}{c}{\textbf{Reference}}
        \\
        \midrule
        \endhead
        \bottomrule
        \endfoot
        %
        %
        %
        %
        %
        \midrule
        \multicolumn{3}{c}{\emph{Bases}}
        \\
        \midrule
        $\bg$
         & Green's basis
         & (\ref{eq:bg})
        \\
        $\bp$
         & polynomial basis
         & (\ref{eq:bp})
        \\
        $\by$
         & spherical harmonic basis
         & (\ref{eq:by})
        \\
        \midrule
        \multicolumn{3}{c}{\emph{Angles \& angular parameters}}
        \\
        \midrule
        $\mathbf{q}$
         & cosine-like parameter of $\valpha$
         & (\ref{eq:q})
        \\
        $\valpha$
         & modified angle along occultor limb
         & (\ref{eq:alpha})
        \\
        $\vlambda$
         & angle along occulted body limb
         & \S\ref{sec:qT}
        \\
        $\pmb{\phi}$
         & angle along occultor limb
         & \S\ref{sec:pT}
        \\
        $\vxi$
         & angle along terminator
         & \S\ref{sec:tT}
        \\
        \midrule
        \multicolumn{3}{c}{\emph{Integrals}}
        \\
        \midrule
        $\rTe$
         & unocculted solution in emitted light
         & \S\ref{sec:starry-review}
        \\
        $\sTe$
         & occultation solution in emitted light
         & \S\ref{sec:starry-review}
        \\
        $\viI$
         & helper integral
         & (\ref{eq:I})
        \\
        $\viJ$
         & helper integral
         & (\ref{eq:J})
        \\
        $\mathbbb{k}$
         & helper integral
         & (\ref{eq:HJK})
        \\
        $\pT$
         & primitive integral
         & (\ref{eq:pT})
        \\
        $\qT$
         & primitive integral
         & (\ref{eq:qT})
        \\
        $\rT$
         & unocculted solution in reflected light
         & (\ref{eq:rT})
        \\
        $\sT$
         & occultation solution in reflected light
         & (\ref{eq:sT})
        \\
        $\tT$
         & primitive integral
         & (\ref{eq:tT})
        \\
        $\viU$
         & helper integral
         & (\ref{eq:U})
        \\
        $\viW$
         & helper integral
         & (\ref{eq:W})
        \\
        \midrule
        \multicolumn{3}{c}{\emph{Other vector quantities}}
        \\
        \midrule
        $\mathbf{a}$
         & vector of albedo values on a discrete surface grid
         & (\ref{eq:Py})
        \\
        $\mathbf{d}$
         & data vector
         & \S\ref{sec:linearity}
        \\
        $\mathbf{f}$
         & vector of flux values
         & \S\ref{sec:linearity}
        \\
        $\mathbf{i}$
         & illumination profile in polynomial basis
         & (\ref{eq:ivec})
        \\
        $\mathbf{x}''$
         & solution to quartic in terminator frame
         & (\ref{eq:quartic})
        \\
        $\mathbf{y}$
         & vector of spherical harmonic coefficients
         & \S\ref{sec:starry-review}
        \\
        $\pmb{\mu}$
         & prior mean
         & \S\ref{sec:linearity}
    \end{longtable}
\end{center}

\vfill
\pagebreak

\begin{center}
    \begin{longtable}{W{c}{1cm} W{l}{9cm} W{l}{2cm}}
        \caption{%
            List of common matrices used in this paper.
        }
        \label{tab:matrices}
        \\
        \toprule
        \multicolumn{1}{c}{\textbf{Symbol}}
         &
        \multicolumn{1}{c}{\textbf{Description}}
         &
        \multicolumn{1}{c}{\textbf{Reference}}
        \\
        \midrule
        \endfirsthead
        \multicolumn{3}{c}%
        {{\bfseries \tablename\ \thetable{}}. (continued from previous page)}
        \\[0.5em]
        \toprule
        \multicolumn{1}{c}{\textbf{Symbol}}
         &
        \multicolumn{1}{c}{\textbf{Definition}}
         &
        \multicolumn{1}{c}{\textbf{Reference}}
        \\
        \midrule
        \endhead
        \bottomrule
        \endfoot
        %
        %
        %
        %
        %
        \midrule
        \multicolumn{3}{c}{\emph{Linear operators}}
        \\
        \midrule
        $\mathbf{A}$
         & change of basis matrix: $\by \rightarrow \bg$
         & \S\ref{sec:starry-review}
        \\
        $\mathbf{A_1}$
         & change of basis matrix: $\by \rightarrow \bp$
         & \S\ref{sec:starry-review}
        \\
        $\mathbf{A_2}$
         & change of basis matrix: $\bp \rightarrow \bg$
         & \S\ref{sec:adapting-starry}
        \\
        $\mathbf{C}$
         & posterior covariance matrix
         & \S\ref{sec:linearity}
        \\
        $\mathbf{I}$
         & illumination operator
         & (\ref{eq:Imat})
        \\
        $\mathbf{P}$
         & pixelization operator
         & (\ref{eq:Py})
        \\
        $\mathbf{P}^+$
         & inverse pixelization operator
         & (\ref{eq:PInv})
        \\
        $\mathbf{R}$
         & rotation matrix: $\mathcal{F}_0 \rightarrow \mathcal{F}$
         & \S\ref{sec:starry-review}
        \\
        $\mathbf{R'}$
         & rotation matrix: $\mathcal{F} \rightarrow \mathcal{F}'$
         & \S\ref{sec:starry-review}
        \\
        $\mathbf{R''}$
         & rotation matrix: $\mathcal{F} \rightarrow \mathcal{F}''$
         & \S\ref{sec:starry-review}
        \\
        $\mathbf{X}$
         & \starry design matrix
         & \S\ref{sec:linearity}
        \\
        $\pmb{\Lambda}$
         & prior covariance matrix
         & \S\ref{sec:linearity}
        \\
        $\pmb{\Sigma}$
         & data covariance matrix
         & \S\ref{sec:linearity}
        \\
        \midrule
        \multicolumn{3}{c}{\emph{Integrals}}
        \\
        \midrule
        $\mathbf{G}$
         & anti-exterior derivative of $\bg$
         & (\ref{eq:G})
        \\
        $\mathbbb{H}$
         & helper integral
         & (\ref{eq:H})
        \\
        $\mathbbb{L}$
         & helper integral
         & (\ref{eq:HJK})
        \\
        $\mathbbb{M}$
         & helper integral
         & (\ref{eq:HJK})
        \\
    \end{longtable}
\end{center}

\begin{center}
    \begin{longtable}{W{c}{1cm} W{l}{11cm} W{l}{1cm}}
        \caption{%
            Tolerance parameters used in the code.
        }
        \label{tab:tolerance}
        \\
        \toprule
        \multicolumn{1}{c}{\textbf{Symbol}}
         &
        \multicolumn{1}{c}{\textbf{Description}}
         &
        \multicolumn{1}{c}{\textbf{Value}}
        \\
        \midrule
        \endhead
        \bottomrule
        \endfoot
        $\epsilon_{0}$
         & If $\xi$ is this close to $\frac{n\pi}{2}$, compute $\mathbb{t}_2$
        in the limit $\xi = \frac{n\pi}{2}$
         & \STARRYANGLETOL
        \\
        $\epsilon_{1}$
         & If $|\sin\theta|$ or $|\cos\theta|$ are less than this value, set
        to this value
         & \STARRYTTOL
        \\
        $\epsilon_{2}$
         & If $k^2$ is within this value of unity, nudge it away
         & \STARRYKTWOONETOL
        \\
        $\epsilon_{3}$
         & If $|b_\mathrm{o} - r_\mathrm{o}|$ is less than this value, nudge $b_\mathrm{o}$
        away from $r_\mathrm{o}$
         & \STARRYBOEQUALSROTOL
        \\
        $\epsilon_{4}$
         & If $b_\mathrm{o}$ is within this amount of $r_\mathrm{o} - 1$, nudge it away
         & \STARRYCOMPLETEOCCTOL
        \\
        $\epsilon_{5}$
         & If $b_\mathrm{o}$ is within this amount of $r_\mathrm{o} + 1$, nudge it away
         & \STARRYNOOCCTOL
        \\
        $\epsilon_{6}$
         & If $|\sin\alpha|$ is less than this value, set to this value
         & \STARRYMINSINALPHA
        \\
        $\epsilon_{7}$
         & If $b$ is within this value of zero, nudge it away
         & \STARRYBZEROTOL
        \\
        $\epsilon_{8}$
         & If $b_\mathrm{o}$ is within this amount of $1 - r_\mathrm{o}$, nudge it away
         & \STARRYGRAZINGTOL
        \\
        $\epsilon_{9}$
         & If two quartic roots are this close, eliminate one of them
         & \STARRYROOTTOLDUP
        \\
        $\epsilon_{10}$
         & If $b$ is within this value of unity, set it to unity
         & \STARRYBONETOL
        \\
        $\epsilon_{11}$
         & If $\theta$ is within this amount of $\frac{\pi}{2}$ when $r_\mathrm{o} = 1$,
        nudge it away
         & \STARRYTHETAUNITRADIUSTOL
    \end{longtable}
\end{center}